\documentclass[12pt]{article}
\usepackage{amssymb,multicol,amscd,amsmath}

\usepackage{hyperref}

\def\theequation{\thesection.\arabic{equation}}

\makeatletter \@addtoreset{equation}{section} \makeatother

\def\ads{AdS_{d}}

\def\be{\begin{equation}}
\def\ee{\end{equation}}

\def\bee{\begin{eqnarray}}
\def\eee{\end{eqnarray}}

\def\bal{\begin{align}}
\def\eal{\end{align}}

\def\ba{\begin{array}}
\def\ea{\end{array}}

\def\nn{\nonumber}

\def\d{\partial}
\def\dps{\displaystyle}
\def\bpsi{\bar{\psi}}

\def\btheta{\bar{\theta}}
\def\bkappa{\bar{\kappa}}

\def\bs{\bar{s}}
\def\bv{\bar{v}}
\def\bta{\bar{\eta}}

\def\bu{ \bar{u}}

\def\rvac{|0\rangle}
\def\lvac{\langle 0|}
\newcommand{\half}{\frac{1}{2}}
\def\lb{\label}
\def\p{\partial}

\def\n{\bar{n}}

\newcommand{\diff}{\mathsf{D}_0}
\newcommand{\extdiff}{{\rm d}}
\newcommand{\Eop}{\mathsf{E}_0}
\newcommand{\der}{\partial}
\newcommand{\ptl}[1]{\frac{\der}{\der#1}}
\newcommand{\pptl}[2]{\frac{\der^2}{\der#1\der#2}}

\def\cA{\mathcal{A}}  
\def\cD{\mathcal{D}} \def\cE{\mathcal{E}} \def\cF{\mathcal{F}}
\def\cG{\mathcal{G}}  
 \def\cK{\mathcal{K}} 
\def\cM{\mathcal{M}}  
\def\cP{\mathcal{P}} \def\cQ{\mathcal{Q}} \def\cR{\mathcal{R}}
\def\cS{\mathcal{S}} \def\cT{\mathcal{T}} 
 \def\cW{\mathcal{W}} 
\def\cY{\mathcal{Y}}

 \newcommand{\oC}{\mathbb{C}}

\newcommand{\ga}{\alpha}
\newcommand{\gb}{\beta}
\newcommand{\gga}{\gamma}   
\newcommand{\gd}{\delta}    \newcommand{\Gd}{\Delta}
\newcommand{\gep}{\epsilon}

\newcommand{\gl}{\lambda}   \newcommand{\Gl}{\Lambda}
\newcommand{\go}{\omega}    \newcommand{\Go}{\Omega}

\newcommand{\um}{{\underline m}}
\newcommand{\un}{{\underline n}}

\begin{document}

\begin{flushright}
{\small FIAN/TD/04-06}
\end{flushright}

\vspace{1mm}

\begin{center}

{\bf \Large Frame-like formulation for free  mixed-symmetry bosonic
massless higher-spin  fields in $AdS_d$}

\vspace{1.2cm}

K.B. Alkalaev\footnote{alkalaev@lpi.ru},
O.V. Shaynkman\footnote{shayn@lpi.ru} and
M.A. Vasiliev\footnote{vasiliev@lpi.ru

\vspace{2mm}
\noindent
This work is partially supported by grants RFBR No 05-02-17654, LSS No
1578.2003.2, INTAS No 03-51-6346. The authors are grateful to R. Metsaev for
the useful discussion.
}

\vspace{5mm}

I.E. Tamm Department of Theoretical Physics, P.N. Lebedev Physical Institute, \\
Leninsky prospect 53, 119991 Moscow, Russia

\vspace{8mm}

\begin{abstract}
In this paper we discuss in detail the frame-like formulation of
free bosonic massless higher-spin fields of general symmetry type in
$AdS_d$, announced recently in \cite{ASV1,ASV2}.
Properties of gauge
invariant and $AdS$ covariant action functionals and their flat limits are carefully analyzed.


\end{abstract}

\end{center}

{\footnotesize \tableofcontents}

\section{Introduction}
\label{Introduction}

 Free field dynamics of massless higher-spin (HS) gauge fields has
 been extensively studied within various approaches. The case of
 totally symmetric massless fields both on the flat and (anti)-de
Sitter backgrounds of any dimensions has been fully investigated
\cite{fronsdal_flat}-\cite{Buchbinder:2004gp}.
Non-symmetric (mixed-symmetry) massless fields were also studied
for decades attracting some more attention in recent years
\cite{AKO}-\cite{Alkalaev:2005kt}.

There are two different approaches to HS massless fields. The metric-like
approach generalizes  the metric formulation of gravity. It was extensively
elaborated both for totally symmetric and for mixed-symmetry HS fields
starting from the original papers of Fronsdal \cite{fronsdal_flat} and
de Wit and Freedman \cite{WF}. The frame-like approach, that generalizes
the Cartan formulation of gravity, was suggested in Ref. \cite{vas_yadfiz}
for $4d$ totally symmetric bosonic and fermionic HS fields and, independently,
in \cite{ardes} for HS fermions.

In this work we consider in detail the frame-like formulation of HS massless field
dynamics presented recently
in \cite{ASV1,ASV2} where manifestly gauge invariant Lagrangian formulation for HS
massless bosonic fields of any symmetry type (\textit{i.e.} of any spin) was announced.
The basic idea of the frame-like formulation is that HS massless fields are described
in terms of differential forms that take values in appropriate
irreducible tensor representations of the $\ads$ algebra $o(d-1,2)$. The formalism of
differential forms provides  clear geometric realization of  $\ads$ HS gauge
symmetries and gauge-invariant HS field strengths. In particular, manifestly $AdS$
covariant and gauge-invariant action functionals for HS fields of any symmetry type are
constructed as specific bilinear combinations of field strengths,
thus generalizing the MacDowell-Mansouri action for gravity \cite{MM} and the
previously known Lagrangian formulation for free totally symmetric HS fields
\cite{V1,LV}.

There are two degenerate cases of bosonic massless fields not considered in
 this paper. First is the case of totally
antisymmetric fields described by $p$-forms including the case of
a scalar field as a zero-form. Here the action does not have a form of
the wedge product of the gauge invariant field strengths, requiring the
Hodge star as is well-known, e.g., from the example of Maxwell theory.
Although this case is also covered by the frame-like formalism
 we do not consider it here because the corresponding model and action is
well known. The second special case is that of self-dual HS fields in $AdS_{1+4k}$.
It requires special consideration that will be given elsewhere.

The frame-like formulation can also
be applied to description of  HS massless fields in Minkowski and de Sitter
backgrounds (for bosons). In the latter case, the $dS_d$ algebra
$o(d,1)$ admits, however,  no
lowest weight (i.e., bounded energy) unitary representations
that implies the instability of the corresponding
theories.  The  frame-like HS theory in Minkowski space can be obtained
in the limit $\Lambda\to 0$ of the
 $\ads$ HS theory, where $\Lambda$ is the cosmological constant.
For mixed-symmetry HS massless fields the flat limit is not
completely trivial because, generically, a given
$\ads$ HS massless field has more degrees of freedom (\textit{i.e.}, less gauge symmetries)
than its flat cousin \cite{Metsaev, BMV}. In other words, a generic irreducible
$AdS$ field decomposes  into a set of massless Minkowski fields in the flat limit.
Alternatively, one can take the flat limit  so that it will exhibit an
enhancement of additional gauge symmetries that gauge away all extra degrees of
freedom of the $AdS$ theory compared to the Minkowski one. As we will
see this flat limit enhancement of gauge symmetries  serves as the guiding
principle that fixes the correct HS Lagrangian in $AdS_d$.

The frame-like formulation is of particular importance
for the study of the non-linear HS theory \cite{fradvas,nonlineq}
 (see also Refs. \cite{V_obz2,V_obz3} for reviews)
because, in first place, it makes nonlinear symmetries
manifest. Since frame-like HS fields are treated  as
gauge connections, they contain information on the structure
of a global HS algebra, which is a specific
infinite-dimensional extension of the $\ads$ spacetime
symmetry. The results of this paper are expected to give an important
information on the structure of an extension
of the nonlinear HS gauge theory of totally symmetric fields
to a HS theory with mixed-symmetry gauge fields. Hopefully,
this analysis will eventually shed light on a symmetric
phase of string theory known to contain mixed-symmetry HS fields which are however massive in its
standard formulation.

The layout of the present paper is as  follows.
In section \ref{sec:review} we review the general structure
of the frame-like formulation of generic massless HS gauge fields in $AdS_d$,
summarizing the main results and ideas.

In section \ref{sec:alg} we collect relevant facts about tensor
representations of (pseudo)\-orthogonal algebras and the compensator formalism.

In section \ref{sec:adsHSfields}  frame-like and metric-like fields are introduced and their
relationship is established. Frame-like fields are described both in anti-de Sitter  $o(d-1,2)$ and Lorentz $o(d-1,1)$
bases. In particular, the dynamical roles of different Lorentz frame-like fields is
explained.

The discussion of general properties of $\ads$ HS action functionals both in metric-like and in frame-like forms is the content of section \ref{sec:HSactions}.
A set of conditions on HS action functionals which single out a physically correct theory are formulated.
The key role in this analysis is played by the analysis of the flat limit $\Lambda=0$ of
the $\ads$ HS field dynamics.

In section \ref{Reconstruction of the action}, $\ads$ HS action is
reformulated in terms of an appropriate fermionic Fock
space.  The problem of finding a free field action is reduced to the analysis of a differential
complex with the derivation $\cQ$ associated with the variation of the action. The key property of  $\cQ$ is
that it is equivalent to a certain de Rham differential.
In subsection \ref{Action from field equations} it is shown how the action
can be reconstructed from field equations in the frame-like formalism.

In section \ref{Higher-Spin action} the field equations are found that describe
correctly the HS gauge fields of general symmetry type in $AdS_d$ and give rise
to the gauge invariant HS action by virtue of the procedure elaborated in subsection
\ref{Action from field equations}.

Explicit check of the flat gauge
symmetry enhancement for the constructed $\ads$ HS
action functional is the content of section \ref{sec:flat}.

Conclusions and outlook are given in section \ref{sec:conc}. The Appendix contains
some relations helpful in the analysis of
the HS field equations.

\vspace{5mm}

Throughout the paper we use the mostly minus signature and adhere notations
$\underline{m},\underline{n} = 0\div d-1\;$ for world indices, $a,b= 0\div d-1$ for tangent Lorentz
$o(d-1,1)$ vector indices and $A,B = 0 \div d $ for tangent $AdS_d$ $o(d-1,2)$ vector
indices. We also use condensed notations of \cite{V1} for a set of antisymmetric or symmetric vector
indices: $a [k]\equiv [a_1 \ldots a_k]$ and $a (k)\equiv (a_1 \ldots a_k)$.
We use the convention that upper (lower) indices denoted by
the same letter are assumed to be symmetrized or antisymmetrized as
$S^{a(2)}=\half (S^{a_1a_2}+S^{a_2a_1})$ or
$A^{a[2]}=\half (A^{a_1a_2}-A^{a_2a_1})$.

\section{Sketch of the frame-like formulation}
\label{sec:review}

$\quad \large{\textmd{F}}$rom the group-theoretical point of view
a free  particle propagating in the $d$-dimensional
anti-de Sitter  spacetime $AdS_d$  corresponds to an irreducible
highest weight unitary module $D(E_0,{\bf s})$ of the $AdS_d$
isometry algebra $o(d-1,2)$. The module is characterized by the
energy $E_0$ and the spin ${\bf s}$ which are
highest weights associated with the maximal compact subalgebra
$o(2)\oplus o(d-1)\subset o(d-1,2)$. The energy $E_0$ is the
weight of $o(2)$ and the spins $\bf s$ are the weights of $
o(d-1)$. The module $D(E_0,{\bf s})$ is induced from the (vacuum)
finite-dimensional weight  $E_0,$ ${\bf s}$ $o(2)\oplus o(d-1)$-module.
Note that, because we do not consider $AdS$ (anti)self-dual fields in
this paper, the corresponding vacuum module
is not necessarily irreducible for $d-1 = 4k$ but
decomposes into the sum of of two submodules with the
positive and negative last weight (if it is different from zero).
Correspondingly, we will assume that
the spins ${\bf s}$ are positive.

In the bosonic case we discuss in this paper, where all spins ${\bf s}$
are integer, vacuum modules can be realized by $o(d-1)$ traceless tensors with
the Young symmetry properties associated with the spins $\bf s$
being  lengths of rows of the corresponding Young tableau.
It is convenient to unify rows of equal lengths into
horizontal blocks as follows

\be \label{summaryvac}
\bigskip
\begin{picture}(100,55)(-1,40)

{\linethickness{.3mm} \put(00,90){\line(1,0){100}}
\put(00,60){\line(1,0){100}} \put(00,40){\line(1,0){60}}
\put(00,00){\line(0,1){90}} \put(100,60){\line(0,1){30}}
\put(60,40){\line(0,1){20}} }

\put(110,72){\footnotesize $p$} \put(70,45){\footnotesize
$\tilde{p}_2$} \put(50,95){\footnotesize $s$}
\put(28,30){\footnotesize $\tilde{s}_2$}

\put(10,30){\circle*{1}} \put(20,30){\circle*{1}}
\put(40,30){\circle*{1}} \put(50,30){\circle*{1}}

\put(10,20){\circle*{1}} \put(20,20){\circle*{1}}
\put(30,20){\circle*{1}} \put(40,20){\circle*{1}}
\put(50,20){\circle*{1}}

\put(10,10){\circle*{1}} \put(20,10){\circle*{1}}
\put(30,10){\circle*{1}} \put(40,10){\circle*{1}}

\put(10,00){\circle*{1}} \put(20,00){\circle*{1}}
\put(30,00){\circle*{1}} \put(40,00){\circle*{1}}


\put(00,80){\line(1,0){100}} \put(00,70){\line(1,0){100}}

\put(10,60){\line(0,1){30}} \put(20,60){\line(0,1){30}}
\put(30,60){\line(0,1){30}} \put(40,60){\line(0,1){30}}
\put(50,60){\line(0,1){30}} \put(60,60){\line(0,1){30}}
\put(70,60){\line(0,1){30}} \put(80,60){\line(0,1){30}}
\put(90,60){\line(0,1){30}}


\put(00,50){\line(1,0){60}}

\put(10,40){\line(0,1){20}} \put(20,40){\line(0,1){20}}
\put(30,40){\line(0,1){20}} \put(40,40){\line(0,1){20}}
\put(50,40){\line(0,1){20}}

\end{picture}
\ee \vspace{0.6cm}

\noindent The uppermost block of length $s$ and height $p$ plays
the distinguished role in the whole analysis.

Massless and
singleton\footnote{Massless gauge fields
correspond to $p<\frac{d-1}{2}$. The case $p=\frac{d-1}{2}$ for
odd $d$ corresponds to singletons which are massless fields on the
boundary of $AdS_d$.} fields on $AdS_d$ are described by UIRs with
lowest energies saturating the unitarity bound $E_0 = E_{0}({\bf
s})$. As shown in \cite{Metsaev}, for bosonic massless gauge fields
\be
\label{unitarybound}
E_0 ({\bf s})= s-p +d -2\;.
\ee
The limiting module $\lim_{E_0 \to E_0({\bf s})} ( D(E_0,{\bf s}))$
necessarily contains null states (that should become negative
states for $E_0 < E_0({\bf s})$) to be factored out to obtain the
irreducible module $D(E_0 ({\bf s}),{\bf s})$. Field-theoretically, this factorization
 manifests gauge symmetry. The irreducible module $D(E_0 ({\bf s}),{\bf s})$
 describes either a gauge massless field with local degrees
of freedom in the $AdS_{d}$ or a singleton field with local degrees
on the boundary of $AdS_{d}$ (i.e., with all bulk degrees of freedom
gauged away \cite{Flato:1986uh}).

In the framework of the frame-like formulation  the dynamics of
massless (gauge) fields is described by a $p$-form
field \cite{ASV1}

\be
\label{summary_omega} \Omega_{(p)}^{I}(x) = \extdiff
x^{\underline{n_1}}\wedge \ldots \wedge  \extdiff
x^{\underline{n}_p} \; \Omega^I_{\underline{n_1}\,\ldots
\,\underline{n}_p}(x)\;
\ee
taking values in the finite-dimensional $o(d-1,2)$-module $I$ described by the $o(d-1,2)$
traceless Young tableau

\be \label{summary_ads}
\bigskip
\begin{picture}(100,65)(-1,40)
{\linethickness{.3mm} \put(00,100){\line(1,0){90}}
\put(00,60){\line(1,0){90}} \put(00,40){\line(1,0){60}}
\put(00,00){\line(0,1){100}} \put(90,60){\line(0,1){40}}
\put(60,40){\line(0,1){20}} }

\put(100,72){\footnotesize $p+1$} \put(70,45){\footnotesize
$\tilde{p}_2$} \put(35,105){\footnotesize $s-1$}
\put(28,30){\footnotesize $\tilde{s}_2$}

\put(10,30){\circle*{1}} \put(20,30){\circle*{1}}
\put(40,30){\circle*{1}} \put(50,30){\circle*{1}}

\put(10,20){\circle*{1}} \put(20,20){\circle*{1}}
\put(30,20){\circle*{1}} \put(40,20){\circle*{1}}
\put(50,20){\circle*{1}}

\put(10,10){\circle*{1}} \put(20,10){\circle*{1}}
\put(30,10){\circle*{1}} \put(40,10){\circle*{1}}

\put(10,00){\circle*{1}} \put(20,00){\circle*{1}}
\put(30,00){\circle*{1}} \put(40,00){\circle*{1}}


\put(00,90){\line(1,0){90}} \put(00,80){\line(1,0){90}}
\put(00,70){\line(1,0){90}}

\put(10,60){\line(0,1){40}} \put(20,60){\line(0,1){40}}
\put(30,60){\line(0,1){40}} \put(40,60){\line(0,1){40}}
\put(50,60){\line(0,1){40}} \put(60,60){\line(0,1){40}}
\put(70,60){\line(0,1){40}} \put(80,60){\line(0,1){40}}


\put(00,50){\line(1,0){60}}

\put(10,40){\line(0,1){20}} \put(20,40){\line(0,1){20}}
\put(30,40){\line(0,1){20}} \put(40,40){\line(0,1){20}}
\put(50,40){\line(0,1){20}}

\end{picture}
\ee \vspace{0.6cm}

\noindent which is obtained from (\ref{summaryvac}) by
 adding the uppermost
(\textit{i.e.} the longest) row of the Young tableau
(\ref{summaryvac}) and then cutting  off the rightmost  (\textit{i.e.}
the shortest) column. Other way around, this rule allows one to
reconstruct  the $o(d-1)$ weights from a $o(d-1,2)$ Young tableau
associated with a given $p$-form gauge field
(\ref{summary_omega}). It is worth to note that since $p\geq
1$, the uppermost horizontal block in
(\ref{summary_ads}) must have at least two rows for the
$o(d-1,2)$ Young tableau to be associated with
one or another massless field.

For example, to describe the spin two massless field, which
corresponds to the $o(d-1)$ tableau
\begin{picture}(12,12)(-1,4)
{\linethickness{0.210mm}
\put(00,10){\line(1,0){10}} 
\put(00,05){\line(1,0){10}} 
\put(00,05){\line(0,1){5}} 
\put(05,05){\line(0,1){5}} 
\put(10,05){\line(0,1){5}}
}
\end{picture}
, one introduces the 1-form gauge field
$\Omega^{AB}_{\underline{n}}(x)=-\Omega^{BA}_{\underline{n}}(x)$
that takes values in the representation
\begin{picture}(12,12)(-1,4)
{\linethickness{0.210mm}
\put(00,15){\line(1,0){5}} 
\put(00,10){\line(1,0){5}} 
\put(00,05){\line(1,0){5}} 
\put(00,05){\line(0,1){10}} 
\put(05,05){\line(0,1){10}} 
}
\end{picture}
of $o(d-1,2)$. This gauge field can be interpreted as the gauge
connection of $o(d-1,2)$. Its decomposition into representations
of the Lorentz algebra $o(d-1,1)\subset o(d-1,2)$ gives rise to
the conventional  frame field and Lorentz spin
connection 1-forms as explained in more detail below.

With $p$-form gauge fields one associates (linearized) curvatures
which are $(p+1)$-forms taking values in the same $o(d-1,2)$-module $I$
\be
\label{summary_curv}
R_{(p+1)}^I=D_0\Omega_{(p)}^I\;, \ee where $D_0T^A={\rm
d}T^A+\Omega_0^A{}_BT^B$ is a $o(d-1,2)$ covariant derivative
evaluated with respect to the background $1$-form connection
$\Omega_0^{AB}$ that satisfies the zero curvature equation
\be
\label{AdS} D_0D_0={\rm
d}\Omega_0^{AB}+\Omega_0^A{}_C\wedge\Omega_0^C{}_B=0
\ee
that can be taken as a definition of $AdS$ space.
Relation (\ref{AdS}) implies that curvatures are invariant under
the gauge transformations  \be \label{summary_trans}
\delta\Omega^I_{(p)}=D_0\xi^I_{(p-1)}\;, \ee where the $(p-1)$-form
$\xi^I_{(p-1)}$ is a gauge parameter. The Bianchi identities take
the form \be \label{summary_bian} D_0R^I_{(p+1)}=0\;. \ee

To elucidate the dynamical content of a theory formulated in terms of the
$p$-form gauge field $\Go_{(p)}^I$, let us decompose   the
$o(d-1,2)$ representation $I$ carried by the tangent indices into
representations of the Lorentz subalgebra $o(d-1,1)\subset
o(d-1,2)$. Schematically,
the result is \be
\label{sumarry_dec} \Omega_{(p)}^I \longrightarrow
\Big(e_{(p)}\;\oplus \;\omega_{(p)}\Big)
\;\oplus\sum\;\;\omega_{(p)}^\prime\;\oplus\sum\;w_{(p)}\;, \ee
where $p$-form gauge fields on the right-hand-side have tangent
indices corresponding to the all possible traceless $o(d-1,1)$
Young tableaux resulting from the $o(d-1,2)$ Young tableau
(\ref{summary_ads}). There is a useful classification of
Lorentz-covariant fields in the decomposition (\ref{sumarry_dec})
according to their different dynamical roles. So we distinguish
between physical field $e_{(p)}$, relevant auxiliary field
$\omega_{(p)}$, irrelevant auxiliary fields $\omega_{(p)}^\prime$,
and extra fields $w_{(p)}$.

The {\it physical field} has tangent indices described by the traceless
$o(d-1,1)$ Young tableau with the minimal possible number of cells
in the decomposition (\ref{sumarry_dec}). This is obtained from
(\ref{summary_ads}) by removing the row of length $s-1$ from the
uppermost horizontal block. Equivalently, the same tableau can be
obtained by removing the column of height $p$ of the Young tableau
(\ref{summaryvac}) so that the length of the uppermost horizontal block
of (\ref{summaryvac}) becomes $s-1$.  The auxiliary fields are
 described by various Young tableaux which differ
from that of the physical field by one additional cell.
The {\it relevant auxiliary field} has an additional cell in the
first column while {\it irrelevant auxiliary fields} have
an additional cell in any other column. All other possible Young
tableaux in the decomposition (\ref{sumarry_dec}) with
 two or more additional cells correspond to {\it extra
fields}.

For the  spin two field example mentioned  above, the
decomposition has the form $\Omega^{AB}\rightarrow e^a \oplus
\omega^{ab}$, where $e^a$ is the frame field (the physical
$1$-form field) and $\omega^{ab}=-\omega^{ba}$ is the Lorentz spin
connection (the relevant auxiliary $1$-form field). Irrelevant and
extra fields are absent in this case.

To extract Lorentz-covariant components
of the $o(d-1,2)$ field in a manifestly $o(d-1,2)$ covariant
manner it is convenient to introduce a compensator field which is an
$o(d-1,2)$ vector $V^A(x)$ normalized as
\be
\label{normal}
V^AV_A=1.
\ee
 The Lorentz
subalgebra $o(d-1,1)\subset o(d-1,2)$ can be identified with the
stability algebra of the compensator, while the
Lorentz-irreducible components can be represented as $o(d-1,2)$
tensors orthogonal to the compensator.

In the case of gravity the decomposition takes the form
$\Omega^{AB} = \omega^{AB}+\lambda (V^AE^B-V^BE^A)$, where
$o(d-1,2)$ covariantized versions of
$e^a$ and $\omega^{ab}$ are defined as $\lambda E^A = D(V^A)$
and $\cD_L V^A\equiv {\rm d}V^A+\omega^{AB}V_B=0$ and $\gl^2=-\Gl$. From the condition (\ref{normal})
it follows  that $E^AV_A=0$. For the linearized gravity the condition $ \omega^{AB} V_A =0$ is also true what
is most evident when $V^A=const$. The decomposition procedure for generic free HS fields is analogous.

The  more traditional metric-like formulation of the HS  field
dynamics results from a partial gauge fixing
of the frame-like formulation considered in the present paper.
In these terms, a mixed-symmetry massless field
is described by a Lorentz-covariant tensor field $\Phi(x)$
that carries Lorentz indices with the symmetry properties of the
 Young tableau (\ref{summaryvac}). It is not traceless however,
satisfying some relaxed tracelessness conditions that generalize the
Fronsdal double tracelessness  conditions
for  symmetric HS gauge fields \cite{fronsdal_flat}.   The metric-like
 field $\Phi(x)$ is a component of
the physical $p$-form gauge field $e_{(p)}$, \textit{i.e.} it is contained
in the tensor
product of the $p$ antisymmetric form (world) indices and the tangent Young tableau indices of $e_{(p)}$

\be \label{summary_metr}
\bigskip
\begin{picture}(390,65)(-1,40)

{\linethickness{.3mm} \put(00,100){\line(1,0){90}}
\put(00,60){\line(1,0){90}} \put(00,40){\line(1,0){60}}
\put(00,00){\line(0,1){100}} \put(90,60){\line(0,1){40}}
\put(60,40){\line(0,1){20}} }

\put(96,76){\footnotesize $p$} \put(65,47){\footnotesize
$\tilde{p}_2$} \put(35,105){\footnotesize $s-1$}
\put(28,30){\footnotesize $\tilde{s}_2$}

\put(10,30){\circle*{1}} \put(20,30){\circle*{1}}
\put(40,30){\circle*{1}} \put(50,30){\circle*{1}}

\put(10,20){\circle*{1}} \put(20,20){\circle*{1}}
\put(30,20){\circle*{1}} \put(40,20){\circle*{1}}
\put(50,20){\circle*{1}}

\put(10,10){\circle*{1}} \put(20,10){\circle*{1}}
\put(30,10){\circle*{1}} \put(40,10){\circle*{1}}

\put(10,00){\circle*{1}} \put(20,00){\circle*{1}}
\put(30,00){\circle*{1}} \put(40,00){\circle*{1}}


\put(00,90){\line(1,0){90}} \put(00,80){\line(1,0){90}}
\put(00,70){\line(1,0){90}}

\put(10,60){\line(0,1){40}} \put(20,60){\line(0,1){40}}
\put(30,60){\line(0,1){40}} \put(40,60){\line(0,1){40}}
\put(50,60){\line(0,1){40}} \put(60,60){\line(0,1){40}}
\put(70,60){\line(0,1){40}} \put(80,60){\line(0,1){40}}


\put(00,50){\line(1,0){60}}

\put(10,40){\line(0,1){20}} \put(20,40){\line(0,1){20}}
\put(30,40){\line(0,1){20}} \put(40,40){\line(0,1){20}}
\put(50,40){\line(0,1){20}}


\put(110,40){$\bigotimes$}


\put(150,40){\footnotesize $p$}

\put(135,22){\line(0,1){40}} \put(145,22){\line(0,1){40}}

\put(135,22){\line(1,0){10}} \put(135,32){\line(1,0){10}}
\put(135,42){\line(1,0){10}} \put(135,52){\line(1,0){10}}
\put(135,62){\line(1,0){10}}


\put(170,42){$=$}

\end{picture}
\begin{picture}(100,65)(190,40)

{\linethickness{.3mm} \put(00,100){\line(1,0){100}}
\put(00,60){\line(1,0){100}} \put(00,40){\line(1,0){60}}
\put(00,00){\line(0,1){100}} \put(100,60){\line(0,1){40}}
\put(60,40){\line(0,1){20}} }

\put(105,76){\footnotesize $p$} \put(64,47){\footnotesize
$\tilde{p}_2$} \put(48,105){\footnotesize $s$}
\put(28,30){\footnotesize $\tilde{s}_2$}

\put(10,30){\circle*{1}} \put(20,30){\circle*{1}}
\put(40,30){\circle*{1}} \put(50,30){\circle*{1}}

\put(10,20){\circle*{1}} \put(20,20){\circle*{1}}
\put(30,20){\circle*{1}} \put(40,20){\circle*{1}}
\put(50,20){\circle*{1}}

\put(10,10){\circle*{1}} \put(20,10){\circle*{1}}
\put(30,10){\circle*{1}} \put(40,10){\circle*{1}}

\put(10,00){\circle*{1}} \put(20,00){\circle*{1}}
\put(30,00){\circle*{1}} \put(40,00){\circle*{1}}


\put(00,90){\line(1,0){100}} \put(00,80){\line(1,0){100}}
\put(00,70){\line(1,0){100}}

\put(10,60){\line(0,1){40}} \put(20,60){\line(0,1){40}}
\put(30,60){\line(0,1){40}} \put(40,60){\line(0,1){40}}
\put(50,60){\line(0,1){40}} \put(60,60){\line(0,1){40}}
\put(70,60){\line(0,1){40}} \put(80,60){\line(0,1){40}}
\put(90,60){\line(0,1){40}}


\put(00,50){\line(1,0){60}}

\put(10,40){\line(0,1){20}} \put(20,40){\line(0,1){20}}
\put(30,40){\line(0,1){20}} \put(40,40){\line(0,1){20}}
\put(50,40){\line(0,1){20}}


\put(110,40){$\bigoplus$} \put(152,50){\footnotesize{other}}
\put(140,40){\footnotesize{components}}

\end{picture}
\ee \vspace{1cm}

\noindent Note that the Young tableaux on the left-hand-side of
(\ref{summary_metr}) are traceless while that one on the right-hand-side,
which is identified with  the metric-like gauge field, is not.
One can see that the generalized Fronsdal double tracelessness
conditions  on the field $\Phi(x)$ that follow from this
construction require that
\begin{itemize}

\item the double contraction of  four
indices of any row of the upper horizontal block is zero;

\item contraction of any two indices that do not belong to
the first horizontal block is zero.

\end{itemize}

For the case of the one-row tableau, that corresponds to a totally symmetric
field, one recovers the usual
Fronsdal double-tracelessness condition \cite{fronsdal_flat}. In the
spin two case, the metric-like field becomes the traceful metric tensor
while the ``other components'' consist of the antisymmetric part of the
frame.

``Other components'' in the tensor product (\ref{summary_metr}) are
compensated by Stueckelberg part of the gauge transformation law
of the physical $p$-form field $e_{(p)}$
 \be
\label{summary_gaugelaw} \delta e_{(p)} = {\cal D}
\varepsilon_{(p-1)} +
\;{\footnotesize{\textrm{Stueckelberg\;\;part}}}
\ee
where ${\cal D}$ is the Lorentz covariant  derivative in the
background $AdS$ gravitational field and  the $(p-1)$-form gauge
parameter $\varepsilon_{(p-1)}$ carries tangent indices
of the same type as the physical $p$-form $e_{(p)}$.
The  $(p-1)$-form gauge parameters of the
 ``Stueckelberg part'' in (\ref{summary_gaugelaw})
carry tangent indices of the same types
as auxiliary  fields, that is with one cell added to the
Young tableau associated with the physical field.
It turns out that all ``other components'' in
(\ref{summary_metr}) can be gauge fixed to zero
by the Stueckelberg gauge transformation so that
the remaining nonzero components belong to the metric-like gauge field $\Phi(x)$.
 The gauge symmetry of the $\Phi(x)$ inherited from
  the transformation law
(\ref{summary_gaugelaw}) is
\be \label{summary_ADSgaug}
\delta\Phi(x) = \Pi ({\cal D} \varepsilon (x))\;,
\ee
where a gauge
parameter $\varepsilon (x)$ carries the indices described by the
Young tableau resulting from that of
 $\Phi(x)$ by cutting a cell of the $p$-th row, and
 $\Pi$ is the projector to the tensor space of $\Phi(x)$.
The gauge transformation law (\ref{summary_ADSgaug})
 is in agreement with the group-theoretical analysis of Metsaev
\cite{Metsaev}.
Note that not all of the components of the parameter $\varepsilon_{(p-1)}$
contribute to $\varepsilon (x)$ because some of them are Stueckelberg
with respect to gauge transformations for gauge parameters
\be \label{summary_transg}
\delta\xi^I_{(p-1)}=D_0\eta^I_{(p-2)}\;.
 \ee

The form of the  action functional for free HS gauge fields
\cite{LV,vf,VD5,ASV1,A2,ASV2,Alkalaev:2005kt} \be \label{summary_act1} {\cal S}_2
= \int_{{\cal M}^d}\; H^{...} \;\underbrace{E_0^{...}\wedge\cdots
\wedge E_0^{...}}_{d-2p-2}\wedge\; R_{(p+1)}^{...} \wedge
\,R_{(p+1)}^{...}\;, \ee
is  analogous to the
MacDowell-Mansouri-Stelle-West action for gravity with the
cosmological term \cite{MM,SW,VD5}. Here
$o(d-1,2)$ covariant coefficients  $H^{...}$,
constructed of the compensator
$V^A$, tangent metric $\eta^{AB}$ and the $o(d-1,2)$ Levi-Civita tensor,
parameterise various ways of index contractions,
$E^{...}_0$ is the 1-form frame field of the
$AdS_d$ background. Any  action of this form
is manifestly invariant under
the gauge transformations (\ref{summary_trans}) because
the field strength $R_{(p+1)}^{...}$ is gauge invariant.

The coefficients in the action functional are to be determined by imposing {\it the
decoupling conditions}
\be \label{summary_decouple} \frac{\delta
{\cal S}_2}{\delta \omega^\prime_{(p)}} \equiv 0\, \quad {\rm
and} \quad \frac{\delta {\cal S}_2}{\delta w_{(p)}} \equiv 0\,.
\ee
The meaning of these conditions is different.

The decoupling condition with respect to the extra fields $w_{(p)}$ implies effectively
that the action is free of higher derivatives of the physical field.
Indeed, once all extra fields are decoupled, the action depends
non-trivially  on the physical and auxiliary fields only. The
auxiliary fields can be expressed by virtue of their equations of
motion in terms of first derivatives of the metric-like gauge
field $\Phi(x)$ modulo pure gauge parts, so that the bosonic HS
equations of motion will be of second-order.

The decoupling condition for the irrelevant
auxiliary fields $\omega^\prime_{(p)}$  guarantees \cite{ASV2} the
correctness of the flat limit $\Lambda\rightarrow0$  of the $AdS_d$ theory
characterized by the enhancement of the additional gauge symmetries in
the flat limit \cite{BMV}. The point is that it is not enough to require
the $AdS_d$ action to be invariant under $AdS_d$ gauge symmetries
and to contain first order derivatives in order to guarantee that
it describes a correct HS dynamics. The correct choice is
dictated by the structure of the kinetic terms in the action
which, for the metric-like field $\Phi(x)$, is given by
${\cS}_2^{flat} \sim \int {\rm d}x^d \d\Phi\d\Phi$ that should result from the $AdS_d$
action ${\cS}_2^{AdS} \sim \int {\rm d}x^d (\cD\Phi\cD\Phi + \Lambda\Phi^2)$ in
the flat limit $\Lambda \to 0$. In the $AdS_d$ space, the mass-like
terms $\Lambda\Phi^2$ break down all the gauge symmetries of the action ${\cS}_2^{flat}$ except
for $\delta\Phi = \cD \varepsilon$ associated with the $AdS_d$
gauge parameter $\varepsilon$. The part
of the Lagrangian that contains two derivatives is not
uniquely fixed by the $AdS$ gauge symmetry and  may describe unwanted
degrees of freedom, if not fine tuned by requiring maximal gauge symmetries in the flat
limit,
\be \label{summary_transflat} \delta\Phi(x) = \d \varepsilon (x) +
\sum_{I>1}\d S_I(x)\;,
\ee
where gauge parameters $S_I(x)$ are described by the Young tableaux resulting from that
of the field $\Phi(x)$ by cutting off  a  cell from the last row of  any $I$-th, ($I>1$) horizontal
block with the convention that $S_1(x)\equiv\varepsilon(x)$.
Note that the  additional gauge symmetry parameters $S_I$, $I>1$ are
absent  for rectangular tableaux. For example, totally symmetric and totally antisymmetric
fields belong to this class.

As explained in section \ref{Reconstruction of the action}, the correct action does exists being fixed
up to an overall factor and total derivatives
by the decoupling conditions (\ref{summary_decouple}) \cite{ASV2}.
Technically,
the problem of finding coefficients  satisfying the decoupling
conditions  gets complicated if operating in terms of  multi-index
tensors.
To simplify  the problem  we reformulate it in terms of
an appropriate fermionic Fock space where
 HS fields are described as Fock vectors. In this setup, a
Fock space version of the action functional (\ref{summary_act1})
reads as
\be \label{summary_act2} {\cal S}_2= \int_{{\cal M}^d}\;
\lvac\, {\mathsf H}\;(\wedge\Eop)^{d-2p-2}\,\wedge {\mathsf
R}_{(p+1)}\wedge {\mathsf R}_{(p+1)}\rvac\,,
\ee
where $\Eop$ and ${\mathsf R}_{(p+1)}$ denote the Fock space
realizations of the background frame field and HS curvatures, respectively.
A nice feature of this formulation is that the variation of
the action (\ref{summary_act2}) has the form
\be
\label{summary_var} \delta {\cal S}_2 = \int_{{\cal M}^d}\;
\lvac\,{\cal Q}{\mathsf H}\;(\wedge\Eop)^{d-2p-1}\, \wedge
{\mathsf R}_{(p+1)}\wedge \delta {\mathsf \Omega}_{(p)}\rvac\,,
\ee
where the operator ${\cal Q}$ satisfies
 \be
\label{summary_Q} {\cal Q}^2=0\;.
\ee
It can be shown that, in a certain basis, the operator  $\cQ$
has the simple form of the de Rham operator.
This observation suggests the following  strategy of
finding action function ${\mathsf H}$. Firstly, we find the
equations of motion in the ${\cal Q}$-closed
form consistent with the decoupling
conditions and then reconstruct the action function ${\mathsf
H}$ that leads to these equations of motion by a homotopy based on
the Poincare lemma. (Note that the ambiguity in adding
${\cal Q}$-exact terms ${\mathsf H} \sim {\cal Q}{\mathsf T}$, that do not
contribute to the field equations, manifests the ambiguity
of the Lagrangian up to total derivatives.) The idea of this approach
is due to the observation that it is  easier to find correct equations of
motion than to analyze the decoupling conditions directly in terms of
the action (\ref{summary_act2}).

\section{Young tableaux and compensator formalism}
\label{sec:alg}

In this section we summarize relevant facts about tensor modules
of the orthogonal algebra which are used in our analysis of
tensor modules of the $AdS_d$ algebra $o(d-1,2)$ its Lorentz
subalgebra $o(d-1,1)$ and massive Wigner little algebra $o(d-1)$.
We also consider the decomposition of a $o(d-1,2)$-module into
$o(d-1,1)$-modules.

\subsection{Tensor modules of orthogonal algebras}

Any irreducible tensor module of the complex Lie algebra $o(M|\oC)$
is defined by a highest weight vector ${\bf l} = (l_1, l_2, ... ,
l_\nu)$, where components $l_i$ are integers that satisfy the
conditions (see, \textit{e.g.}, \cite{BarutRonchka})
\begin{align}
M=2\nu\;&: &&l_1\geq l_2 \geq \cdots \geq l_{\nu-1}\geq
|l_\nu|\;,\\
M=2\nu+1\;&: &&l_1\geq l_2 \geq \cdots \geq l_{\nu-1}\geq l_\nu
\geq 0\;.
\end{align}
The weights ${\bf l}$ (modulo a sign) can be depicted as a Young
tableau

\be \label{YT}
\bigskip
\begin{picture}(100,75)(0,45)

\put(-10,102){\footnotesize $s_1$} \put(-10,92){\footnotesize
$s_2$} \put(-10,82){\footnotesize $s_3$}
\put(-10,72){\footnotesize $s_4$} \put(-10,62){\footnotesize
$s_5$}

\put(02,114){\footnotesize $h_1$} \put(12,114){\footnotesize
$h_2$} \put(22,114){\footnotesize $h_3$}
\put(32,114){\footnotesize $h_4$} \put(42,114){\footnotesize
$h_5$}

\put(-6,55){\circle*{1}} \put(-6,50){\circle*{1}}
\put(-6,45){\circle*{1}} \put(55,115){\circle*{1}}
\put(60,115){\circle*{1}} \put(65,115){\circle*{1}}

\put(10,50){\circle*{1}} \put(20,50){\circle*{1}}
\put(30,50){\circle*{1}} \put(40,50){\circle*{1}}
\put(50,50){\circle*{1}} 

\put(10,40){\circle*{1}} \put(20,40){\circle*{1}}
\put(30,40){\circle*{1}} \put(40,40){\circle*{1}}

\put(10,30){\circle*{1}} \put(20,30){\circle*{1}}
\put(30,30){\circle*{1}} 


\put(00,110){\line(1,0){100}} \put(00,100){\line(1,0){100}}
\put(00,90){\line(1,0){100}} \put(00,80){\line(1,0){100}}
\put(00,70){\line(1,0){60}} \put(00,60){\line(1,0){60}}


\put(00,30){\line(0,1){80}} \put(10,60){\line(0,1){50}}
\put(20,60){\line(0,1){50}} \put(30,60){\line(0,1){50}}
\put(40,60){\line(0,1){50}} \put(50,60){\line(0,1){50}}
\put(60,60){\line(0,1){50}} \put(70,80){\line(0,1){30}}
\put(80,80){\line(0,1){30}} \put(90,80){\line(0,1){30}}
\put(100,80){\line(0,1){30}}

\end{picture}
\ee where the $i$-th row consists of $s_i=|l_i|$ cells and
$j$-th  column consists of $h_j$ cells. Proper realization of the
irreducible module corresponding to the highest weight ${\bf l}$
can be given in terms of the complex rank-$P$ $o(M)$-tensors
$T^{ab\ldots}$, $a,b=1,\ldots M$ where $P=s_1+\ldots+s_{\nu}$ is
the total number of cells in (\ref{YT}). The irreducibility
conditions on T are
\begin{itemize}
\item Young symmetry conditions discussed in subsections \ref{sec_sym_bas_YT} and
\ref{sec:antibasis};
\item tracelessness conditions
\be\label{tlc} \eta_{ab}T^{\ldots a\ldots b\ldots} =0\,; \ee
\item
(anti-)selfduality conditions \be\label{asdc} {}^*T=\pm T\,, \ee
where ${}^*$ is the Hodge automorphism, for the case of even $M$
and  $l_\nu \neq 0$.
\end{itemize}

For the real form $o(M-t,t)$ of  $o(M|\oC)$, tensor
representations are also characterized by different Young
tableaux (\ref{YT}). The $o(M-t,t)$ irreducible module corresponding
to (\ref{YT}) can be realized as the space of  rank-$P$ real
tensors satisfying the irreducibility conditions described above with
(anti-)selfduality conditions to be imposed if $\l_\nu\neq 0$ and
$M-2t\equiv 0\quad{\rm mod}\,\, 4$.

\subsubsection{Symmetric basis}
\label{sec_sym_bas_YT}

Let $l_q$, $q\leq \nu$ be the last nonzero component in the weight vector {\bf l}. We
group  indices of a rank-$P$ tensor $T$ into the sets
corresponding to the rows of the Young tableau (\ref{YT}),
 \be
\label{sym_T} T^{a_1(s_1),\;a_2(s_2),\; \ldots\;, a_q(s_q)}\,,
\qquad s_i=|l_i|
\ee
and require $T$ to be symmetric in each group of indices $a_i
(s_i)$ and to satisfy the conditions that symmetrization of all
indices of a $i$-th group with any index from a $j$-th group gives
zero for $i<j$.



In the symmetric basis, it may be convenient to characterize a
Young tableaux by horizontal blocks. Namely, combining rows of
equal length into horisontal blocks, the Young tableau (\ref{YT})
can be described by a
set of pairs of positive integers $(\tilde{s}_I,p_I)$,
$I=1,\ldots,k$ with $\tilde{s}_1 > \tilde{s}_2 > \cdots >
\tilde{s}_k >0$ and $p_I$ such that $p_1+ \ldots + p_k=q$. The
result can be depicted as \be \label{YT_s_b}
\bigskip
\begin{picture}(100,75)(0,40)

\put(105,90){$(\tilde s_1,p_1)$} \put(65,60){$(\tilde s_2,p_2)$}
\put(80,45){\circle*{1}} \put(80,40){\circle*{1}}
\put(80,35){\circle*{1}}

\put(10,50){\circle*{1}} \put(20,50){\circle*{1}}
\put(30,50){\circle*{1}} \put(40,50){\circle*{1}}
\put(50,50){\circle*{1}} 

\put(10,40){\circle*{1}} \put(20,40){\circle*{1}}
\put(30,40){\circle*{1}} \put(40,40){\circle*{1}}

\put(10,30){\circle*{1}} \put(20,30){\circle*{1}}
\put(30,30){\circle*{1}} 

{\linethickness{.3mm} \put(00,110){\line(1,0){100}}
\put(00,80){\line(1,0){100}} \put(00,30){\line(0,1){80}}
\put(60,60){\line(0,1){20}} \put(100,80){\line(0,1){30}} }

\put(00,100){\line(1,0){100}} \put(00,90){\line(1,0){100}}
\put(00,70){\line(1,0){60}} \put(00,60){\line(1,0){60}}


\put(10,60){\line(0,1){50}} \put(20,60){\line(0,1){50}}
\put(30,60){\line(0,1){50}} \put(40,60){\line(0,1){50}}
\put(50,60){\line(0,1){50}} \put(60,80){\line(0,1){30}}
\put(70,80){\line(0,1){30}} \put(80,80){\line(0,1){30}}
\put(90,80){\line(0,1){30}}

\end{picture}
\ee where an  $I$-th horizontal block has length $\tilde s_I$ and
height $p_I$. The exact identification of rows of equal length  in
(\ref{YT}) and horizontal blocks in (\ref{YT_s_b})  reads as
\be
\label{block_notation1} \tilde{s}_1= \underbrace{s_1= \ldots =
s_{p_1}}_{p_1} > \tilde{s}_2=\underbrace{s_{p_1+1} = \ldots =
s_{p_1+p_2}}_{p_2}
> \ldots >\tilde{s}_k = \underbrace{s_{p_1+ \ldots +p_{k-1}+1} = \ldots =
s_q}_{p_k}\;. \ee

It is worth to note that, as a consequence of its Young
symmetry properties, a horizontal block $(\tilde s_I,p_I)$ is
invariant  with respect
to exchange of its rows up to a sign factor $(-1)^{\tilde s_I}$.

\subsubsection{Antisymmetric basis}
\label{sec:antibasis}

Let us group the indices of a rank-$P$ tensor $T$ into the sets
corresponding to the columns of the Young tableau (\ref{YT})
\be
\label{asym_T} T^{a_1[h_1],\;a_2[h_2],\; \ldots\;,
a_{s_1}[h_{s_1}]}\,, \ee
where  the length of the first row $s_1$ is
the number of  columns in (\ref{YT}).
In the antisymmetric basis, $T$ is required
to be antisymmetric in each group of indices $a_i [h_i]$ and to
satisfy the conditions that antisymmetrization of all indices of a
$i$-th group with any index of a  $j$-th group gives zero if $i<j$.
One should note that tensors (\ref{sym_T}) and (\ref{asym_T})
corresponding to the same Young tableau form isomorphic modules
of the orthogonal algebra defined in different
(namely, symmetric and antisymmetric) Young bases. Let
$Y_{M-t,t}(s_1,\ldots,s_\nu)$ denote the space of $o(M-t,t)$
tensors satisfying the Young symmetry conditions either in symmetric or
in antisymmetric basis.


In the antisymmetric basis, it may be convenient to characterize
Young tableaux by vertical blocks. Namely, combining columns of
equal height into vertical blocks, the Young tableau (\ref{YT})
can be described by a
set of pairs of positive integers $(m_I,\tilde h_I)$ with
$\tilde{h}_1 > \tilde{h}_2 > \cdots > \tilde{h}_k >0$ and $m_I$
such that $m_1+ \ldots + m_k=s_1$. The result can be depicted as \be
\label{YT_a_b}
\bigskip
\begin{picture}(100,75)(0,45)

\put(63,115){$(m_k,\tilde h_k)$} \put(50,118){\circle*{1}}
\put(45,118){\circle*{1}} \put(40,118){\circle*{1}}

\put(10,50){\circle*{1}} \put(20,50){\circle*{1}}
\put(30,50){\circle*{1}} \put(40,50){\circle*{1}}
\put(50,50){\circle*{1}} 

\put(10,40){\circle*{1}} \put(20,40){\circle*{1}}
\put(30,40){\circle*{1}} \put(40,40){\circle*{1}}

\put(10,30){\circle*{1}} \put(20,30){\circle*{1}}
\put(30,30){\circle*{1}} 

{\linethickness{.3mm} \put(00,110){\line(1,0){100}}
\put(60,80){\line(1,0){40}} \put(00,30){\line(0,1){80}}
\put(60,60){\line(0,1){50}} \put(100,80){\line(0,1){30}} }

\put(00,100){\line(1,0){100}} \put(00,90){\line(1,0){100}}
\put(00,80){\line(1,0){60}} \put(00,70){\line(1,0){60}}
\put(00,60){\line(1,0){60}}


\put(10,60){\line(0,1){50}} \put(20,60){\line(0,1){50}}
\put(30,60){\line(0,1){50}} \put(40,60){\line(0,1){50}}
\put(50,60){\line(0,1){50}} \put(70,80){\line(0,1){30}}
\put(80,80){\line(0,1){30}} \put(90,80){\line(0,1){30}}

\end{picture}
\ee where $I$-th vertical block has length $m_I$ and height
$\tilde{h}_I$. The exact identification of  columns  of equal height in
(\ref{YT}) and vertical blocks in (\ref{YT_a_b})  reads as \be
\label{block_notation3} \ba{c} \hspace{-30mm}\tilde{h}_1=
\underbrace{h_1= \ldots = h_{m_1}}_{m_1} >
\tilde{h}_2=\underbrace{h_{m_1+1} = \ldots = h_{m_1+m_2}}_{m_2}
> \ldots
\\
\\
\hspace{60mm}\ldots >\tilde{h}_k = \underbrace{h_{m_1+ \ldots
+m_{k-1}+1} = \ldots = h_{s_1}}_{m_k}\;. \ea \ee

Note that a number of vertical blocks of any
 Young tableau  equals to the number of its horizontal blocks. Also note
that, as a consequence of its Young symmetry properties,
 a vertical block is invariant under exchange of
its columns.

\subsection{Tracelessness conditions}
\label{Trace conditions on the metric-like field}
%
%
%
In this section we define  subspaces of the spaces
of traceful tensors that, apart from Young symmetry conditions,
satisfy specific tracelessness conditions which extend the
Fronsdal double tracelessness condition for massless  symmetric fields to
mixed-symmetry massless fields of general type.

Let $B^{d-1,1}_m(s_1,...,s_q,0,...,0)$ be the linear space of
tensors which have the Young properties of the type $Y_{d-1,1}(s_1,
\ldots,  s_q,0,...,0 )$ and satisfy the tracelessness  conditions
\be \label{tr1} \eta_{a_i a_i} \eta_{a_i
a_i} T^{a_1(s_1), ...\,,\, a_{q}(s_{q})}=0\;, \qquad 0<  i \leq
m\, \ee
 and
 \be \label{tr2} \eta_{a_i a_i} T^{a_1(s_1),\, ...\,,\,
a_{q}(s_{q})}=0\;, \qquad m<  i \leq q\,.
\ee
 Note that $B^{d-1,1}_m(s_1, \ldots ,
s_q,0,...,0 ) \subset B^{d-1,1}_n(s_1, \ldots , s_q,0,...,0)$
for $m<n$.

$B^{d-1,1}_0(s_1, \ldots , s_q,0,...,0)$ is the space of
traceless tensors with the $Y_{d-1,1}(s_1, \ldots, s_q,0,...,0)$
Young properties.

The following lemmas are simple consequences of the definition
of $B^{d-1,1}_m(s_1,...,s_q,0,...,0)$.

\vspace{3mm}

{\it  \underline{Lemma 1}}

Contraction of $\eta_{(a_ia_j}\eta_{a_ka_l)}$ with any four
symmetrized indices of a tensor from $B^{d-1,1}_m(s_1, \ldots ,
s_q,0,...,0 )$ gives zero.

Lemma 1 is a corollary of (\ref{tr1}) and the Young symmetry
properties, which guarantee that any group of symmetrized indices
can be placed in the first row.

{\it  \underline{Lemma 2}}

{}From Lemma 1 it follows that \be \label{tr3} \eta_{a_i (a_j}
\eta_{a_k a_l)} T^{a_1(s_1),\, ...\,,\, a_{q}(s_{q})}=0\;, \qquad
\forall \;i,j,k,l\,, \ee \textit{i.e.} any double trace gives
zero provided that any three of the contracted indices  are
symmetrized.

This is because $\eta_{ab} \eta_{cd}$ belongs to the symmetric
part of the tensor product \be \Big(\begin{picture}(12,12)(-1,4)
{\linethickness{0.210mm}
\put(00,10){\line(1,0){10}} 
\put(00,05){\line(1,0){10}} 
\put(00,05){\line(0,1){5}} 
\put(05,05){\line(0,1){5}} 
\put(10,05){\line(0,1){5}}
}
\end{picture}
\otimes
\begin{picture}(12,12)(-1,4)
{\linethickness{0.210mm}
\put(00,10){\line(1,0){10}} 
\put(00,05){\line(1,0){10}} 
\put(00,05){\line(0,1){5}} 
\put(05,05){\line(0,1){5}} 
\put(10,05){\line(0,1){5}}
}
\end{picture}
\Big)_{\rm sym}
 =
\begin{picture}(22,12)(-1,4)
{\linethickness{0.210mm}
\put(00,10){\line(1,0){20}} 
\put(00,05){\line(1,0){20}} 
\put(00,05){\line(0,1){5}} 
\put(05,05){\line(0,1){5}} 
\put(10,05){\line(0,1){5}}
\put(15,05){\line(0,1){5}} 
\put(20,05){\line(0,1){5}} 
}
\end{picture}
\oplus
\begin{picture}(13,12)(-1,1)
{\linethickness{0.210mm} \put(00,10){\line(1,0){10}}
\put(00,05){\line(1,0){10}} \put(00,00){\line(1,0){10}}
\put(00,00){\line(0,1){10}} \put(05,00){\line(0,1){10}}
\put(10,00){\line(0,1){10}} }
\end{picture}\,,
\ee
so that the symmetrization of any three indices of $\eta_{ab}\eta_{cd}$ implies the total symmetrization.
Therefore nonzero traces in $B^{d-1,1}_m(s_1, \ldots , s_q,0,...,0 )$
 can only appear when all elementary contractions hit
different rows.

{\it  \underline{Lemma 3}}

The condition (\ref{tr2}) along with Lemma 2 mean that
contraction of any $m+1$ pairs of indices of $T^{a_1(s_1),\,
...\,,\, a_{q}(s_{q})}\in B^{d-1,1}_m(s_1, \ldots , s_q,0,...,0
)$ gives zero.

\vspace{0.2cm}

Recall that a rectangular block is invariant (up to a sign) under
exchange of its rows. As a result it follows

\vspace{0.2cm} {\it  \underline{Lemma 4}}

Once  (\ref{tr2}) is true for one of the rows of a rectangular
block it is true for the entire block, i.e. $B^{d-1,1}_m(s_1,
\ldots , s_q,0,...,0 )=B^{d-1,1}_n(s_1, \ldots , s_q,0,...,0 )$
if $s_{m+1} = s_{n+1}$.

Therefore, it is sufficient to impose the trace condition
(\ref{tr2}) for any row inside a horizontal block ({\it e.g.}, the
upper row).

\subsection{Dimensional reduction and compensator}
\label{dim_comp}

Let us now address the question what is a pattern  of the
decomposition of a given $o(d-1,2)$-module described by a Young
tableau (\ref{YT}) into modules of the subalgebra $o(d-1,1)\subset
o(d-1,2)$.
%
For our purposes it is convenient to describe this decomposition
in a manifestly $o(d-1,2)$ covariant manner. To this end, let us
introduce a compensator $V^A$, which is an $o(d-1,2)$ vector
normalized as $V_A V^A=1$. (Note that when discussing gauged
orthogonal algebras the compensator vector becomes an $x$-dependent
field $V^A(x)$). The role of the compensator in
the decomposition procedure has clear geometrical interpretation
since the Lorentz algebra
$o(d-1,1)\subset o(d-1,2)$  can be  identified
as the stability algebra of the
compensator. This is most evident in the standard form of the
compensator $V^A=(0,\ldots,0,1)=\gd^A_{d}$.

Let a traceless $o(d-1,2)$ tensor representation of some symmetry type
\\$Y_{d-1,2}(s_1, ... ,s_q,0\ldots,0)$
be considered either in symmetric basis
$T^{A_1(s_1),\ldots,A_q(s_q)}$ or in antisymmetric basis
$T^{A_1[h_1],\ldots,A_{s_1}[h_{s_1}]}$.
The decomposition results from the following
procedure. Every $o(d-1,2)$
index of $T$ has one component along $V^A$ and $d$ components
orthogonal to $V^A$. In the former case we cancel a cell
of the corresponding $o(d-1,2)$ Young tableau, while in the latter
we keep it. A number of indices along $V^A$ cannot exceed
$s_1$  because the symmetrization of more than $s_1$ indices in
$T$ gives zero by the defining property of the Young tableau
$Y_{d-1,2}(s_1,\ldots,s_q,0,\ldots ,0)$. Moreover,
no two cut cells can belong to the same column as is obvious from
the realization of a Young tableau in the antisymmetric basis
because the tensor $V^A V^B$ is  symmetric.
Otherwise, any set of indices
0 to $s_1$ can be aligned along $V^A$. Therefore any
number of cells from 0 to $s_1$ can be cut under the condition that
no two cells are cut from  the same column.
Of course only such
cuts are allowed that give rise to a proper Young tableau.
(Otherwise the resulting tensor is identically zero.)

The particular $o(d-1,1)$  tensors result from contractions of
some of the indices of the original $o(d-1,2)$ tensor with the
compensator $V^A$ followed by the proper (anti)symmetrizations
and projecting to the $V^A$-transversal components with respect to the rest of
indices. Note that, in the symmetric basis, all contractions with the
compensator are equivalent to some its contractions with indices of the first
row because $V^{A}V^BV^C\ldots$ is a totally symmetric tensor.

The  resulting list of the $o(d-1,1)$ components consists of the Young
tableaux drawn in bold on the right hand side of the decomposition

\vspace{5mm} \be \label{dimred}
\begin{picture}(130,55)(-1,40)


{\linethickness{.3mm}
\put(00,90){\line(1,0){100}}
\put(00,50){\line(1,0){100}}
\put(100,50){\line(0,1){40}}
\put(00,50){\line(0,1){40}}
} \put(00,70){\line(1,0){100}} \put(00,80){\line(1,0){100}}
\put(00,60){\line(1,0){100}} \put(00,50){\line(1,0){100}}
\put(10,50.0){\line(0,1){40}} \put(20,50.0){\line(0,1){40}}
\put(30,50.0){\line(0,1){40}} \put(40,50.0){\line(0,1){40}}
\put(50,50.0){\line(0,1){40}} \put(60,50.0){\line(0,1){40}}
\put(70,50.0){\line(0,1){40}} \put(80,50.0){\line(0,1){40}}
\put(90,50.0){\line(0,1){40}}



{\linethickness{.3mm}
\put(00,10){\line(1,0){70}}
\put(70,10){\line(0,1){40}}
\put(00,10){\line(0,1){40}}
}

\put(60,10){\line(0,1){40}}
\put(50,10){\line(0,1){40}}
\put(40,10){\line(0,1){40}}
\put(30,10){\line(0,1){40}}
\put(20,10){\line(0,1){40}}
\put(10,10){\line(0,1){40}}

\put(00,20){\line(1,0){70}}
\put(00,30){\line(1,0){70}}
\put(00,40){\line(1,0){70}}


{\linethickness{.3mm}
\put(00,-30){\line(0,1){40}}
}

\put(10,00){\circle*{2}} \put(20,0){\circle*{2}}
\put(30,00){\circle*{2}} \put(40,00){\circle*{2}}
\put(50,00){\circle*{2}} \put(60,00){\circle*{2}}

\put(10,-10){\circle*{2}} \put(20,-10){\circle*{2}}
\put(30,-10){\circle*{2}} \put(40,-10){\circle*{2}}
\put(50,-10){\circle*{2}}

\put(10,-20){\circle*{2}} \put(20,-20){\circle*{2}}
\put(30,-20){\circle*{2}} \put(40,-20){\circle*{2}}



{\linethickness{.3mm}
\put(00,-30){\line(1,0){40}}
\put(00,-70){\line(1,0){40}}
\put(40,-70){\line(0,1){40}}
\put(00,-70){\line(0,1){40}}
}

\put(10,-70){\line(0,1){40}}
\put(20,-70){\line(0,1){40}}
\put(30,-70){\line(0,1){40}}

\put(00,-60){\line(1,0){40}}
\put(00,-50){\line(1,0){40}}
\put(00,-40){\line(1,0){40}}

\end{picture}
\begin{picture}(180,55)(-80,40)


{\linethickness{.3mm}
\put(00,90){\line(1,0){100}}
\put(00,50){\line(1,0){80}}
\put(100,60){\line(0,1){30}}
\put(00,50){\line(0,1){40}}
\put(80,50){\line(0,1){10}}
\put(80,60){\line(1,0){20}}
}

\put(100,50){\line(0,1){40}}

\put(00,70){\line(1,0){100}} \put(00,80){\line(1,0){100}}
\put(00,60){\line(1,0){100}} \put(00,50){\line(1,0){100}}
\put(10,50.0){\line(0,1){40}} \put(20,50.0){\line(0,1){40}}
\put(30,50.0){\line(0,1){40}} \put(40,50.0){\line(0,1){40}}
\put(50,50.0){\line(0,1){40}} \put(60,50.0){\line(0,1){40}}
\put(70,50.0){\line(0,1){40}} \put(80,50.0){\line(0,1){40}}
\put(90,50.0){\line(0,1){40}}

\put(93,53){\scriptsize o} \put(83,53){\scriptsize o}



{\linethickness{.3mm}
\put(00,10){\line(1,0){60}}
\put(70,20){\line(0,1){30}}
\put(00,10){\line(0,1){40}}
\put(60,10){\line(0,1){10}}
\put(60,20){\line(1,0){10}}
}

\put(60,10){\line(0,1){40}}
\put(50,10){\line(0,1){40}}
\put(40,10){\line(0,1){40}}
\put(30,10){\line(0,1){40}}
\put(20,10){\line(0,1){40}}
\put(10,10){\line(0,1){40}}

\put(00,20){\line(1,0){70}}
\put(00,30){\line(1,0){70}}
\put(00,40){\line(1,0){70}}

\put(70,10){\line(0,1){10}}
\put(60,10){\line(1,0){10}}
\put(63,13){\scriptsize o}


{\linethickness{.3mm}
\put(00,-30){\line(0,1){40}}
}

\put(10,00){\circle*{2}} \put(20,0){\circle*{2}}
\put(30,00){\circle*{2}} \put(40,00){\circle*{2}}
\put(50,00){\circle*{2}} \put(60,00){\circle*{2}}

\put(10,-10){\circle*{2}} \put(20,-10){\circle*{2}}
\put(30,-10){\circle*{2}} \put(40,-10){\circle*{2}}
\put(50,-10){\circle*{2}}

\put(10,-20){\circle*{2}} \put(20,-20){\circle*{2}}
\put(30,-20){\circle*{2}} \put(40,-20){\circle*{2}}



{\linethickness{.3mm}
\put(00,-30){\line(1,0){40}}
\put(00,-70){\line(1,0){10}}
\put(10,-60){\line(1,0){30}}
\put(40,-60){\line(0,1){30}}
\put(10,-70){\line(0,1){10}}
\put(00,-70){\line(0,1){40}}
}

\put(10,-70){\line(0,1){40}}
\put(20,-70){\line(0,1){40}}
\put(30,-70){\line(0,1){40}}
\put(40,-70){\line(0,1){40}}

\put(00,-60){\line(1,0){40}}
\put(00,-50){\line(1,0){40}}
\put(00,-40){\line(1,0){40}}
\put(00,-70){\line(1,0){40}}

\put(13,-67){\scriptsize o} \put(23,-67){\scriptsize o}
\put(33,-67){\scriptsize o}

\put(-60,10){$\Longrightarrow\quad \bigoplus\limits_{\{r_I\}}\qquad$}%
\put(-8,100){\footnotesize $o(d-1,1)\;\; {\sf Young\;\; tableaux}$}%
\put(-215,100){\footnotesize $o(d-1,2)\;\; {\sf Young\;\; tableau}$}%

\end{picture}
\ee

\vspace{4.5cm}

\noindent where  $r_I$ denotes a number of cells cut from
the last row of the $I$-th horizontal block of the
original $o(d-1,2)$ Young tableau \be \label{r} 0\leq  r_I \leq
(\tilde{s}_{I}-\tilde{s}_{I+1}) \;, \qquad 1\leq I\leq k\;, \ee
with the convention that $\tilde{s}_{k+1}=0$. In the $o(d-1,1)$
tableaux (\ref{dimred}), the indices contracted with the
compensator are denoted by $\begin{picture}(10,10)(-1,5)
{\linethickness{0.210mm}
\put(00,12){\line(1,0){7}} 
\put(00,05){\line(1,0){7}} 
\put(00,05){\line(0,1){7}} 
\put(07,05){\line(0,1){7}} 
} \put(1.6,06.5){\scriptsize o}
\end{picture}$.
Taking into account that the compensator is $o(d-1,1)$ invariant,
they disappear from the $o(d-1,1)$ tableau
according to the decomposition procedure described above.

It is sometimes convenient to use another parametrization by introducing the
parameters
\be \label{t} t_I \equiv
(\tilde{s}_{I}-\tilde{s}_{I+1}) - r_I\;: \qquad 0\leq  t_I \leq
(\tilde{s}_{I}-\tilde{s}_{I+1}) \;.
\ee
Note that the $o(d-1,1)$
tableau with the minimal number of cells corresponds to all
$t_I=0$. The parameters $t_I$ measure a deviation from
the minimal number of cells.

%
%
%

As an illustration, let us consider the
example of a $o(d-1,2)$ traceless tensor $T^{A(2),\,B}$
 with the symmetry of the three-cell "hook" Young
tableau $Y_{d-1,2}(2,1,0,\ldots,0)$, using for definiteness the
symmetric basis. Its $o(d-1,1)$ decomposition gives
 four traceless tensors having the Young
symmetries of $Y_{d-1,1}(2,1,0,\ldots,0)$,
$Y_{d-1,1}(2,0,\ldots,0)$, $Y_{d-1,1}(1,1,0,\ldots,0)$ and
$Y_{d-1,1}(1,0,\ldots,0)$ \be T^{A(2),\,B} = A^{A(2),\,B} \oplus
B^{A(2)}\oplus C^{A,\,B}\oplus D^A \,,\ee
that are
$V^A$-transversal
 \be \label{trans}
A^{A(2),\,B}V_A=0\;, \qquad A^{A(2),\,B}V_B=0\;, \quad
B^{A(2)}V_A=0\;, \quad C^{A,\,B}V_B=0\;, \quad D^A V_A=0\;. \ee
Note that the second condition in (\ref{trans}) follows from the
first one by virtue of the Young symmetry properties of
$A^{A(2),\,B}$. The explicit form of the projectors to these
$o(d-1,1)$ components reads as

\be
\ba{l}
D^A=T^{B(2),A}V_{(B}V_{B)}\,,
\\
\\
B^{A(2)}=T^{A(2),B}V_B+D^{(A}V^{A)}\,,
\\
\\
\dps
C^{A,B}=\Big(T^{AC,B}-T^{BC,A}\Big)V_C+\frac{3}{2}\Big(D^AV^B-D^BV^A\Big)\,,
\\
\\
\dps
A^{A(2),B}=T^{A(2),B}-\Big(B^{A(2)}V^B-B^{B(A}V^{A)}\Big)+C^{B,(A}V^{A)}+
\Big(D^{(A}V^{A)}V^B-D^B V^{(A}V^{A)}\Big)\,.
\ea
\ee

\section{Higher-spin gauge fields in $AdS_d$}
\label{sec:adsHSfields}

In this section we introduce following to \cite{ASV1} the
frame-like fields  and related metric-like fields for HS fields of
 general symmetry type.

\subsection{Example of gravity}
Many
features of the frame-like formulation are
illustrated by the example of Einstein
gravity with the cosmological term in the formulation of MacDowell and
Mansouri
\cite{MM,SW,VD5}. Here
the gravitational field is described by the $1$-form \be
\label{ads_con} \Omega^{AB}(x)=-\Omega^{BA}(x)={\rm d}
x^{\underline{n}}\:\Omega^{AB}_{\underline{n}}(x)\;, \ee which
is the gauge connection of the $\ads$
algebra $o(d-1,2)$. By introducing the $o(d-1,2)$ covariant
derivative $D$
\be
DT^A=\extdiff T^A+\Go^A{}_BT^B \ee one defines the gauge
transformation law as \be \label{comp_def} \delta\Omega^{AB} = D
\xi^{AB}\equiv {\rm d}\xi^{AB}+\Omega^{A}{}_{C}\xi^{CB}
+\Omega^B{}_C\xi^{AC}\;,
\ee
where ${\rm d}= {\rm d}x^{\underline{m}}\d_{\underline{m}}$ is the exterior
differential and $\xi^{AB}(x)=-\xi^{BA}(x)$ is a $0$-form  gauge
parameter.

To establish a precise relationship between the
metrics $g_{\um\un}$ and the connection $\Omega^{AB}$,
 the latter should be decomposed into Lorentz $o(d-1,1)\subset o(d-1,2)$
components which are the frame 1-form $e^a$ and the Lorentz
connection $\omega^{ab}$. To make this decomposition
$o(d-1,2)$ covariant it is convenient to use the compensator formalism
described in section \ref{dim_comp}. Namely,
$o(d-1,2)$ covariant versions of the frame field and Lorentz
connection are defined as follows \cite{compensator}
\be \label{s2_def_lor}
\lambda E^A = D\,V^A\equiv {\rm d} V^A+\Omega^{AB}V_B\;, \quad
\omega^{AB} = \Omega^{AB} -\lambda\,(E^{A}\,V^{B}-E^{B}\,V^{A})\;,
\ee
where the dimensionful parameter $\gl$ is introduced to make the frame field dimensionless.
Conventional Lorentz-covariant fields $e^a$ and $\omega^{ab}$
result from these formulae with the compensator in the standard gauge
$V^A=\delta^A_{d}$.  The
 metrics
 \be \label{metr}
g_{\um\un}=\eta_{AB}E^A_{\underline{m}}E^B_{\underline{n}}\;,
\ee
is the $o(d-1,2)$ invariant extension of the standard formula
$g_{\underline{mn}}(x)=\eta_{ab}e^a_{\underline{m}}(x)e^b_{\underline{n}}(x)$,
which is (\ref{metr}) in the standard gauge.
Note that the definitions (\ref{s2_def_lor})
comply with Lorentz invariance of the compensator  ${\cal
D}V^A\equiv {\rm d}V^A+\omega^{AB}V_B=0$.

The curvature associated with the gauge connection
(\ref{ads_con}) is  \be
\label{s2_ads_curv} R^{AB}={\rm
d}\Omega^{AB}+\Omega^{A}{}_C\wedge \Omega^{CB}\;. \ee The
remarkable property of the formulation of
gravity in terms of $o(d-1,2)$ connection
is that the anti-de
Sitter geometry is described by a nondegenerate flat connection
$\Omega_0^{AB}=(h^A,\omega_0^{AB})$ that satisfies
\bee \label{nond}
{\rm rank}(h_\um^A)=d\,,\\
\label{zer} R^{AB}(\Omega_0)=0\;. \eee
(The notation $h^A$ is used for the background $AdS$ frame.)
In terms of the covariant derivative $D_0$ with respect to the background
connection $\Omega_0$, the zero-curvature condition (\ref{zer}) implies
 \be
 \label{zerocurv} D_0^2=0\;.
\ee

Consider the perturbation expansion
$\Omega^{AB}=\Omega_0^{AB}+\Omega_1^{AB}$, where $\Omega_1$
describes dynamical
fluctuations around the background connection $\Omega_0$.
From (\ref{comp_def}) and (\ref{s2_ads_curv}) it follows that the
linearized gauge transformation and curvature have the form \be
\label{s2_def_gaug+curv}
\delta_0\Omega_1^{AB}=D_0\xi^{AB}\;,\qquad\qquad
R_1^{AB}=D_0\Omega_1^{AB}\,.
\ee
{}From (\ref{zerocurv}) it follows then
that the linearized curvature $R_1$
is gauge invariant
\be
\delta_0R_1^{AB}=0\,.
\ee

Assuming that the compensator is of order zero,
the dynamical frame and Lorentz
connection are
 \be
 \label{2}
 \Omega_1^{AB}
=\omega_1^{AB}+\lambda\,(E_1^{A}\,V^{B}-E_1^{B}\,V^{A})\;,
\ee
where $E_1^{A}V_A=0$ and $\omega_1^{AB}V_B=0$.

{}From the formula (\ref{metr}), one finds that the fluctuational
part $g_1{}_{\um\un}$ of the metrics is \be
\label{metr_vs_frame} g_1{}_{\um\un} =
\eta_{AB}\Big(E_1{}_{\um}^Ah_\un^B+E_1{}_{\un}^Ah_\um^B\Big)=
E_{1\um;\un}+E_{1\un;\um}\;, \ee where
$E_{1\um;\un}=\eta_{AB}E_{1\um}^Ah_\un^B$.
The inverse
frame $h^\un_A$ \be h^\un_Ah^A_\um=\gd^\un_\um \ee exists due
to the nondegeneracy condition (\ref{nond}). Its form is fixed
uniquely by requiring
 \be h^\un_Ah_\un^B=\gd_A^B-V_AV^B. \ee
One can rewrite (\ref{metr_vs_frame}) as \be\label{3}
g_{1AB}=E_{1A;B}+E_{1B;A}\,, \ee where
$g_{1A,B}=g_{1\un\um}h^\un_A h^\um_B$ and $E_{1A;B}=E_{1\un
B}h^\un_A$. The gauge transformation law for $E_{A;B}$, that
follows from (\ref{s2_def_gaug+curv}), has the form
\be
\label{1}
\gd E_{1A;B}=h^\un_A\cD_{\un}\varepsilon_B+\varepsilon_{AB}\,,
\ee
where $\cD$ is Lorentz covariant derivative evaluated with
respect to $\go_0^{AB}$ while $\varepsilon_A$, $\varepsilon_{AB}$
are Lorentz components of the gauge parameter $\xi^{AB}$, {\it
i.e.}
$\xi^{AB}=\varepsilon^{AB}+\gl(\varepsilon^AV^B-\varepsilon^BV^A)$,
$\varepsilon^AV_A=0$, $\varepsilon^{AB}V_B=0$\,.

That the metric fluctuation
is described by the symmetric component of the frame field
({\it i.e.} $E_{1A;B}+E_{1B;A}$) is consistent with the fact
that, as follows  from (\ref{1}), the antisymmetric component of the
frame $E_{1A;B}-E_{1B;A}$ is compensated by local Lorentz
transformations with the gauge transformation parameter $\varepsilon^{AB}$. The
gauge transformation of the symmetric component $E_{1A;B}+E_{1B;A}$
induced from (\ref{1}) reproduces the linearized diffeomorphism
with the gauge parameter $\varepsilon^A$.

The formulae (\ref{s2_def_gaug+curv}), (\ref{2}), (\ref{3}) for the
gravitational field admit a straightforward generalization to
higher spins. HS fields $\Phi_{\underline{m},
\underline{n}, \underline{k}, ...}(x)$ which generalize the
fluctuational part of the metrics in gravitation will be referred
to as {\it metric-like} HS fields. Analogously, their $p$-form
cousins $\Omega_{(p)}{}^{A,B,C,\,...}(x)$ will be referred to as
\textit{frame-like} HS fields. As a preparation to the general case,
let us consider bosonic symmetric HS fields.

\subsection{Symmetric higher-spin gauge fields}
\label{sec:tot_sym}

The metric-like approach to  totally symmetric bosonic massless
fields of all spins  was developed by Fronsdal
both in flat \cite{fronsdal_flat} and $AdS$ space \cite{fronsdal_ads}.
Here an integer spin $s$ massless field is described by a totally symmetric rank $s$
$o(d-1,1)$ tensor
\be
\label{frf} \Phi^{a_1\ldots a_s}(x)\equiv
\Phi^{a(s)}(x)
\ee subject to the Fronsdal double tracelessness
condition \cite{fronsdal_flat,fronsdal_ads}
\be \label{frtr}
\eta_{b_1b_2}\eta_{b_3b_4}\Phi^{b_1b_2b_3b_4 a(s-4)}=0\,,
\ee
which is nontrivial for $s\geq 4$. The  HS gauge transformation is
\be
\delta\Phi^{a(s)}(x)={\cal D}^{a}
\varepsilon^{a(s-1)}(x)\,,\qquad {\cal D}^{a}
=h^{\underline{n}\,a} {\cal D}{}_{\underline{n}}
\ee
where the parameter $\varepsilon^{a(s-1)}$ is a rank  $s-1$ symmetric
traceless $o(d-1,1)$ tensor and ${\cal D}{}_{\underline{n}}$ is
the background Lorentz derivative.

The frame-like formulation operates in terms of a $1$-form
frame-like HS gauge field \cite{vas_yadfiz,V1,LV}
\be
\label{ph}
e^{a(s-1)}=\textrm{d}x^\un\, e_{\un}{}^{a(s-1)}
\ee
that is traceless in the tangent indices
\be
\label{etr}
e_b{}^{ba(s-3)}=0\,. \ee
The HS gauge transformation law is
 \be \label{ltr} \delta e^{a(s-1)} = \cD\varepsilon^{a(s-1)}
+ h_b\,\varepsilon^{a(s-1),\,b}\,, \ee where $h^a$ is the
background frame 1-form. The totally symmetric traceless 0-form
gauge parameter $\varepsilon^{a(s-1)}(x)$ is   equivalent to
that of the Fronsdal's formulation. The
0-form gauge parameter $\varepsilon^{a(s-1),\,b}(x)$ is also
traceless having the symmetry type $Y_{d-1,1}(s-1,1,0, ..., 0)$
which means that  $\varepsilon^{a(s-1),\,a}(x) =0$.
It is the HS generalization of the parameter $\varepsilon^{a,b}$
of the local Lorentz transformations in gravity ($s=2$).  The
Lorentz type gauge ambiguity related to $\varepsilon^{a(s-1),\,b}$
can be fixed by requiring the frame-like HS gauge field to be totally
symmetric by setting
\be
\label{gfix}
h^{\un;\, b}e_\un^{a(s-1)}= \Phi^{a(s-1)b}\,,
\ee
where the symmetric tensor field
\be
\Phi^{a(s)}=h^{\un;\, a}e_\un^{a(s-1)}
\ee
identifies with the metric-like field of the Fronsdal formulation. Note that
$\Phi^{a(s)}$ is  double traceless as a
consequence of (\ref{etr}).

The Lorentz-like HS symmetry with the parameter
$\varepsilon^{a(s-1),\,b}$ assumes a HS Lorentz connection-like 1-form
$\go^{a(s-1),\,b}$. {}The analysis of its transformation law shows
 \cite{V1,LV} that, for $s>2$, some additional gauge
connections and symmetry parameters have to be introduced. As a
result, the full set of HS frame-like fields associated with a
spin $s$ massless field consists of the $1$-forms \be
\label{sym_fr_fields} \Upsilon^{a(s-1),\, b(t)}(x) =
\textrm{d}x^\un \Upsilon_\un{}^{a(s-1),\, b(t)}(x)\;,
\ee that
take values in  all traceless tensor representations of the
Lorentz algebra $o(d-1,1)$ described by the  Young tableaux
$Y_{d-1,1}(s-1,t,0,\ldots,0)$ with at most two rows, such that the
upper row has length $s-1$
\bigskip
\vskip -5mm \be \label{dia}
\begin{picture}(20,50)
\put(-120,25){\footnotesize$o(d-1,1)\;:$}
\put(20,45){\footnotesize$s-1$}
\put(00,40){\line(1,0){70}} \put(00,30){\line(1,0){70}}
\put(50,30){\line(0,1){10}} \put(60,30){\line(0,1){10}}
\put(70,30){\line(0,1){10}}
\put(00,20){\line(1,0){50}} \put(00,20){\line(0,1){20}}
\put(10,20){\line(0,1){20}} \put(40,20){\line(0,1){20}}
\put(30,20){\line(0,1){20}} \put(20,20){\line(0,1){20}}
\put(50,20){\line(0,1){20}} \put(25,10){\footnotesize$t$}\,
\end{picture}
\ee
The field $\Upsilon^{a(s-1)}$ with $t=0$ identifies
with the physical spin $s$ frame-like field $e^{a(s-1)}$. The
case of $t=1$ corresponds to the auxiliary Lorentz-like field
$\go^{a(s-1),\,b}$. The remaining fields
(\ref{sym_fr_fields}) with $t\geq 2$ are called {\it extra fields}
\cite{V1,LV}.

The frame-like formalism  works both in the $\ads$
and in the flat space. In the $\ads$ case it is  convenient to use the
observation of \cite{VD5} that the  set of the HS  1-forms
$ \Upsilon{}^{a(s-1), b(t)}$ with all $0\leq t \leq s-1$
results from the 1-form \be
\label{frO} \Omega_{(1)}{}^{A(s-1),\, B(s-1)} =\textrm{d}x^\un\,
\Omega_{\un}{}^{A(s-1),\, B(s-1)} \ee
that
 carries the traceless
tensor representation of  $o(d-1,2)$ described by the length $s-1$ two-row
rectangular Young tableau
\bigskip
\vskip -5mm \be \label{dia2}
\begin{picture}(20,50)
\put(-120,25){\footnotesize$o(d-1,2)\;:$}
\put(25,45){\footnotesize$s-1$}
\put(00,40){\line(1,0){70}} \put(00,30){\line(1,0){70}}
\put(60,20){\line(0,1){20}} \put(70,20){\line(0,1){20}}
\put(00,20){\line(1,0){70}} \put(00,20){\line(0,1){20}}
\put(10,20){\line(0,1){20}} \put(40,20){\line(0,1){20}}
\put(30,20){\line(0,1){20}} \put(20,20){\line(0,1){20}}
\put(50,20){\line(0,1){20}}

\end{picture}
\ee
{\it i.e.}  symmetrization of any $s$ indices of $\Omega_{(1)}{}^{A(s-1), B(s-1)}$
gives zero.

The Lorentz-irreducible HS fields $ \Upsilon{}^{a(s-1),\, b(t)}$ result
 from the field $\Omega_{(1)}{}^{A(s-1), B(s-1)}$ by means
of the reduction procedure described in section
\ref{dim_comp}. In particular, the component of
$\Omega_{(1)}{}^{A(s-1), B(s-1)}$, that is most parallel to the
compensator $V^A$, is the physical frame-like field \be
\gl^{s-1}e^{A(s-1)} = \Omega_{(1)}^{A(s-1), B(s-1)} V_B \cdots
V_B\;.
\ee
(Note that a contraction of $s$  or more
indices of $\Omega_{(1)}{}^{A(s-1), B(s-1)}$
with the compensator  gives zero by the Young properties.)
The less $V^A$-longitudinal components identify with the
other fields in the set (\ref{sym_fr_fields}).

The linearized curvature is defined as the $o(d-1,2)$ covariant
derivative of the HS connection $1$-form  (\ref{frO}) \be
\label{R1A} R_{(2)}{}^{A(s-1),\, B(s-1)} = D_0
\Omega_{(1)}{}^{A(s-1),\, B(s-1)}\,. \ee
Due to the
zero-curvature condition (\ref{zerocurv}), the curvature
(\ref{R1A}) is  invariant under the linearized HS gauge
transformations
\be
\label{litr}
\delta \Omega_{(1)}{}^{A(s-1),
B(s-1)} = D_0 \xi_{(0)} {}^{A(s-1), B(s-1)}
\ee
with the traceless 0-form gauge parameter of the Young symmetry (\ref{dia2}).

Being decomposed into Lorentz components, the gauge transformation
low (\ref{litr}) reproduces gauge transformations for the fields
$\Upsilon$. In particular, the maximally $V$-tangential part of
(\ref{litr}) reproduces the gauge transformation (\ref{ltr}) for
the frame-like field $e^{A(s-1)}$.

The formulation in terms  of the Lorentz 1-forms
$\Upsilon$ is equivalent to that in terms
of the 1-form field $\Go_{(1)}$. The advantage of the latter formulation
is that it has simple algebraic meaning,
operating with the HS connection that takes values in a
single irreducible representation of the $AdS$ algebra $o(d-1,2)$.

\subsection{Mixed-symmetry higher-spin gauge fields
 in $AdS$ basis}
\label{sec_frame-like HS fields}

The  described approach  admits a straightforward
generalization to  massless HS fields of any symmetry
type. Consider a $\ads$  spin ${\bf
s}=(\underbrace{s,\ldots,s}_{p},s_{p+1},\ldots,s_q,0,\ldots,0)$
massless  field characterized by
the following $o(d-1)$ Young tableau of the lowest energy (vacuum) state
\be \label{summary_vac_}
\bigskip
\begin{picture}(100,55)(-1,40)

\put(110,72){\footnotesize $p$} \put(67,50){\footnotesize
$s_{p+1}$} \put(50,95){\footnotesize $s$}
\put(48,40){\footnotesize $s_{p+2}$} \put(25,0){\footnotesize
$s_{q}$}

\put(10,30){\circle*{1}} \put(10,20){\circle*{1}}
\put(20,30){\circle*{1}} \put(30,30){\circle*{1}}
\put(20,20){\circle*{1}}


\put(10,0){\line(0,1){10}} \put(20,0){\line(0,1){10}}
\put(00,0){\line(1,0){20}} \put(00,10){\line(1,0){20}}


\put(00,90){\line(1,0){100}} \put(00,80){\line(1,0){100}}
\put(00,70){\line(1,0){100}} \put(00,60){\line(1,0){100}}

\put(00,80){\line(0,1){10}} \put(00,70){\line(0,1){10}}

\put(00,00){\line(0,1){90}}

\put(10,60){\line(0,1){30}} \put(20,60){\line(0,1){30}}
\put(30,60){\line(0,1){30}} \put(40,60){\line(0,1){30}}
\put(50,60){\line(0,1){30}} \put(60,60){\line(0,1){30}}
\put(70,60){\line(0,1){30}} \put(80,60){\line(0,1){30}}
\put(90,60){\line(0,1){30}} \put(100,60){\line(0,1){30}}


\put(00,50){\line(1,0){60}} \put(00,40){\line(1,0){40}}

\put(10,40){\line(0,1){20}} \put(20,40){\line(0,1){20}}
\put(30,40){\line(0,1){20}} \put(40,40){\line(0,1){20}}
\put(50,50){\line(0,1){10}} \put(60,50){\line(0,1){10}}

\end{picture}
\ee \vspace{0.6cm}

\noindent that has the upper block of length $s$ and height $p$.
As shown in  \cite{ASV1}, in the frame-like formulation, its field
dynamics  can be described by  a $p$-form gauge
field \be \label{ads_field}
\Omega_{(p)}{}^{A_0(s-1),\,\ldots\,,A_p(s-1),\,A_{p+1}(s_{p+1}),\,
\ldots\,,A_{q}(s_q)}(x)\,, \ee
that takes values in the traceless
$o(d-1,2)$  tensor module corresponding to the Young
tableau
$Y(\underbrace{s-1,\ldots,s-1}_{p+1},s_{p+1},\ldots,s_q,0,\ldots,0)$

\be \label{pic:ads_YT}
\bigskip
\begin{picture}(100,55)(-1,40)

\put(100,77){\footnotesize $p+1$} \put(67,50){\footnotesize
$s_{p+1}$} \put(40,104){\footnotesize $s-1$}
\put(48,40){\footnotesize $s_{p+2}$} \put(25,0){\footnotesize
$s_{q}$}

\put(10,30){\circle*{1}} \put(10,20){\circle*{1}}
\put(20,30){\circle*{1}} \put(30,30){\circle*{1}}
\put(20,20){\circle*{1}}


\put(10,0){\line(0,1){10}} \put(20,0){\line(0,1){10}}
\put(00,0){\line(1,0){20}} \put(00,10){\line(1,0){20}}

\put(00,100){\line(1,0){90}} \put(00,90){\line(1,0){90}}
\put(00,80){\line(1,0){90}} \put(00,70){\line(1,0){90}}
\put(00,60){\line(1,0){90}}

\put(00,80){\line(0,1){10}} \put(00,70){\line(0,1){10}}

\put(00,00){\line(0,1){100}}

\put(10,60){\line(0,1){40}} \put(20,60){\line(0,1){40}}
\put(30,60){\line(0,1){40}} \put(40,60){\line(0,1){40}}
\put(50,60){\line(0,1){40}} \put(60,60){\line(0,1){40}}
\put(70,60){\line(0,1){40}} \put(80,60){\line(0,1){40}}
\put(90,60){\line(0,1){40}}


\put(00,50){\line(1,0){60}} \put(00,40){\line(1,0){40}}

\put(10,40){\line(0,1){20}} \put(20,40){\line(0,1){20}}
\put(30,40){\line(0,1){20}} \put(40,40){\line(0,1){20}}
\put(50,50){\line(0,1){10}} \put(60,50){\line(0,1){10}}

\end{picture}
\ee \vspace{0.6cm}

\noindent A simple mnemonic rule is that, to obtain the $\ads$
tensor representation carried by a HS gauge connection
(\ref{ads_field}),  one adds the longest row to the
$o(d-1)$  Young tableau
of the vacuum energy representation under consideration
(\ref{summary_vac_}) and then cuts the shortest column. The
resulting gauge field is a $p$-form where $p$ is the height of
the rightmost column of the original vacuum representation.

The gauge transformation is
\be
\label{adsgaugetr}
\delta\Omega_{(p)}{}^{A_0(s-1),\, \ldots\, ,\,A_{q}(s_q)} = D_0
\xi_{(p-1)}{}^{A_0(s-1),\, \ldots\, ,\,A_{q}(s_q)}\;,
\ee
where a traceless tensor
$\xi_{(p-1)}{}^{A_0(s-1),\,\ldots\,,\,A_{q}(s_q)}$ is a
$(p-1)$-form gauge parameter that takes values in the
same representation of $o(d-1.2)$. There is a set of level-$l$
($1\leq l \leq p-1$) gauge parameters and gauge transformations
 \be \delta\xi_{(p-l)}{}^{A_0(s-1),\, ...\,
,\,A_{q}(s_q)} = D_0 \xi_{(p-l-1)}{}^{A_0(s-1),\, ...\,
,\,A_{q}(s_q)}\;, \qquad l=1,\ldots,p-1\,. \ee From the
zero-curvature condition (\ref{zerocurv}) it follows that the $(p+1)$-form curvature
associated with the $p$-form gauge field
\be \label{adscurv}
R_{(p+1)}{}^{A_0(s-1),\, \ldots\, ,\,A_{q}(s_q)} = D_0
\Omega_{(p)}{}^{A_0(s-1),\, \ldots\, ,\,A_{q}(s_q)}\, \ee
is invariant under the gauge transformations
(\ref{adsgaugetr})
\be
\delta R_{(p+1)}=0
\ee
 and
satisfies the Bianchi identities \be\label{Bianchi} D_0
R_{(p+1)}{}^{A_0(s-1),\, \ldots\, ,\,A_{q}(s_q)} = 0\;.
\ee

Also, we shall use the $p$-form gauge field (\ref{ads_field}) rewritten
in the antisymmetric basis \be \label{ads_field2}
\Omega_{(p)}{}^{A_1[\tilde{h}_1],\,...\,,A_{s-1}[\tilde{h}_{s-1}]}\,.
\ee
Here $\tilde{h}_1\geq\cdots\geq\tilde{h}_{s-1}\geq p+1$ are
the heights of the columns of the $o(d-1,2)$ Young tableau
(\ref{pic:ads_YT}). In the sequel we shall switch freely
between the symmetric and antisymmetric descriptions of the
frame-like HS fields.

\subsection{Mixed-symmetry higher-spin gauge fields in Lorentz basis}
\label{sec: lorframelike}

The dynamical content of the frame-like formulation
of massless HS fields is most conveniently analyzed
in terms of Lorentz-tensor components of a HS field.
To this end, the $o(d-1,2)$-module carried by the $p$-form field  $\Go_{(p)}$ should be decomposed
into  $o(d-1,1)$-modules. According to section \ref{dim_comp},
the result of the decomposition of the $p$-form field $\Go_{(p)}$
with tangent indices associated with
traceless $AdS$  Young tableau (\ref{pic:ads_YT}) into a set of
Lorentz-covariant $p$-form fields is
\be \label{lor_col}
\Omega_{(p)} = \bigoplus_{(t_1, ..., \,t_k)}
\lambda^{(s-1-\sum\limits_{I} t_I)} \Upsilon_{(p)}{}^{(t_1, ...,
\,t_k)}\;,
\ee
where  fields $\Upsilon_{(p)}{}^{(t_1, ...,
\,t_k)}$,
parameterized by integers $t_I$ (\ref{t}), have tangent
indices corresponding to various traceless $o(d-1,1)$ Young tableaux of
the form (\ref{dimred}). The $\lambda$-dependent factors are introduced to
highlight  that different Lorentz-covariant fields will
 have different mass dimensions compatible with the flat limit
 $\lambda \to 0$.

 The following
classification of Lorentz-covariant $p$-form fields is motivated by
their different dynamical roles \cite{ASV1,ASV2}.
The field of the set (\ref{lor_col}) with the minimal number of
cells is called \textit{physical}.
Auxiliary fields are those with the Lorentz-covariant components
$\{\Upsilon_{(p)}{}^{(0, ...,0,1,\,0, ...,\,0)}\}$.
 In other words, auxiliary
fields have one more Lorentz index  compared to
the physical field. We distinguish between
\textit{relevant}  auxiliary field, that has an additional cell in
the first column, and \textit{irrelevant} auxiliary fields
that have additional cells in any other columns.
\textit{Extra fields} are those that  have two or
more Lorentz indices  compared to the physical field.
More precisely,

\begin{itemize}

\item Physical field $\Upsilon_{(p)}{}^{(0, ..., \,0)}\equiv e_{(p)}$ has tangent Lorentz
indices described by the $o(d-1,1)$ Young tableau of the form

\be \label{pic:phys}
\bigskip
\begin{picture}(100,55)(-1,40)

\put(100,73){\footnotesize $p$} \put(67,50){\footnotesize
$s_{p+1}$} \put(40,96){\footnotesize $s-1$}
\put(48,40){\footnotesize $s_{p+2}$} \put(25,0){\footnotesize
$s_{q}$}

\put(10,30){\circle*{1}} \put(10,20){\circle*{1}}
\put(20,30){\circle*{1}} \put(30,30){\circle*{1}}
\put(20,20){\circle*{1}}


\put(10,0){\line(0,1){10}} \put(20,0){\line(0,1){10}}
\put(00,0){\line(1,0){20}} \put(00,10){\line(1,0){20}}


\put(00,90){\line(1,0){90}} \put(00,80){\line(1,0){90}}
\put(00,70){\line(1,0){90}} \put(00,60){\line(1,0){90}}

\put(00,80){\line(0,1){10}} \put(00,70){\line(0,1){10}}

\put(00,00){\line(0,1){90}}

\put(10,60){\line(0,1){30}} \put(20,60){\line(0,1){30}}
\put(30,60){\line(0,1){30}} \put(40,60){\line(0,1){30}}
\put(50,60){\line(0,1){30}} \put(60,60){\line(0,1){30}}
\put(70,60){\line(0,1){30}} \put(80,60){\line(0,1){30}}
\put(90,60){\line(0,1){30}}


\put(00,50){\line(1,0){60}} \put(00,40){\line(1,0){40}}

\put(10,40){\line(0,1){20}} \put(20,40){\line(0,1){20}}
\put(30,40){\line(0,1){20}} \put(40,40){\line(0,1){20}}
\put(50,50){\line(0,1){10}} \put(60,50){\line(0,1){10}}

\end{picture}
\ee \vspace{0.6cm}

\noindent It is identified with the maximally
$V$-tangential component of the $\ads$ field (\ref{pic:ads_YT})
obtained by contracting some its $s-1$ indices with $V^A$,
\textit{i.e.} the $o(d-1,2)$ covariant expression for the
physical field is
 \be \label{phys_field}
\lambda^{s-1}e_{(p)}{}^{A_1(s-1),\,\, ...\,, A_p(s-1),\,
A_{p+1}(s_{p+1}),\,...\,, A_{q}(s_{q})} =\underbrace{V_{A_0}
\ldots V_{A_0}}_{s-1} \Omega_{(p)}{}^{A_0(s-1),\,A_1(s-1),\,
...\, ,\,A_{q}(s_q)}\,.
\ee
Recall that contraction of any $s$ indices of $\Omega_{(p)}{}^{A_0(s-1),\, ...\, ,\,A_{q}(s_q)}$
with $V^A$ gives zero because of the Young properties of
$\Omega_{(p)}{}^{A_0(s-1),\, ...\, ,\,A_{q}(s_q)}$. This $V^A$
transversality  effectively means
that all indices of the physical field $e_{(p)}$
(\ref{phys_field}) are Lorentz indices.

In the antisymmetric basis
(\ref{ads_field2}), the analog of (\ref{phys_field})
 is
  \be \label{phys}
\lambda^{s-1}e_{(p)}{}^{A_1[\tilde{h}_1-1],\,\ldots\,,A_{s-1}[\tilde{h}_{s-1}-1]}
=V_{A_1} \ldots V_{A_{s-1}}
\Omega_{(p)}{}^{A_1[\tilde{h}_1],\,A_2[\tilde{h}_2],\,\ldots\,,\,A_{s-1}[\tilde{h}_{s-1}]}\,.
\ee

\end{itemize}

\begin{itemize}
\item Relevant auxiliary field $\Upsilon_{(p)}{}^{(0, ...,0, \,1)}\equiv \omega_{(p)}$ is described by the  $o(d-1,1)$ Young
tableau of the form

\be \label{pic:aux}
\bigskip
\begin{picture}(100,55)(-1,40)

\put(100,73){\footnotesize $p$} \put(67,50){\footnotesize
$s_{p+1}$} \put(40,96){\footnotesize $s-1$}
\put(48,40){\footnotesize $s_{p+2}$} \put(25,0){\footnotesize
$s_{q}$}

\put(10,30){\circle*{1}} \put(10,20){\circle*{1}}
\put(20,30){\circle*{1}} \put(30,30){\circle*{1}}
\put(20,20){\circle*{1}}


\put(10,0){\line(0,1){10}} \put(20,0){\line(0,1){10}}
\put(00,0){\line(1,0){20}} \put(00,10){\line(1,0){20}}


\put(00,90){\line(1,0){90}} \put(00,80){\line(1,0){90}}
\put(00,70){\line(1,0){90}} \put(00,60){\line(1,0){90}}

\put(00,80){\line(0,1){10}} \put(00,70){\line(0,1){10}}

\put(00,00){\line(0,1){90}}

\put(10,60){\line(0,1){30}} \put(20,60){\line(0,1){30}}
\put(30,60){\line(0,1){30}} \put(40,60){\line(0,1){30}}
\put(50,60){\line(0,1){30}} \put(60,60){\line(0,1){30}}
\put(70,60){\line(0,1){30}} \put(80,60){\line(0,1){30}}
\put(90,60){\line(0,1){30}}


\put(00,50){\line(1,0){60}} \put(00,40){\line(1,0){40}}

\put(10,40){\line(0,1){20}} \put(20,40){\line(0,1){20}}
\put(30,40){\line(0,1){20}} \put(40,40){\line(0,1){20}}
\put(50,50){\line(0,1){10}} \put(60,50){\line(0,1){10}}


\put(0,-10){\line(0,1){10}} \put(10,-10){\line(0,1){10}}
\put(0,-10){\line(1,0){10}}
\end{picture}
\ee \vspace{0.6cm}

\noindent It has an additional cell in the first column compared to
the Young tableau of the physical field (\ref{pic:phys}). The
$o(d-1,2)$ covariant
expression for the relevant auxiliary field
in the antisymmetric basis is
\be \ba{c}
\label{true_auxiliary} \dps
\gl^{s-2}\go_{(p)}{}^{A_1[\tilde{h}_1],A_2[\tilde{h}_2-1],\,\ldots\,,
A_{s-1}[\tilde{h}_{s-1}-1]}= V_{A_2}\cdots V_{A_{s-1}}
\Omega_{(p)}{}^{A_1[\tilde{h}_1],A_2[\tilde{h}_2],\,\ldots\,,A_{s-1}[\tilde{h}_{s-1}]}
\\
\\
- \tilde{h}_1V^{A_1}V_{A_2} \ldots V_{A_{s-1}}V_B
\,\Omega_{(p)}{}^{BA_1[\tilde{h}_1-1],\,\ldots\,,A_{s-1}[\tilde{h}_{s-1}]}\;.
\ea \ee
It is easy to see that the tensor on the right hand side of
(\ref{true_auxiliary})  has the correct Young
symmetry and is $V^A$-transversal.
The $o(d-1,2)$ covariant expression for the relevant auxiliary field
in the symmetric basis is not given here because it
is more involved, requiring explicit Young symmetry projectors.
Note that in the case of gravity, the  formula
(\ref{true_auxiliary}) gives, as expected, $\omega_1^{AB} = \Go_1^{AB}
-\lambda\,(E_1^{A}\,V^{B}-E_1^{B}\,V^{A})$ (\ref{2}).

\item Irrelevant auxiliary fields $\{\Upsilon_{(p)}{}^{(0, ...,0,1,\,0, ...,\,0)}\}\equiv\{\omega^\prime_{(p)}\}$
have an additional cell compared to the Young tableau of the physical
field, that is situated in any column except for the first one.
 Generally, if a field under consideration is described by a
$\ads$ tangent Young tableau which consists of $N$ blocks, it gives rise to
 $N-1$ irrelevant auxiliary fields. In
particular, for the case of totally symmetric fields $N=1$ and,
in agreement with \cite{LV}, there is only one auxiliary field,
namely, the relevant one, while the
 irrelevant auxiliary fields do not appear.
 The same  is true for any HS field described in the frame-like formalism
 by a $p$-form that takes values in some rectangular Young
tableau.

\end{itemize}

\begin{itemize}

\item Extra fields $\{\Upsilon_{(p)}{}^{(t_1, ..., \,t_k)},\;
\sum\limits_I t_I \geq 2 \}\equiv w_{(p)}$ have two or more additional cells
compared to Young tableau of the physical field.

\end{itemize}

\noindent
HS curvatures and gauge parameters admit the
decompositions analogous to (\ref{lor_col})
\be \label{lor_col_cur} R_{(p+1)} =
\bigoplus_{(t_1, ..., \,t_k)} \lambda^{(s-1-\sum\limits_{I} t_I)}
{\cal R}_{(p+1)}{}^{(t_1, ..., \,t_k)}\,,
\ee
  \be \label{lor_col_gaug} \xi_{(p-1)}= \bigoplus_{(t_1,
..., \,t_k)} \lambda^{(s-1-\sum\limits_{I} t_I)}
\varepsilon_{(p-1)}{}^{(t_1, ..., \,t_k)}\;. \ee
From these
decompositions and (\ref{adscurv}) it follows that
Lorentz-covariant components of the HS curvature ${\cal R}_{(p+1)}{}^{(t_1, ...,
\,t_k)}$ have the form
\be
\label{curv_sigma}
{\cal R}_{(p+1)}{}^{(t_1, ..., \,t_k)}={\cal
D}(\Upsilon_{(p)}){}^{(t_1, ..., \,t_k)}+
\sum_{I=1}^k\Big(\sigma^{(I)}_-(\Upsilon_{(p)}){}^{(t_1, ...,
\,t_k)} +\lambda^2\,\sigma^{(I)}_+ (\Upsilon_{(p)}){}^{(t_1, ...,
\,t_k)} \Big)\,,
\ee
where the sigma-operators have the following structure
\be \label{sigma-}
\sigma^{(I)}_- (\Upsilon_{(p)}){}^{(t_1, ..., t_i,..., \,t_k)}
\equiv \alpha_I (t) {\cal P}_-^{(I)} \Big(h\wedge
\Upsilon_{(p)}{}^{(t_1, ..., t_I+1,..., \,t_k)}\Big)\;, \ee \be
\label{sigma+} \sigma_+^{(I)} (\Upsilon_{(p)}){}^{(t_1, .
..,
t_I,..., \,t_k)}\equiv \beta_I (t) {\cal P}_+^{(I)}\Big(h\wedge
\Upsilon_{(p)}{}^{(t_1, ..., t_I-1,..., \,t_k)}\Big)\;. \ee
Here the background frame 1-form  $h$ contracts an extra index
of $\Upsilon_{(p)}{}^{(t_1, ..., t_I+1,..., \,t_k)}$ in (\ref{sigma-})
and adds a missed index of $\Upsilon_{(p)}{}^{(t_1, ..., t_I-1,..., \,t_k)}$
 in (\ref{sigma+}), ${\cal P}_\pm^{(I)}$ are the projectors to the
 irreducible $o(d-1,1)$ module carried by
 $\Upsilon_{(p)}){}^{(t_1, ..., t_I,..., \,t_k)}$ and $\alpha_I (t)$
 and $\beta_I (t)$ are some coefficients.

The gauge transformations  have analogous form
\be
\label{gtr}
\delta\Upsilon_{(p)}{}^{(t_1, ..., \,t_k)}={\cal
D}(\varepsilon_{(p-1)}){}^{(t_1, ..., \,t_k)}+
\sum_{I=1}^k\Big(\sigma^{(I)}_-(\varepsilon_{(p-1)}){}^{(t_1, ...,
\,t_k)} +\lambda^2\,\sigma^{(I)}_+ (\varepsilon_{(p-1)}){}^{(t_1, ...,
\,t_k)} \Big)
\ee

As a consequence of the flatness of $AdS$ covariant derivative,
$D^2_0=0$, the sigma-operators satisfy the  relations \be
\label{flat_cond_sig} {\cal D}^2 + \lambda^2\,\sum\limits_{I=0}^k
\{\sigma^{(I)}_-,\sigma_+^{(I)}\} =0\;, \qquad
\{\sigma^{(I)}_\pm,\sigma^{(J)}_\pm\}=0\;, \qquad \{{\cal
D},\sigma^{(I)}_\pm\}=0\;. \ee
Other way around, these conditions
determine the  coefficients $\alpha_i (t)$ and $\beta_i
(t)$ of sigma-operators (\ref{sigma-}) and (\ref{sigma+}) up to
free parameters which manifest the  rescaling ambiguity
of $\Upsilon_{(p)}){}^{(t_1, ..., t_I,..., \,t_k)}$.

\subsection{Frame versus metric}

To explain  how  a metric-like HS field of a given
spin is encoded in the corresponding physical  frame-like
HS field let us analyze  its gauge
transformation law. Introducing schematic notations for gauge
parameter associated with the physical field,
$\varepsilon_{(p-1)} $ and with the  auxiliary fields,
$\varepsilon_{(p-1)}^{(I)}$, from (\ref{gtr}) we obtain
\be
\label{gaugelaw}
\delta e_{(p)} = {\cal D} \varepsilon_{(p-1)}{} +
\sum_{I=1}^{k}\sigma_-^{(I)}\varepsilon_{(p-1)}^{(I)}\;.
\ee
The part of the gauge transformation with the parameters
$\varepsilon_{(p-1)}^{(I)}$ is Stueckelberg, allowing
to gauge away some of components of the physical field.
The metric-like field is the part of the physical field
invariant under the Stueckelberg symmetries, pretty much like
the usual metric tensor can be understood as a part of the frame
invariant under the Lorentz transformations.

To find out  which components of the physical field are invariant
under the Stueckelberg shift transform
it is convenient to convert all world indices of the physical
field and gauge parameters in (\ref{gaugelaw}) into the
tangent ones.  For the physical field the result is
\be
\label{phys_field_conv}
\ba{c} \hspace{-50mm} e^{[n_1 ...
n_p];\;a_1(s-1),\,\, ...\,, a_p(s-1),\, a_{p+1}(s_{p+1}),\,...\,,
a_{q}(s_{q})}=
\\
\\
\hspace{20mm}=h_{\underline{m}_1}{}^{n_1} \ldots
h_{\underline{m}_p}{}^{n_p}\;
e^{[\underline{m}_1...\underline{m}_p];\;}{}^{a_1(s-1),\,\,
...\,, a_p(s-1),\, a_{p+1}(s_{p+1}),\,...\,, a_{q}(s_{q})}\;. \ea
\ee
Conversions of the gauge parameters is analogous.

The metric-like HS field $\Phi^{a(s), ... , a_q(s_q)}(x)$
is the following component of the physical field
\be \label{fo} \Phi^{a_1(s),\,
...\,,a_p(s)\,,a_{p+1}(s_{p+1})\,,..., a_{q}(s_{q})} = e^{[a_1
...a_p];\;}{}^{a_1(s-1),\,\, ...\,, a_p(s-1), \,
a_{p+1}(s_{p+1}),\,...\,, a_{q}(s_{q})}\,,
\ee
\textit{i.e.} it results from
symmetrization of the form indices  with
tangent indices of the first $p$ rows of $e_{(p)}$.

To check the invariance of the metric-like field defined according to the formula (\ref{fo})
under Stueckelberg shift symmetry (\ref{gaugelaw}) with the  gauge parameter
$\varepsilon_{(p-1)}^{(I)}$
\be
\label{stuc0}
\ba{c}
\hspace{-50mm}\delta_{(I)} \Phi^{a_1(s),\, ...\,,a_p(s)\,,a_{p+1}(s_{p+1})\,,..., a_{q}(s_{q})}\equiv
\\
\\
\hspace{20mm}\equiv\delta_{(I)} e^{[a_1 ...a_p];\;}{}^{a_1(s-1),\,\, ...\,, a_p(s-1), \,
a_{p+1}(s_{p+1}),\,...\,, a_{q}(s_{q})} = 0
\ea
\ee
one rewrites (\ref{gaugelaw})  as
\be
\label{stuc}
\ba{c}
\hspace{-50mm}\delta_{(I)} e^{n[p];\;a_1(s-1),\,\, ...\,, a_p(s-1),\, a_{p+1}(s_{p+1}),\,...\,,
a_{q}(s_{q})}=
\\
\\
\hspace{20mm}=\cP_{(I)}\Big(\varepsilon^{(I)\;\;n[p-1];\,a_1(s-1),\,\, ...\,, a_p(s-1),\, a_{p+1}(s_{p+1}),\,...\,,n a_I(s_I),\, ...\,,
a_{q}(s_{q})}\Big)\;,
\ea
\ee
where $\cP_{(I)}$ projects r.h.s. of (\ref{stuc}) on the Young tableau associated with the indices
of the field $e_{(p)}$. Substituting the expression (\ref{stuc}) into the variation of the metric-like field
(\ref{stuc0}) one obtains that the index $n$ from the $I$-th row of the Stueckelberg
gauge parameter in the formula (\ref{stuc}) is symmetrized with all indices from a row of its uppermost
rectangular block what gives zero by virtue of Young symmetry properties.

Also the invariance (\ref{stuc0}) can be proved by checking that all remaining
components of the physical field $e_{(p)}$ are Stueckelberg
with respect to the gauge parameters $\varepsilon_{(p-1)}^{(I)}$.
This can be achieved by comparing the contents of the tensor
product of the antisymmetric world Lorentz modules associated with
the differential forms with the tangent Lorentz modules carried,
respectively by the physical $p$-form on the one side
and the auxiliary  $p-1$ form gauge parameters on the other side
(modulo the level 2 gauge symmetries).

As a consequence  of the formula (\ref{fo}),
the metric-like field $\Phi$ satisfies some
tracelessness conditions because the physical frame-like field
is traceless in the tangent Lorentz indices. Namely, it follows
that the tensor $\Phi^{a_1(s), ...\,,\, a_{q}(s_{q})}$
satisfies the tracelessness conditions (\ref{tr1}) and  (\ref{tr2})
\cite{ASV1}, {\it i.e.}
 the metric-like field $\Phi$ belongs to the
tensor space $B_p^{d-1,1}(s,\ldots,s_q,0,\ldots,0)$ of section
\ref{Trace conditions on the metric-like field}.
The Fronsdal double tracelessness condition for totally
symmetric fields
$
\Phi^{a(s)} \in
B_1^{(d-1,1)}(s,0,...,0)\;,
$
is the particular case of the tracelessness conditions for
a general mixed-symmetry field.

As an illustration, let us consider a spin $(2,1,0,...,0)$
mixed-symmetry field described by the three-cell "hook" tableau (for more examples, see
\cite{ASV1,A2}).
Its frame-like description gives rise to the
physical $1$-form  field $e_{\underline{m}}{}^{a,b}$
which is an antisymmetric Lorentz tensor. There is just one auxiliary
field carrying  three antisymmetrized indices. The corresponding gauge
transformation  (\ref{gaugelaw}) is
\be
\label{gauge_hook1} \delta e_{\underline{m}}{}^{a,b} = {\cal
D}_{\underline{m}} \,  \varepsilon^{a,b} +
h_{\underline{m};\,c}\,\varepsilon^{a,b,c}\;, \ee
or, converting world indices  into tangent ones,
\be \label{gauge_hook2}
\delta e^{m;\;a,b} = {\cal D}^{m} \,  \varepsilon^{a,b} +
\varepsilon^{m,b,c}\;. \ee
A set of traceless $o(d-1,1)$
tensor components contained in the physical field is given by the
following tensor product

\be \label{tensorprod}
\begin{picture}(12,12)(-1,4)

\put(00,17){\line(1,0){10}} \put(00,7){\line(1,0){10}}
\put(00,-3){\line(1,0){10}} \put(0,-3){\line(0,1){20}}
\put(10,-3){\line(0,1){20}}

\end{picture}
\;\;\; \otimes\;\;
\begin{picture}(12,12)(-1,4)

\put(00,13){\line(1,0){10}} \put(00,3){\line(1,0){10}}
\put(10,3){\line(0,1){10}} \put(0,3){\line(0,1){10}}

\end{picture}
\;\; = \;
\begin{picture}(12,12)(-1,4)

\put(00,17){\line(1,0){20}} \put(00,7){\line(1,0){20}}
\put(00,-3){\line(1,0){10}} \put(0,-3){\line(0,1){20}}
\put(10,-3){\line(0,1){20}} \put(20,7){\line(0,1){10}}

\end{picture}
\;\;\;\;\oplus \;\;
\begin{picture}(12,12)(-1,4)

\put(00,23){\line(1,0){10}} \put(00,13){\line(1,0){10}}
\put(00,3){\line(1,0){10}} \put(00,-7){\line(1,0){10}}
\put(0,-7){\line(0,1){30}} \put(10,-7){\line(0,1){30}}

\end{picture}
\;\;\oplus\;\;
\begin{picture}(12,12)(-1,4)

\put(00,13){\line(1,0){10}} \put(00,3){\line(1,0){10}}
\put(10,3){\line(0,1){10}} \put(0,3){\line(0,1){10}}

\end{picture}
\ee \vspace{1mm}

\noindent By virtue of the Stueckelberg symmetry part of
(\ref{gauge_hook2}) with the antisymmetric parameter
$\varepsilon^{a,b,c}$  the second component in
(\ref{tensorprod}) can be gauged away.
The remaining first and third
components form a tracefull tensor $\Phi^{ab,c}$
given by (\ref{fo}),
\be
\Phi^{ab,c}= \frac{1}{2}(e^{a;\,b,c}+e^{b;\,a,c})\;. \ee
The gauge transformation law for the field $\Phi^{ab,c}$
that results from (\ref{gauge_hook2}) is
\be
\delta \Phi^{ab,c}  = \cD^a \varepsilon^{b,c} + \cD^b \varepsilon^{a,c}\;.
\ee

Other way around, the metric-like formulation can be taken
as a starting point of the derivation of the
frame-like formulation of HS fields. Namely, given a metric-like
HS gauge field, one adds  Stueckelberg components so that
together with the original metric-like field they
form a $p$-form gauge
field which carries tangent Lorentz indices associated with some
irreducible $o(d-1,1)$-module.
This $p$-form field is the physical field and its
transformation law is postulated to be (\ref{gaugelaw}). The
frame-like machinery  evolves further by introducing the new gauge
fields associated with Stueckelberg shift parameters. These will
be auxiliary fields with the  gauge transformation law that contains
 the Lorentz derivative acting on the
Stueckelberg shift parameters. In addition, there will be
some new shift parameters in the gauge transformation of the
auxiliary fields. In their turn, these new shift
parameters require new gauge fields which are extra fields.
This procedure continues further to obtain a full set of physical,
auxiliary and extra fields necessary to construct
curvature $(p+1)$-forms
manifestly invariant under the full set of gauge
symmetries. As explained in  section \ref{sec: lorframelike}, the resulting set forms
 a $p$-form gauge field taking values in the irreducible
 $o(d-1,2)$-module described by the Young tableau (\ref{pic:ads_YT}).

\section{General properties of a higher-spin action}
\label{sec:HSactions}

\subsection{Background}
\label{Background}

As shown in \cite{Metsaev,BMV} generic massless fields in $AdS_d$
are different from the massless fields in Minkowski space in the
sense that an irreducible gauge field in $AdS_d$ reduces
in the flat limit to a number of massless fields of different types
in Minkowski space. This effect has
clear interpretation in terms of representations of the $AdS_d$
algebra $o(d-1,2)$.

As discussed in section \ref{Introduction},
the space of single-particle states of a given relativistic
field with the energy bounded from below forms a lowest weight
 $o(d-1,2)$-module, $D(E_0,{\bf s})$. Its lowest weight is defined in
terms of the lowest energy $E_0$ and  spin ${\bf s}$ associated with the
weights of the maximal compact subalgebra $o(2)\oplus o(d-1)
\subset o(d-1,2)$. For quantum-mechanically consistent fields, the
modules $D(E_0,{\bf s})$  correspond to unitary representations of
$o(d-1,2)$. Unitarity requires  $E_0 \geq E_0 ({\bf s})$ where
$E_0 ({\bf s})$ is some function of the spin $s$ found for the case of
$d=4$ in \cite{Evans} and for the general case in
\cite{Metsaev}.
 At $E_0 = E_0 ({\bf s})$ null states appear that form a submodule to be
factored out. These signal a gauge symmetry of the system. The
corresponding fields are the usual massless fields.

Modules with
lower energies $E< E_0({\bf s})$ correspond to nonunitary (ghost)
massive fields. At certain singular values $E_i ({\bf s})$ of $E$ the
corresponding nonunitary module may contain a submodule that again
signals a gauge symmetry in the field-theoretical description.
Such modules correspond either to the partially massless fields
\cite{DW,Zin,SV} or to ``non-unitary massless fields". More specifically,
one can see that a $o(d-1)$ tensor module carried by the vacuum space of
 a submodule (called singular space) corresponds to the
$o(d-1,1)$ tensor module carried by the associated gauge symmetry
parameter while a level, at which the singular space
appears, equals to a highest order of derivatives that act on the
gauge parameter in the gauge transformation law. The parameters
that enter the gauge transformation law with one derivative correspond
to different massless fields in $AdS_d$. Those that enter the gauge
transformation law with two or more  derivatives correspond to partially
massless fields.

Since singular  energies $E_i ({\bf s})$ are scaled in units of $AdS_d$
curvature $\lambda$, {\it i.e.}
\be
E_i({\bf s}) = \lambda e_i ({\bf s})\,,
\ee
where $e_i$ is $\lambda$-independent, all special energies
tend to zero in the flat limit $\lambda \to 0$ so that all gauge
symmetries are inherited by  one massless theory in Minkowski space.
More precisely, different gauge symmetries of unitary and non-unitary
 massless theories
in $AdS_d$ become different gauge symmetries of the same massless theory
in Minkowski space while the flat limit of partially massless
gauge symmetries seem to correspond to particular flat massless gauge
symmetries with the gauge parameters expressed via derivatives of some
other tensors identified with the gauge parameters of
partially massless models in $\ads$.

The gauge parameter associated with the unitary massless field is most
antisymmetric, resulting from cutting a cell from the shortest column
of the $o(d-1)$ Young tableau of the vacuum space of $D(E_0,{\bf s})$. All
other gauge symmetries resulting from different cell cuts correspond to
``non-unitary massless fields". These are absent in a consistent $AdS_d$
massless theory but may re-appear in its flat limit.

\subsection{Degrees of freedom}
\label{sec:flat_sym}

An action functional that describes one or another dynamical system should
exhibit appropriate global symmetries and describe a correct number of degrees of
freedom. An irreducible dynamical
system carries  a minimal possible number of degrees of freedom.
 The reduction of a number of degrees of freedom is achieved either
via gauge symmetries or via constraints.

Generally, gauge symmetries
kill more degrees of freedom than constraints because they require $q+1$
gauge conditions for a gauge transformation $\delta \phi = \partial^q\epsilon$
that contains $q$ time derivatives of the gauge parameter. This fact
is most obviously seen in terms of the Dirac constraint dynamics
\cite{dirac,GT},
 where gauge symmetries correspond to first-class constraints
while constraint field equations correspond to second-class constraints.
The general situation can be illustrated by the example of massive and massless
spin one field in flat space.

The Proca equation for a massive spin one field is
\be
\label{proca}
\Box A_\mu(x) - \d_\mu\d^\nu A_\nu(x) + m^2 A_\mu(x) = 0\;,
\qquad
\nu = 0, ..., d-1\,.
\ee
For $m\neq 0$, the  model has no gauge symmetry and describes $d-1$ physical
degrees of freedom. Indeed, taking divergency of the left hand side of this equation
with $m\neq 0$, one obtains the Lorentz condition
\be
\label{lor}
m^2 \d^\nu A_\nu(x) =0\,.
\ee
As a result, the Lagrangian equation (\ref{proca}) is equivalent to
\be
\label{s1m}
\Box A_\mu(x)  + m^2 A_\mu(x) = 0\;,
\ee
and (\ref{lor}), thus describing $d-1$ degrees of freedom in $d$ dimensions.

Maxwell equations have the form  (\ref{proca}) with $m=0$
\be
\label{max}
\Box A_\mu(x) - \d_\mu\d^\nu A_\nu(x) = 0\;.
\ee
In this case, differentiating the left hand side by $\p^\mu$ one obtains
the Bianchi identity which manifests the spin one gauge symmetry
\be
\delta A_\mu = \p_\mu \epsilon\,.
\ee
This gauge symmetry kills two degrees of freedom so that a
massless spin one particle carries $d-2$ degrees of freedom.
Indeed, let us impose the Coulomb gauge condition
\be
\label{coul}
\p^i A_i =0 \qquad i=1,2\ldots, d-1 \,.
\ee
Then the equation (\ref{max}) with $\mu =0$ gives the constraint
\be
\label{coull}
\triangle A_0 =0\,,
\ee
where $\triangle = \p_i \p^i$ is the $(d-1)$-dimensional Laplace operator, which does not
contain time derivatives. For $A_0$ vanishing at space infinity this implies
\be
A_0=0\,.
\ee
Together with (\ref{coul}), (\ref{coull}) reduces a number of degrees of freedom
of the Maxwell theory to $d-2$. (Note that, in presence of
currents, the charge density appears on the right hand side of
the equation (\ref{coull}) which  reconstructs
the electric potential in terms of electric charge.)

We see that in the both cases it is crucial that
the highest derivative part of the field equations
has the correct form. If a relative coefficient between the two
second-order derivative terms in (\ref{proca})  was changed, both the
constraint in the massive case and the gauge symmetry in the massless
case would be lost and the system would describe some more degrees of
freedom (usually ghost-like).

More generally, the condition that a system
describes a minimal possible number of degrees of freedom is that
the term with highest derivatives should satisfy the Bianchi identities
that, for a Lagrangian system, is equivalent to the condition that it is
gauge invariant. This can be equivalently formulated in the form that
a limit of a theory with all dimensionful parameters tending to zero
(defined so that only the highest derivative terms survive) should
give a gauge theory with a maximal possible number of gauge symmetries.
For a theory in a curved space with a dimensionful
parameter $\lambda$ which characterizes its curvature, the analogous condition
should be imposed on the part of the action that survives in the flat limit
$\lambda \to 0$.
So, the condition that a theory in curved space should exhibit enhancement
of gauge symmetries in a flat limit
\be
\label{fl_gaug}
\delta_{\varepsilon}\Phi(x) = \sum\limits_{I} \d \varepsilon_I(x)\;,
\ee
is analogous to the gauge symmetry enhancement of the Proca equation in the massless limit.
The gauge parameters $\varepsilon_I(x)$ in (\ref{fl_gaug}) are described by the Young tableaux resulting from that
of the field $\Phi(x)$ by cutting off  a  cell from the last row of  any $I$-th horizontal
block.

More precisely, a HS action in $AdS_d$ has the form
\be
\label{adsf}
\dps{\cS}_2^{AdS} \sim \int \cD\Phi\cD\Phi + \lambda^2\Phi^2\,.
\ee
The  mass-like terms $\lambda^2\Phi^2$ break down all the gauge
symmetries (\ref{fl_gaug}) of the action
${\cS}_2^{flat}$ except for that associated with the $AdS_d$ gauge parameter.
However, in the flat limit $\lambda \to 0$,  the action
$\dps{\cS}_2^{flat} \sim \int \d\Phi\d\Phi$
exhibits enhancement of gauge symmetries.

Let us stress that the precise form of the kinetic term of the action
(\ref{adsf}) is not uniquely fixed by the gauge invariance condition
with respect to the true $AdS_d$ gauge symmetry. So the
requirement of the flat limit gauge
symmetry enhancement is an important additional condition
that determines the structure of the action.
Of course, when there are several different terms of subleading orders
 of derivatives, these should also be adjusted in a way implying the maximal
 reduction of degrees of freedom, i.e. preserving a maximal possible number of
 gauge symmetries and constraints.

Once the correct model of an irreducible  massless HS theory
of  a given spin $\bf s$ is formulated in terms of a field $\Phi(x)$,
it should be possible to impose various covariant irreducibility
tracelessness and  Lorentz-type conditions
\be
\label{ircon}
{\rm tr}\, \Phi(x) =0\;,
\qquad
{\cal D} \Phi(x) =0\;,
\ee
which are either gauge conditions for the gauge symmetry transformations
of the model
\be
\label{adstheory2}
\delta_\varepsilon \Phi(x) = {\cal D} \varepsilon(x)
\ee
or constraints which follow from the field equations as the spin one
Lorentz condition follows from the Proca equation.
As a result, the remaining field equations get the Klein-Gordon form
\be
\label{KG}
({\cal D}^2 + \lambda^2 \,M(d,{\bf s}))\,\Phi(x) = 0\;,
\ee
where the explicit form of $M(d,\bf s)$
\be
\label{met}
M(d,{\bf s})=\sum_{i=1}^{q}s_i-(s-p-1)(s-p-2+d)\;
\ee
was found by Metsaev in \cite{Metsaev}. This formula provides one more
check whether or not a theory under consideration describes properly
a massless field of a given type.

In fact, (\ref{KG}) fixes the quadratic Casimir operator of the
$\ads$ algebra realized on Lorentz-covariant
tensor fields satisfying the constraints (\ref{ircon}).
Note that the equation (\ref{KG}) possesses a leftover gauge
symmetry with the parameter
$\varepsilon(x)$ satisfying the irreducibility conditions
\be
{\rm tr}\, \varepsilon(x) =0\,,\qquad {\cal D} \varepsilon(x)=0\,
\ee
along with  certain differential conditions of the type (\ref{KG}).
The gauge parameters of this  symmetry form a
 $o(d-1,2)$-module  which is  the singular submodule
to be factored out ({\it i.e.,} gauged away) to
obtain the irreducible massless module.

\subsection{Higher-spin action in the frame-like formalism}
\label{sec:dec}

The form of a HS action is to large extent fixed by the gauge
symmetry principle. The frame-like formalism is convenient in
first place because it allows us to have true HS symmetries
manifest.

It is natural to search for a HS action for a
given gauge HS field in the form \cite{V1,LV,VD5}
\be
\label{act_general}
{\cal S}_2 = \gl^{-2(s-1)}\int_{{\cal M}^d}\;\sum H^{...} \;\underbrace{E_0^{...}\wedge\cdots \wedge
E_0^{...}}_{d-2p-2}\wedge\; R_{(p+1)}^{...} \wedge
\,R_{(p+1)}^{...}\;.
\ee
Here factor $\gl^{-2(s-1)}$ is introduced to provide correct flat limit
of (\ref{act_general}) (see section \ref{sec:flat}) and $H^{...}$ are some coefficients built of the Levi-Civita tensor and
the compensator $V^A(x)$, which parameterize various possible
contractions of tangent indices.
Since the curvatures $R_{(p+1)}$ (\ref{adscurv})
are by construction invariant under the gauge transformations
(\ref{adsgaugetr}) of the $p$-form gauge
fields $\Omega_{(p)}$, the action
${\cal S}_2$ is gauge invariant for any choice of the coefficients $H^{...}$.

Any action of the form (\ref{act_general}) is manifestly
 invariant both under diffeomorphisms (because of using the exterior algebra
 formalism) and
under the local $o(d-1,2)$ symmetry that acts on the tangent indices
$A,B,\ldots =0,1\ldots $. Since the the compensator $V^A$ and
background gravitation fields, which
enter the covariant derivative $D_0$, are supposed to take some fixed values,
these  symmetries are broken.
(Note that, in a full interacting theory, this breakdown will be spontaneous,
induced by the vacuum expectation values of the gravitational and
compensator fields.) However, the action (\ref{act_general}) turns out to be
invariant under the global $o(d-1,2)$ symmetry which is a combination of
diffeomorphisms and local $o(d-1,2)$ transformations  that leaves the background gravitation fields
and the compensator invariant. (For more detail we refer the reader to \cite{V_obz3}.)

Let us now discuss which additional conditions
should be imposed on the coefficients
$H^{...}$ to guarantee that the action (\ref{act_general}) indeed describes
a given massless HS field.
The correct action should

\vspace{0.2cm}
\noindent
{(i)}
be expressible in terms of the physical frame-like field and its first derivatives,

\vspace{0.2cm}
\noindent
{(ii)}
exhibit  gauge symmetry enhancement  in the flat limit.
\vspace{0.2cm}

Note that, taking into account the HS gauge invariance of the action and that
the metric-like field identifies with the frame-like physical field modulo
Stueckelberg gauge symmetries, the first of these conditions means that the
action is expressible in terms of the metric-like field and its first
derivatives.

The condition (i) is not automatically satisfied because the HS curvatures
depend on physical, auxiliary and extra fields in the terminology of section
(\ref{sec: lorframelike}). It will be  fulfilled, however, if the
coefficients $H^{...}$ are such that the  action is independent of
the extra fields
\be
\label{dc_ex}
\frac{\delta{\cal S}_2}{\delta w_{(p)}} \equiv 0\;,
\ee
and the auxiliary fields satisfy algebraic field equations that express them
in terms of first derivatives of the physical field. As we shall see, the latter
condition turns out to be automatically true once (\ref{dc_ex}) is satisfied.
The extra fields should not appear at the free field level because
they bring in extra degrees of freedom both if treated as
independent fields and if expressed via derivatives of the
physical fields by some constraints. In the latter case, carrying two more Lorentz
indices compared to the physical field, extra fields can only be expressed via
second derivatives of the physical field thus bringing higher derivatives
into the action if (\ref{dc_ex}) is not true.

To summarize, the extra field decoupling condition (\ref{dc_ex})
guarantees that extra fields enter the action
through the total derivative terms
and do not contribute to nontrivial  equations of motion.

It turns out however that the extra field decoupling condition
(\ref{dc_ex}) alone does not fix the coefficients uniquely,
admitting a $N$-parametric family of $\ads$
gauge-invariant actions, where $N$ is a number of
different auxiliary 1-form connections. This means that there
exist $N$ different  actions expressible in terms of
the metric-like field and its first derivatives which are all
gauge invariant under the necessary $AdS_d$ gauge symmetry.
This is the manifestation of the fact that the $AdS_d$ gauge symmetry
alone does not fix uniquely a form of the kinetic term for a generic
mixed-symmetry field. To determine which particular combination of these
$N$ actions is the correct one, the condition (ii) should be imposed.

It turns out \cite{ASV2} that the condition (ii) is satisfied if all irrelevant
auxiliary fields also do not contribute to the action, {\it i.e.}
\be
\label{dc_ir}
\frac{\delta{\cal S}_2}{\delta \omega^\prime_{(p)}} \equiv 0\;.
\ee
The proof of this fact is given in section \ref{sec:flat}.

The extra field decoupling condition (\ref{dc_ex})
together with the irrelevant auxiliary field decoupling condition
(\ref{dc_ir}) fix uniquely the
coefficients $H^{\cdots}$ in the action (\ref{act_general}),
which has the following structure
\be
{\cal S}_2 \sim \int_{{\cal M}^d}  \Big(\omega_{(p)} ({\cal D} e_{(p)} - \omega_{(p)}) + \lambda^2 e_{(p)} e_{(p)}\Big)\;,
\ee
where $e_{(p)}$ is the physical field and $\omega_{(p)}$ is the relevant auxiliary field.
From this  form of the action (\ref{act_general}) it follows that the relevant
auxiliary field $\omega_{(p)}$
is expressed by virtue of its equation of motion through the first
derivatives of the physical field (modulo pure gauge parts). Plugging
the resulting expression back into the action gives rise to the
second-order action for the metric-like HS field that has necessary gauge
symmetry and the correct kinetic term.

\section{Action and $\cQ$-complex}
\label{Reconstruction of the action}
\subsection{Fock space notations}
\label{sec:Fock}

To simplify the analysis of the HS action
 it is convenient to
reformulate  the problem in terms of a certain  Fock space \cite{ASV2}.
This approach generalizes that applied to totally
symmetric fields in \cite{LV, vf}.

The analysis of the HS dynamics is somewhat simpler
in the antisymmetric basis for Young tableaux where
the expression (\ref{true_auxiliary}) for the relevant
auxiliary field is  particularly simple.
To have antisymmetries manifest,
let us introduce the set of fermionic oscillators
$$\psi_\alpha{}^A =(\psi_i{}^A, \psi^{j\,A})
\quad
{\rm and}
\quad
\bpsi_\alpha{}^A = (\bpsi_i{}^A,
\bpsi^{j\,A})\;,
$$
where $i,j =1\div (s-1)$, $\;\alpha=1\div 2(s-1)$. These oscillators satisfy the anticommutation relations
\be
\label{psi_anticommutator}
\{\psi_i{}^A, \bpsi^{jB}\}=\delta_i^j\;\eta^{AB}\;,
\qquad
\{\psi^{iA}, \bpsi_j{}^B\}=\delta^i_j\;\eta^{AB}
\ee
with all other anticommutators equal to zero.

Also we introduce fermionic oscillators $\theta^A$ and $\btheta^B$
that satisfy  anticommutation relations
\be
\label{theta}
\{\theta^A,\btheta^B\}=\eta^{AB}\,, \quad \{\theta^A,
\theta^B\} = 0\,, \quad \{\btheta^A, \btheta^B\} = 0
\ee
and anticommute with $\psi_\ga{}^A$ and $\bpsi_\ga{}^A$.

Let us define the left and right Fock vacua by
\begin{align}
\label{Rfock}
&\lvac\psi_\ga^A=0\;,&
&\lvac\btheta^A=0\,,\\
\label{Lfock}
&\bpsi_\ga^A\rvac=0\;,&
&\btheta^A\rvac=0
\end{align}
along with
\begin{align}
\label{LC_sym}
\lvac \,\theta^{A_1}\cdots\theta^{A_{d+1}}\rvac=\gep^{A_1\cdots A_{d+1}}\,,
&& \lvac \,\theta^{A_1}\cdots\theta^{A_{k}}\rvac=0\,\;\;\textrm{for}\;\;  k\neq d+1\,.
\end{align}
The oscillators $\theta$ provide a convenient way to introduce the $o(d-1,2)$
Levi-Civita tensor via formula (\ref{LC_sym}).

In our construction, a $p$-form $o(d-1,2)$ gauge field will be described as Fock vectors of two types
$|\hat{{\mathsf \Omega}}_{(p)}\rangle=\hat{{\mathsf \Omega}}_{(p)}\rvac$ or
$|\breve{{\mathsf \Omega}}_{(p)}\rangle=\breve{{\mathsf \Omega}}_{(p)}\rvac$, where
\be
\label{operators}
\ba{c}
\hat{{\mathsf \Omega}}_{(p)}=\Omega_{(p)}{}^{A_1[\tilde{h}_1],...,A_{s-1}[\tilde{h}_{s-1}]}
(\psi^1_{A_1})^{\tilde{h}_1}\cdots(\psi^{s-1}_{A_{s-1}})^{\tilde{h}_{s-1}}\,,
\\
\\
\breve{{\mathsf \Omega}}_{(p)}=\Omega_{(p)}{}_{A_1[\tilde{h}_1],...,A_{s-1}[\tilde{h}_{s-1}]}
(\psi_1^{A_1})^{\tilde{h}_1}\cdots (\psi_{s-1}^{A_{s-1}})^{\tilde{h}_{s-1}}\,.
\ea
\ee
More generally,  operators $\hat{{\mathsf A}}_{(m)}$ and $\breve{{\mathsf A}}_{(m)}$ will be assumed to be
analogously constructed from a $m$-form $A_{(m)}$ instead of $\Omega_{(p)}$.

Note that the proposed  approach is  different from that  used for
the totally symmetric HS fields \cite{LV}. Indeed, in Ref.
\cite{LV}, HS fields were considered as elements of left and right
Fock modules, \textit{i.e.} $\langle {\mathsf \Omega}|$ and
$|{\mathsf \Omega}\rangle$. In our approach  HS fields are
elements of the tensor product $|\hat{{\mathsf \Omega}} \otimes
\breve{{\mathsf \Omega}}\rangle$.

The Young symmetry and tracelessness conditions on the $p$-form gauge fields
imply
\begin{align}
\label{irr1}
l^i{}_j|\hat{{\mathsf \Omega}}_{(p)}\rangle&=0\,,\;\;i<j\,,&\bs_{ij}|\hat{{\mathsf \Omega}}_{(p)}\rangle&=0\,,\\
\label{irr2}
l_i{}^j|\breve{{\mathsf \Omega}}_{(p)}\rangle&=0\,,\;\;i<j\,,&\bs^{ij}|\breve{{\mathsf \Omega}}_{(p)}\rangle&=0\,,\\
\label{irr3}
l^i{}_i |\hat{{\mathsf \Omega}}_{(p)}\rangle& = \tilde{h}_i|\hat{{\mathsf \Omega}}_{(p)}\rangle,\;\;\; & l_i{}^i|\breve{{\mathsf \Omega}}_{(p)}\rangle& =
\tilde{h}_i|\breve{{\mathsf \Omega}}_{(p)}\rangle\;,
\end{align}
where
\be
\label{operSL}
l_{\alpha\beta}=\eta_{AB}\,\psi_\alpha^A\bpsi_\beta^B\,,\quad
\bar{s}_{\alpha\beta}=\eta_{AB}\,\bpsi_\alpha^A\bpsi_\beta^B\,.
\ee
The linearized curvatures (\ref{adscurv})  in the antisymmetric basis are
\be
\label{curv}
|\hat{{\mathsf R}}_{(p+1)}\rangle = \hat{{\mathsf R}}_{(p+1)}\rvac=\diff|\hat{{\mathsf \Omega}}_{(p)}\rangle\,,
\qquad |\breve{{\mathsf R}}_{(p+1)}\rangle =\breve{{\mathsf R}}_{(p+1)}\rvac=
\diff|\breve{{\mathsf \Omega}}_{(p)}\rangle\,.
\ee
Here the $o(d-1,2)$ covariant background derivative is
\be
\diff=\extdiff+\Go_0{}^A{}_B\psi^i_A\bpsi_i^B+\Go_0{}_A{}^B\psi_i^A\bpsi^i_B+
\Go_0{}^A{}_B\theta_A\btheta^B\,,\qquad \diff^2=0\,,
\ee
where $\Go_0{}^{AB}$ is the background $\ads$ gauge field satisfying  the zero-curvature condition (\ref{zerocurv}).
The gauge transformations (\ref{adsgaugetr}) and the Bianchi identities (\ref{Bianchi}) are
\begin{align}
\label{gauge}
&\dps \delta|\hat{\Omega}_{(p)}\rangle = \diff|\hat{\xi}_{(p-1)}\rangle\,, &
&\delta|\breve{\Omega}_{(p)}\rangle = \diff|\breve{\xi}_{(p-1)}\rangle\,,\\
\label{bianchi}
&\diff |\hat{R}_{(p+1)}\rangle = 0\,,&
&\dps\diff |\breve{R}_{(p+1)}\rangle  = 0\,.
\end{align}

In the sequel we make use of the following operators
\be
\label{defoper}
\bar{\eta}_\ga=\bpsi_\ga^A\theta_A\;,
\quad
\bar{v}_\ga=\bpsi_\ga^AV_A\;,
\quad
\chi=\theta^AV_A\;,
\quad
\Eop=E_0^A\theta_A\;,
\ee
where $V^A$ and $E_0^A$ are the compensator and the background frame field, respectively.

Let us introduce a notion of weak equality.
Two polynomials ${\mathsf A}(\bs,\bta,\bv)$ and ${\mathsf B}(\bs,\bta,\bv)$
are  weakly equivalent,  ${\mathsf A}\sim {\mathsf B}$, if
\be
\label{weak}
\lvac(\wedge\Eop)^{d-m-n}\chi\Big({\mathsf A}-{\mathsf B}\Big)\wedge\hat{A}_{(m)}
\wedge \breve{B}_{(n)}\rvac =0\,
\ee
for any fields $A_{(m)}$ and $B_{(n)}$ that satisfy the
Young symmetry and tracelessness  conditions
(\ref{irr1})-(\ref{irr3}).
In other words, the weak equivalence of two functions means that they
differ by terms proportional to Young symmetrizers and trace operators
which are zero by (\ref{irr1})-(\ref{irr3}).
A generic weakly zero function $\cW=\cW(\bs,\bta,\bv)$ can be cast into the following  form
\be
\label{weakly_zero}
\ba{c}
\dps
\cW=\sum_{i,j=1}^{s-1}\cW^{ij}\bs^{ij}+\sum_{i,j=1}^{s-1}\cW_{ij}\bs_{ij}+
\sum_{i,j=\!1,\,i<j}^{s-1}[\cW_i{}^j,l_i{}^j]
\\
\\
\dps
+\sum_{i,j=\!1,\,i<j}^{s-1}[\cW^i{}_j,l^i{}_j]
+\sum_{i=1}^{s-1}\Big([\cW_i,l_i{}^i]-\tilde{h}_i\cW_i\Big)+
\sum_{i=1}^{s-1}\Big([\cW^i,l^i{}_i]-\tilde{h}_i\cW^i\Big)\,,
\ea
\ee
where $\cW_{ij}$, $\cW^{ij}$, $\cW_i{}^j$, $\cW^i{}_j$, $\cW_i$ and
$\cW^i$ are arbitrary functions of $\bs$, $\bta$ and  $\bv$. Here the first two terms are weakly
zero due to the tracelessness condition  and the other terms are weakly zero
due to the Young conditions (\ref{irr1}), (\ref{irr2}), (\ref{irr3}). Note that any operator
$\cY$ acting on polynomials of $\bs$, $\bta$ and $\bv$, that
commutes with $\bs_{ij}$, $\bs^{ij}$,
$l_i{}^j$ and $l^i{}_j$
\be\label{Ycommutes}
[\bs_{ij},\cY]=[\bs^{ij},\cY]=[l_i{}^j,\cY]=[l^i{}_j,\cY]=0\,,
\ee
preserves the form (\ref{weakly_zero}) and thus maps weakly
zero polynomials to weakly zero polynomials, {\it i.e.}
\be\label{wzero}
\cY\cW\sim 0\qquad \forall\quad \cW\sim 0\,.
\ee

\subsection{General ansatz for a higher-spin action}
\label{sec:HSactionFock}

The Fock space form of the frame-like action
(\ref{act_general}) is
\be
\label{action}
\cS_2=\gl^{-2(s-1)}\int_{\cM^d}\lvac(\wedge\Eop)^{d-2p-2}\chi\,{\mathsf H}(\bar{s},\bar{\eta},\bar{v})
\wedge\hat{{\mathsf R}}_{(p+1)}\wedge \breve{{\mathsf R}}_{(p+1)}\rvac\,,
\ee
where ${\mathsf H}(\bar{s},\bar{\eta},\bar{v})$ is some polynomial of
the commuting variables $\bar{s}_{\ga\gb}$,
$\bar{\eta}_{\ga}$ and anticommuting variables $\bar{v}_\ga$.
Oscillators $\theta$ contained
in ${\mathsf H}$ add up to $d-2p-1$ oscillators $\theta$ in $(\Eop)^{d-2p-2}\chi$ to
generate the Levi-Civita  tensor via (\ref{LC_sym}).
{}From (\ref{LC_sym})
it follows that the number of $\theta$'s in ${\mathsf H}$ is $2p+2$ (otherwise the
action (\ref{action}) is zero) since these oscillators enter
${\mathsf H}$ through the operators $\bta_\alpha$, it follows that
\be
\label{degn}
\bta_\alpha \frac{\d }{\d \bta_\alpha}{\mathsf H} = (2p+2){\mathsf H}
\ee
\textit{i.e.}, ${\mathsf H}$ is a  degree $2p+2$ homogeneous
polynomial of $\bta_\alpha$.

The operators $\bar{s}_{\alpha\beta}$, $\bar{\eta}_\ga$, $\bar{v}_\ga$, $\chi$, and $\Eop$
are responsible for the following contractions of indices
in the  action

\begin{itemize}

\item the operator $\bar{s}_{\alpha\beta}$ contracts two indices of $\hat{{\mathsf A}}_{(m)}$
and/or $\breve{{\mathsf A}}_{(n)}$
placed in $\alpha$-th and $\beta$-th columns;

\item the operator $\bar{\eta}_\ga$ contracts an index of the Levi-Civita tensor with an index in $\alpha$-th
column of $\hat{{\mathsf A}}_{(m)}$ or $\breve{{\mathsf A}}_{(n)}$;

\item the operator $\bar{v}_\ga$ contracts the compensator $V^A$ with an index in $\alpha$-th column
of $\hat{{\mathsf A}}_{(m)}$ or $\breve{{\mathsf A}}_{(n)}$,

\item the operator $\chi$ contracts the compensator $V^A$ with the Levi-Civita tensor;

\item the operator $\Eop$ contracts the frame $E_0^A$ with the Levi-Civita tensor.

\end{itemize}

\noindent This list  exhausts all possible independent contractions between
the constituents of the action. One can see that
all other contractions either are zero (like $E_0^A\,V_A=0$)
or can be reduced to the contractions listed above. For example, the
term $\eta_{AB}E_0^A\bpsi^B_\alpha$ responsible for contracting the index of frame
with an index of $\hat{{\mathsf A}}_{(m)}$ or $\breve{{\mathsf A}}_{(n)}$
 reduces to the $\Eop$-type terms by virtue of the following identity
\be
\label{identity}
E_0^A \wedge (\wedge\Eop)^{m}\rvac = \frac{1}{m+1}\,\btheta^A\,(\wedge\Eop)^{m+1}\rvac
\ee
along with the property that $\btheta^A$ annihilates both left and right Fock vacua
(\ref{Rfock}), (\ref{Lfock}).

Using the symmetry of (\ref{action}) with respect to
the exchange of the $(p+1)$-form HS curvatures,
we require ${\mathsf H}(\bs,\bta,\bv)=
{\mathsf H}(\bs_{\alpha\beta},\bta_\ga,\bv_i,\bv^i)$  to satisfy the symmetry property
\be\label{u_symmetry}
{\mathsf H}(\bs_{\alpha\beta},\bta_\ga,\bv_i,\bv^i)=
(-1)^{p+P+1}{\mathsf H}(-\bs_{\alpha\beta},\bta_\ga,\bv^i,\bv_i)\,,
\ee
where $P$ is defined by $\hat{{\mathsf \Omega}}_{(p)}(\psi)=(-1)^P\hat{{\mathsf \Omega}}_{(p)}(-\psi)$.

\subsection{$\cQ$-complex}
\label{sec:Q-complex}

Taking into account the definitions of the
curvature (\ref{curv}) and the compensator (\ref{comp_def}), applying
the Bianchi identities (\ref{bianchi}) and making use of the
 identity (\ref{identity}), the variation of the action can be reduced to
 the following form
\begin{align}
\label{var}
\gd\cS_2&=2\gl^{2(s-1)}(-1)^{d+p}\int_{\cM^d}\lvac\diff\Big((\wedge\Eop)^{d-2p- 2}\chi{\mathsf H}\Big)\wedge\hat{{\mathsf R}}_{(p+1)}\wedge \delta\breve{{\mathsf \Omega}}_{(p)}\rvac=\nn
\\
&=2(-1)^p\frac{\gl^{2s-1}}{d-2p-1}\int_{\cM^d}\lvac(\wedge\Eop)^{d-2p-1}\chi\cQ{\mathsf H} \wedge\hat{{\mathsf R}}_{(p+1)}\wedge \delta\breve{{\mathsf \Omega}}_{(p)}\rvac  \,,
\end{align}
where \be \label{Q}
\cQ=\Big(d-1+\bv^\ga\ptl{\bv^\ga}-\bta^\ga\ptl{\bta^\ga}\Big)\bv^\gb\ptl{\bta^\gb}+
\bs^{\ga\gb}\pptl{\bv^\ga}{\bta^\gb}\,. \ee
Note that $\cQ$ satisfies (\ref{Ycommutes}). As a result,
$\cQ\cW\sim 0$ for any weakly zero function $\cW \sim 0$.

The important fact is
that
\be
\label{Q_nilpotent}
\cQ^2=0\;
\ee
as one can check directly.
(Essentially, (\ref{Q_nilpotent})  is a  consequence of $\diff^2=0$).
A natural guess is that, in an appropriate representation, $\cQ$ can be rewritten as a de Rham operator.
Indeed, one can see that
\be
\label{deRahm}
\delta =\bv^\ga\ptl{\bta^\ga}=A^{-1}\cQ A\,,
\ee
where
\be
\label{A}
A=\Big(d-1+\bv^\ga\ptl{\bv^\ga}-\bta^\ga\ptl{\bta^\ga}\Big)!!
\exp\Big(\frac{1}{2}\bs^{\ga\gb}\pptl{\bv^\ga}{\bv^\gb}\Big)\;.
\ee


The properties of the operator $\cQ$ allow us to analyse
the action and the decoupling conditions in terms  of the
$\cQ$-complex. {}From  the form of variation (\ref{var})
it follows in particular that total derivative terms in the action
are described by various $\cQ$-closed functions ${\mathsf H}$
(modulo weakly zero terms).

Suppose now that
the variation of the action has the form
\be
\label{fieldEq}
\gd \cS_2=\gl^{2s-1}\lvac(\wedge\Eop)^{d-2p-1}\chi\cE(\bs,\bta,\bv)\wedge\hat{{\mathsf R}}_{(p+1)} \wedge \delta\breve{{\mathsf \Omega}}_{(p)}
\rvac
\ee
where  $\cE(\bs,\bta,\bv)$ is some polynomial function which is $\cQ$-closed
modulo weakly zero terms
\be\label{ccon}
\cQ\cE\sim 0\,.
\ee
The question is whether it is possible to reconstruct an action that leads
to the field equations (\ref{fieldEq}), {\it i.e.} to represent $\cE$ in the form
\be
\label{Qexact}
\cE\sim \cQ {\mathsf H}\,
\ee
with some ${\mathsf H}(\bs,\bta,\bv)$.
The answer is yes and an explicit formula for ${\mathsf H}(\bs,\bta,\bv)$
in terms of $\cE$ is given in subsection \ref{Action from field equations}.
This result will allow us in section \ref{Higher-Spin action} to analyze the
decoupling conditions at the level of field equations, that is relatively simple,
reconstructing the action afterwards.

Analogously, one can show that the total derivative terms in the
Lagrangian are described by $\cQ$-exact coefficient function
\be
\label{total}
{\mathsf H}(\bs,\bta,\bv)=\cQ {\mathsf T}(\bs,\bta,\bv)
\ee
for some ${\mathsf T}(\bs,\bta,\bv)$.

\subsection{Action from field equations}
\label{Action from field equations}

To reconstruct the action function ${\mathsf H}$ (\ref{action}) through the function $\cE$
in the variation (\ref{fieldEq}) we have to solve the equation (\ref{Qexact})
for the known function $\cE$ satisfying the consistency condition
(\ref{ccon}). This can be easily done in the
basis (\ref{A}) where $\cQ$ has a form of the standard de Rham operator $\delta$
\be
\qquad
\delta=\bv^\ga\ptl{\bta^\ga}\;,
\qquad
\delta^2 =0\;.
\ee
Introducing ${\mathsf H}'=A^{-1}{\mathsf H}$, $\cE'=A^{-1}\cE$ and taking into account that the
operators $A$, $A^{-1}$ map weakly zero polynomials to weakly zero polynomials
(see the arguments in the end of subsection \ref{sec:Fock})
one rewrites equation (\ref{Qexact}) as
\be\label{delta_exact}
\gd{\mathsf H}'\sim\cE'\,.
\ee

Let us solve first the strong equation
\be
\label{cohEq}
\delta \cF = \cG\;,
\ee
where
$\cF=\cF(\bs,\bta,\bv)$, $\cG=\cG(\bs,\bta,\bv)$ are some polynomials
and $\cG$ satisfies the compatibility condition
\be\label{compcon}
\gd\cG=0\,.
\ee

Consider the operator $\gd^*$
\be
\gd^*=\bta^\ga\ptl{\bv^\ga}\;, \qquad \gd^{*\,2}=0\,.
\ee
Acting by $\gd^*$ on the both sides of (\ref{cohEq})
one obtains
\be
\Gd\cF=\gd^*\cG+\gd\gd^*\cF\,,
\ee
where the operator
\be
\Gd\equiv \{\gd,\gd^*\}=\bta^\ga\ptl{\bta^\ga}+\bv^\ga\ptl{\bv^\ga}\,
\ee
 commutes with $\gd$ and $\gd^*$. As a result, neglecting a $\delta$-exact term,
 a  partial solution of
(\ref{cohEq}) is
\be\label{K__}
\cF=\Gd^{-1}\gd^*\cG\,.
\ee
The operator $\Gd^{-1}$ admits the following integral realization
\be\label{Delta-1}
\Gd^{-1}\cA(\bs,\bta,\bv)=\int_0^1 \!\frac{{\rm d}t}{t}\cA(\bs,t\bta,t\bv)\,
\ee
for a function $\cA(\bs,\bta,\bv)$ such that $t^{-1}A(\bs, t\bta,t\bv)$ is
 polynomial in $t$
 (in the cases of interest this is always true because of (\ref{degn})).
Substituting (\ref{Delta-1}) into (\ref{K__}) one obtains for the general solution of (\ref{cohEq})
\be\label{gen_sol}
\cF(\bs,\bta,\bv)=\int_0^1 \!\frac{{\rm d}t}{t}\bta^\ga\ptl{\bv^\ga}
\cG(\bs,t\bta,t\bv)+\bv^\ga\ptl{\bta^\ga}\cT(\bs,\bta,\bv)\,,
\ee
with an arbitrary polynomial $\cT$.
It is worth to note that since the operator $\Gd^{-1}\gd^*$ satisfies (\ref{Ycommutes})
the equation (\ref{cohEq}) with a weakly zero polynomial $\cG\sim 0$
always admits a  weakly zero solution.

Now we are in a position to solve the weak equation (\ref{delta_exact}) which has
 the form
\be
\gd{\mathsf H}'=\cE'+\cK\,,
\ee
where $\cK\sim 0$.
From (\ref{gen_sol}) one obtains
\be\label{gen_sol_}
{\mathsf H}'(\bs,\bta,\bv)=\int_0^1 \!\frac{{\rm d}t}{t}\bta^\ga\ptl{\bv^\ga}
\Big(\cE'(\bs,t\bta,t\bv)+\cK(\bs,t\bta,t\bv) \Big)
+\bv^\ga\ptl{\bta^\ga}\cT(\bs,\bta,\bv)\,.
\ee
Acting on the both hand sides of (\ref{gen_sol_}) by the operator $A$
and neglecting the weakly zero term resulting from $\cK$ and $\cQ$-exact
term resulting from $\cT$, we obtain
\be
\label{H}
{\mathsf H}(\bs,\bta,\bv)\sim A\Big[\int_0^1 \!\frac{{\rm d}t}{t}\bta^\ga\ptl{\bv^\ga}
\Big(A^{-1}\cE\Big)(\bs,t\bta,t\bv)\Big]\,.
\ee
Substituting (\ref{H}) into the action (\ref{action}), we finally obtain
the following expression for the action functional that gives rise to the
variation (\ref{fieldEq})
\be
\label{actionFIN}
\cS_2=\gl^{-2(s-1)}\int_{\cM^d}\lvac(\wedge\Eop)^{d-2p-2}\chi A\Big[
\int_0^1 \!\frac{{\rm d}t}{t}\bta^\gb\ptl{\bv^\gb}
\Big(A^{-1}\cE\Big)(\bs,t\bta,t\bv)\Big]\wedge\hat{R}_{(p+1)}\wedge \breve{R}_{(p+1)}\rvac\,.
\ee
It remains to find the function $\cE (\bs,\bta,\bv)$ that gives rise to
the correct field equations of a HS gauge field
to reconstruct its action by this formula.

\section{Higher-spin equations of motion}
\label{Higher-Spin action}

In this section we find the function $\cE$ that satisfies equation (\ref{ccon})
and  the decoupling conditions (\ref{dc_ex}) and (\ref{dc_ir}) which
require that the variation (\ref{fieldEq}) with respect to extra fields
$w_{(p)}$ and irrelevant auxiliary fields $\go'_{(p)}$
should be identically zero.  This condition fixes the dependence
of $\cE(\bs,\bta,\bv)$ on $\bv$ as follows
\be
\label{functE}
\cE(\bs,\bta,\bv) = \Big(\bta_1\frac{\d}{\d \bv_1}
-\bta^1\frac{\d}{\d \bv^1}\Big)\tilde{\cE}(\bs,\bta)\bv^{2(s-1)}\;,
\ee
where
\be
\bv^{2(s-1)}=\bv_1\cdots\bv_{s-1}\,\bv^1\cdots\bv^{s-1}\;.
\ee
Indeed, substituting (\ref{functE}) into (\ref{fieldEq}) one finds that the first
term of (\ref{functE})
contains $(s-1)$ (maximal possible number) compensators  contracted
with $\delta\Omega_{(p)}$
and therefore corresponds to the variation with respect to the physical field (cf. (\ref{phys})).
Analogously, the second term contains $(s-2)$ compensators contracted with
all columns of $\delta\Omega_{(p)}$ except for the first one and therefore
corresponds to the variation with respect to the relevant auxiliary field
(cf. (\ref{true_auxiliary})). (It is this place where the antisymmetric
basis turns out to be most convenient). In the both cases,
the remaining index in the first column of either $R_{(p+1)}$ or $\delta\Omega_{(p)}$
is contracted with the Levi-Civita tensor via $\bta_1$ or $\bta^1$.
The relative coefficient in (\ref{functE})
is fixed by the symmetry property of  ${\mathsf H}$ (\ref{u_symmetry}).

The next step is to find such a  function $\tilde {\cE}(\bs,\bta)$
that $\cE(\bs,\bta,\bv)$ of (\ref{functE}) satisfies the weak
closedness condition (\ref{ccon}).

Using the antisymmetric basis,
let us arrange tangent indices of the $p$-form field $\Go_{(p)}$
 into vertical blocks $(m_I,\tilde h_I)$, $I=1,..., N$ as explained in
section \ref{sec:antibasis}. Let $\mu_I$ be a number of the first
column of the $I$-th vertical block.
Since the length of the uppermost row of $\Go_{(p)}$ is $s-1$,
we have $\sum\limits_{I=1}^{N}\;m_I = s-1\;, \;\; \mu_{I+1}-\mu_I=m_I$, $I=1,\ldots,N-1$.

Now we use the Young symmetries of the HS connections to choose
a particular basis in the space of different terms that contribute
to the action. Algebraically, this is equivalent to choosing a
representative of $\tilde{\cE}(\bs,\bta)$ modulo weakly zero-terms.
The important fact is that, by
adding weakly zero terms in the variation (\ref{fieldEq}),
the function $\tilde{\cE}$ in (\ref{functE}) can always be chosen in the form
\be
\label{newE}
\tilde{\cE}(\bs,\bta) = \tilde{\cE}(\bu,\n)\,,
\ee
where the operators
\be
\label{newvar}
\bu_i=\bs_i{}^i\,,
\qquad
\n_I=\bta_{\mu_I}\bta^{\mu_I}\,,
\ee
(no sums over repeated indices) realize column-to-column contractions between
$R_{(p+1)}$ and $\gd\Go_{(p)}$ and contractions of the Levi-Civita tensor with the
first columns of $I$-th vertical blocks of  $R_{(p+1)}$
and $\gd\Go_{(p)}$.

It follows that the
variables (\ref{newvar}) enter the function (\ref{newE})
through the combinations of the form
\be
\Big(\prod_{i=\mu_I}^{\mu_{I}+m_I-1}(\bu_i)^{\tilde{h}_i-1}\Big)
\Big(\frac{\n_I}{\bu_{\mu_I}}\Big)^k
\ee
for some $k\leq p$. In terms of tensors this means
that the indices of, say, $\gd\Go_{(p)}$
in $I$-th vertical block of length $m_I$ and height $\tilde h_I$
are contracted as follows.
One index from every column is contracted
with the compensator (or with the Levi-Civita tensor for the second term
in (\ref{functE}) and $i=1$). The contractions of
the remaining $(\tilde{h}_I-1)\times m_I$ indices
depend on whether they belong to the first column of $I$-th block or not.
The remaining indices
of columns $\mu_I+1,\ldots,\mu_I+m_I-1$ are column-to-column contracted with
the corresponding indices of $R_{(p)}$.
The remaining indices of the first column of the $I$-th block $\mu_I$ are
either contracted with $k_I$ indices of the Levi-Civita tensor or  contracted with
$\tilde{h}_I-1 -k_I$ indices of the
 $\mu_I$-th column of $R_{(p+1)}$. Note that, due to (\ref{LC_sym}),
a function
$\cE$  in the variation (\ref{fieldEq}) should have
$2p$ operators $\bta$ to contract the remaining $2p$ indices of
the  Levi-Civita tensor.
These indices are contracted with $k_I$ indices of
 the first columns of the $i$-th vertical block of both $R_{(p+1)}$ and
$\gd\Go_{(p)}$ so that $\sum_{I=1}^Nk_I=p$.

As a result, the function $\tilde{\cE}$ has the  form
\begin{align}
\label{ee}
&\tilde{\cE}(\bu,\n)=\Big(\prod_{i=1}^{s-1} (\bu_i)^{\tilde{h}_i-1}\Big)\tilde e(t)\,,\\
\label{eee}
&\tilde e(t)=\sum_{\stackrel{\scriptstyle k_I\geq 0,\; I=1\div N}
{\stackrel{\scriptstyle k_1+\cdots +k_N=p}{}}}\!\!\!
\rho(k_1,\ldots,k_N)t_1^{k_1}\cdots
t_N^{k_N}\,,
\end{align}
where
\be
\label{var-t}
t_I = \frac{\n_I}{\bu_{\mu_I}}\;,
\qquad
I=1,\ldots, N\,.
\ee
The coefficients $\rho(k_1,\ldots,k_N)$ parameterize various
types of contractions between
$2p$ indices of the Levi-Civita tensor and those of $R_{(p+1)}$ and  $\gd\Go_{(p)}$.

For example, for the gauge field $\Go_{(p)}$ with tangent indices described
by  a rectangular  $o(d-1,2)$ Young tableau,
the function $\tilde \cE$ is
\be
\label{tEblock}
\tilde{\cE}=\rho\prod_{i=1}^{s-1} (\bu_i)^{\tilde{h}-1}\;t_1{}^p
\ee
with an arbitrary constant $\rho$. As can be easily checked,
the corresponding function $\cE$
 (\ref{functE}) satisfies equation (\ref{ccon}).

In the rest of this section we show that the coefficient function
$\rho(k_1,\ldots,k_N)$, is uniquely fixed by
the compatibility condition
(\ref{ccon}) on the function $\cE$ and has the form
\be
\label{gensol}
\ba{l}
\rho(k_1,..., k_N) =
\\
\dps
=\frac{\rho\,\delta(p-\sum\limits_{I=1}^{N}k_I)}{(k_1!)^2(k_1+1)(h_1-k_1-1)!}
\prod_{I=2}^{N}\frac{(k_I+m_I-1)!}{(k_I!)^2\,(h_I-k_I-1)!}
\frac{\Big(\vartheta(I)-\sum\limits_{J=I}^{N}k_J\Big)!}
{\Big(\vartheta(I)+m_I-\sum\limits_{J=I+1}^{N}k_J\Big)!}\,,
\ea
\ee
where $\rho$ is an arbitrary constant and
\be
\label{vartheta}
\vartheta(I) = s-\mu_I-m_I+\tilde h_I-1\;,
\quad
\vartheta(I)\geq p\,.
\ee

The formula (\ref{gensol}) is obtained as follows.
Using the explicit form (\ref{Q}) of the operator $\cQ$,
equation (\ref{ccon}) gives for
the function $\cE$ of the
form (\ref{functE})
\be
\ba{c}
\lb{QE}
\dps
\left(\bs_i{}^j\Big(\bta_1\pptl{\bv_i}{\bv_1}-
\bta^1\pptl{\bv_i}{\bv^1}\Big)\bv^{2(s-1)}\ptl{\bta^j}-
\bs_j{}^i\Big(\bta_1\pptl{\bv^i}{\bv_1}-
\bta^1\pptl{\bv^i}{\bv^1}\Big)\bv^{2(s-1)}\ptl{\bta_j}-\right.
\\
\\
\dps
\left.{}-\bs_1{}^i\pptl{\bv^i}{\bv_1}\bv^{2(s-1)}-
\bs_i{}^1\pptl{\bv_i}{\bv^1}\bv^{2(s-1)}\right)\tilde \cE(\bu,\n)\sim 0\,.
\ea
\ee
Naively, the operators $\bs_i{}^j$ at $i\neq j$ spoil the
column-to-column character of contractions encoded in the
form of function
(\ref{newE}). However, by using the Young symmetry properties ({\it i.e.,} by adding
proper weakly zero terms) it is possible to rewrite (\ref{QE}) in terms of
the variables $\bu$ and $\n$. Namely,
 after some algebra based on identities (\ref{A1})-(\ref{A4}) given in Appendix,
equation (\ref{QE}) takes the form
\begin{align}
\label{transformed}
&\left( \frac{(\bar{N}_1+2)(\bar{N}_I+m_I)}{\bar{U}_1(\bar{N}_I+1)}
\bu_{\mu_1}\ptl{\n_1}
+\sum_{J=2}^{I-1}\;\frac{(\bar{N}_J+1)(\bar{N}_I+m_I)}
{\bar{U}_J(\bar{N}_I+1)}\bu_{\mu_J}\ptl{\n_J}+\nn\right.\\
&\left.+\Big(\sum_{J=I+1}^{n}\frac{\bar{N}_J}{\bar{U}_I}
-\frac{s-\mu_I-m_I}{\bar{U}_I}-1\Big)\bu_{\mu_I}\ptl{\n_I}\right) \tilde{\cE}(\bu,\n)=0\,,
\qquad 2\leq I\leq n\,,
\end{align}
where
\begin{align}
\bar{N}_I=\n_I\ptl{\n_I}\,,\qquad \bar{U}_I=\bu_{\mu_I}\ptl{\bu_{\mu_I}}\;,\quad
\hbox{no summation}\,.
\end{align}

Plugging the ansatz (\ref{ee}) into (\ref{transformed})
one obtains the equation
\be
\label{triangle}
\sum_{J=2}^{I-1}\;A_J\,z_J + B_I\,z_I = -z_1\;,\quad I=2,\ldots, N\;,
\ee
where we use  notations
\be
\label{new-t}
\bar{T}_I=t_I\frac{\d}{\d t_I}\;,
\ee
\be
\label{new-t2}
-\frac{(\bar{T}_1+2)}{(\tilde{h}_1-\bar{T}_1-1)}\;\ptl{t_1} \tilde{e}(t)=z_1 \;,
\qquad
\ptl{t_I}\tilde{e}(t)=z_I  \;,
\quad
I=2,\ldots, N\;,
\ee
\be
A_I= \frac{(\bar{T}_I+1)}{(\tilde{h}_I-\bar{T}_I-1)}\;,
\quad
\dps B_I= A_I\frac{\Big(\sum\limits_{J=I}^{N}\,\bar{T}_J-\vartheta(I)\Big)}{(\bar{T}_I+m_I)}\;.
\ee
Equation (\ref{triangle}), which  has the triangle form,  implies
\be
\label{solution1}
z_I = (-)^{I+1}\;\frac{\prod\limits_{J=2}^{I-1}(A_J-B_J)}{\prod\limits_{J=2}^{I}B_J}\;z_1\;.
\ee
This gives the following equations on the function $\tilde{e}(t)$:
\be
\label{solution2}
\ptl{t_I} \tilde{e}(t)= \frac{(\bar{T}_I-\tilde{h}_I+1)(\bar{T}_I+m_I)(\bar{T}_1+2)}
{(\tilde{h}_1-\bar{T}_1-1)(\bar{T}_I+1)}\;
\frac{\prod\limits_{J=2}^{I-1}\Big(\sum\limits_{K=J+1}^{N}\bar{T}_K - \vartheta(J)-m_J\Big)}
{\prod\limits_{J=2}^{I}\Big(\sum\limits_{K=J}^{N}\bar{T}_K - \vartheta(J)\Big)}\;
\ptl{t_1} \tilde{e}(t)\;,
\ee
for $I=2,\ldots, N$.

The substitution of  $\tilde{e}(t)$ (\ref{eee}) into (\ref{solution2})
gives the following equations on the coefficient
function $\rho(k_1, ... , k_N)$
\begin{align}
\label{recurrent1}
&\rho(k_1-1, ... ,k_I+1,..., k_N)= G_I(k_1,\ ...,
k_I,...,k_N)\,\rho(k_1,...,k_I,..., k_N)\,,\;\; I=2\div
N\,,\\
&\rho(p,0,\ldots,0)=\rho\,,
\end{align}
where $\rho$ is an arbitrary constant and
\begin{align}
\label{Green} &G_I(k_1,..., k_n)=
\frac{(k_I-\tilde{h}_I+1)(k_I+m_I)}{(k_I+1)^2}\frac{k_1(k_1+1)}{\tilde{h}_1-k_1}
\frac{\prod\limits_{J=2}^{I-1}\Big(\sum\limits_{K=J+1}^{N}k_K-\vartheta(J)-m_J\Big)}
{\prod\limits_{J=2}^{I}\Big(\sum\limits_{K=J}^{N}k_K -
\vartheta(J)\Big)}\,,\\
\label{range}
&k_1\geq 1\;,\qquad k_I\geq 0\;,\;\; I=2\div N\;,\qquad k_1+ ... + k_N=p\;,
\end{align}

The equations (\ref{recurrent1})-(\ref{range})  give
\be
\rho(k_1, ... , k_N) = \rho\prod_{I=2}^N\;\rho_I(p-\sum_{J=I+1}^N k_J,k_2, ...,k_I,0,...,0)\;,
\ee
where $\rho_I(p-\sum_{J=I+1}^N k_J,k_2, ...,k_I,0,...,0)$
is a solution of (\ref{recurrent1}) with fixed $I$ and $k_J=0$ at $J>I$.
This gives the final result (\ref{gensol}).
It is elementary to check that  (\ref{gensol}) does solve the system
(\ref{recurrent1})-(\ref{range}).

To summarize, the expressions  (\ref{functE}), (\ref{ee}), (\ref{eee}), (\ref{gensol})
determine the function $\cE$ that satisfies the weak
$\cQ$-closedness equation (\ref{ccon}) and gives rise to the action (\ref{actionFIN})
satisfying the decoupling conditions (\ref{dc_ex}), (\ref{dc_ir}).
The constructed function $\cE$ is (weakly) unique up to a normalization factor $\rho$.

\section{Gauge symmetry enhancement at $\lambda=0$}
\label{sec:flat}

As explained in section \ref{sec:flat_sym}, to describe correctly
dynamics of a mixed-symmetry gauge field,  the $\ads$ HS action
should admit  additional  gauge symmetries (\ref{fl_gaug}) in the
flat limit $\gl=0$. In this section we show that the $\ads$ action
(\ref{actionFIN}) constructed with the function $\cE$
(\ref{functE}), (\ref{ee}), (\ref{eee}), (\ref{gensol})
indeed exhibits the flat space   gauge
symmetry enhancement with traceless gauge parameters
$S_I$, $I=1,\ldots, N$.

To analyze the flat limit we choose the compensator in the
standard form $V^A=\delta^A_{d}$ so that  $A=(a,d+1)$ where
$a$ is a $o(d-1,1)$ vector index while value $d+1$ corresponds to the last $(d+1)$-th direction. Choosing
Cartesian coordinate system with background frame
field $E_{0\un}^a=\gd_\un^a$ and Lorentz spin connection
$\omega^{ab}_{\un}=0$, we replace the Lorentz covariant
derivative $\cD$ by the flat derivative $\d$. In the
subsequent analysis we identify the world and tangent Lorentz
indices with the help of the background frame field $\gd^a_\un$.

\subsection{Extended  Fock space notations}

Let us introduce new fermionic oscillators $\kappa^\bullet_a$, $\bkappa_\bullet^a$ and
$\kappa_\bullet^a$, $\bkappa^\bullet_a$, which anticommute
with the previously introduced oscillators
$\psi, \bpsi$ (\ref{psi_anticommutator}) and $\theta,\btheta$ (\ref{theta}),
and satisfy the anticommutation relations
\be
\label{kappa}
\{\bkappa_\bullet^a,\kappa^{\bullet b}\}=\eta^{ab}\,,\qquad
\{\bkappa^{\bullet a},\kappa_\bullet^ b\}=\eta^{ab}\,,
\ee
with other anticommutators being zero. The left and right Fock vacua are defined by
(\ref{Rfock})-(\ref{LC_sym}) and
\begin{align}
\lvac\kappa_\bullet^a&=0\;,&
\lvac\kappa^{\bullet a}&=0\,,\\
\bkappa_{\bullet}^a\rvac&=0\;,&
\bkappa^{\bullet a}\rvac&=0\;.
\end{align}

We contract new oscillators either with the world indices of forms or with the $s$-th column of  the metric-like field $\Phi$, flat gauge parameter $S_I$,
etc...
For example, the operators $\hat{{\mathsf A}}_{(p)}$, $\breve{{\mathsf A}}_{(p)}$
associated with the $p$-form $A_{(p)}$  now read as
\be
\label{operA}
\ba{c}
\hat{{\mathsf A}}_{(p)}=A^{a_1[\tilde{h}_1],...,a_{s-1}[\tilde{h}_{s-1}];\;m[p]}\,
(\psi^1_{a_1})^{\tilde{h}_1}\cdots(\psi^{s-1}_{a_{s-1}})^{\tilde{h}_{s-1}}
(\kappa^\bullet{}_m)^p\,,
\\
\\
\breve{{\mathsf A}}_{(p)}=A_{a_1[\tilde{h}_1],...,a_{s-1}[\tilde{h}_{s-1}];\;m[p]}\,
(\psi_1^{a_1})^{\tilde{h}_1}\cdots (\psi_{s-1}^{a_{s-1}})^{\tilde{h}_{s-1}}
(\kappa_\bullet{}^m)^p\,.
\ea
\ee
For the metric-like fields the associated operators $\hat{\Phi}$, $\breve{\Phi}$ read as
\be
\label{operF}
\ba{c}
\hat{{\Phi}}=\Phi^{a_1[\tilde{h}_1],...,a_{s-1}[\tilde{h}_{s-1}],a_s[p]}\,
(\psi^1_{a_1})^{\tilde{h}_1}\cdots(\psi^{s-1}_{a_{s-1}})^{\tilde{h}_{s-1}}
(\kappa^\bullet{}_{a_s})^p\,,
\\
\\
\breve{{\Phi}}=\Phi_{a_1[\tilde{h}_1],...,a_{s-1}[\tilde{h}_{s-1}],a_s[p}\,
(\psi_1^{a_1})^{\tilde{h}_1}\cdots (\psi_{s-1}^{a_{s-1}})^{\tilde{h}_{s-1}}
(\kappa_\bullet{}^{a_s})^p\,.
\ea
\ee

We extend the sets of
operators $\bs_{\alpha\beta}$ and $l_{\alpha\beta}$ (\ref{operSL}) by the operators
\begin{align}
\bs^i{}^\bullet&=\bpsi^i_a\bkappa^\bullet_b\eta^{ab}\,,&
\bs_i{}_\bullet&=\bpsi_i^a\bkappa_\bullet^b\eta_{ab}\,,\\
\bs^i{}_\bullet&=\bpsi^i_a\bkappa_\bullet^a\,,&
\bs_i{}^\bullet&=\bpsi_i^a\bkappa^\bullet_a\,,&
\bu_\bullet=\bs_\bullet{}^\bullet=\bkappa_\bullet^a \bkappa^\bullet_a\,,
\end{align}
\begin{align}
l^i{}_\bullet&=\psi^i_a\bkappa_\bullet^a\,,&
l_i{}^\bullet&=\psi_i^a\bkappa^\bullet_a\,,&
l^\bullet{}_i&=\kappa^\bullet_a\bpsi_i^a\,,\\
l_\bullet{}^i&=\kappa_\bullet^a\bpsi^i_a\,,&
l^\bullet{}_\bullet&=\kappa^\bullet_a\bkappa_\bullet^a\,,&
l_\bullet{}^\bullet&=\kappa_\bullet^a\bkappa^\bullet_a
\end{align}
and introduce derivative operators
\begin{align}
D_i&=\psi_i^a\d_a\,,&
D^i&=\psi^i_a\d^a\,,&
D_\bullet&=\kappa_\bullet^a\d_a\,,&
D^\bullet&=\kappa^\bullet_a\d^a\,,\\
\bar{D}_i&=\bpsi_i^a\d_a\,,&
\bar{D}^i&=\bpsi^i_a\d^a\,,&
\bar{D}_\bullet&=\bkappa_\bullet^a\d_a\,,&
\bar{D}^\bullet&=\bkappa^\bullet_a\d^a\,.
\end{align}
Finally let $\nu_I$ denote the number of the last column of the $I$-th vertical
block. Recall that by $\mu_I$ we denote  the number of
the first column in $I$-th vertical block. Thus
\be
\nu_I = \mu_{I+1}-1\;,
\qquad
I=1,\ldots, N-1\,.
\ee

\subsection{Variation of the flat higher-spin action}

Taking into account the form of the function $\cE$ (\ref{functE}), the
variation of $\ads$ HS action constructed by the formula (\ref{actionFIN}) is
\begin{align}
\label{flat_variation}
\gd\cS_2\propto\int_{\cM^d}\Big(&\lvac(\wedge\Eop)^{d-2p-1}\,\chi \tilde{\cE}(\bu,\bta)\,
\bta^1\,\wedge \hat{r}_{(p+1)}\wedge \delta\breve{\omega}_{(p)}\rvac+\nn\\
{}+(-)^s&\lvac(\wedge\Eop)^{d-2p-1}\,\chi \tilde{\cE}(\bu,\bta)\,\bta_1\,\wedge\hat{\cR}_{(p+1)}\wedge
\delta\breve{e}_{(p)}\rvac\Big)\,.
\end{align}
Here Lorentz-covariant $p$-forms
\be\label{f_var}
\ba{c}
\gl^{s-1}\breve{e}_{(p)}\rvac=\bar{v}^1\cdots\bar{v}^{s-1}\breve{{\mathsf \Omega}}_{(p)}\rvac\,,
\\
\\
\gl^{s-2}\breve{\omega}_{(p)}\rvac=(1-v_1\bar{v}^1)\bar{v}^2\cdots
\bar{v}^{s-1}\breve{{\mathsf \Omega}}_{(p)}\rvac\
\ea
\ee
are the physical and relevant auxiliary fields (cf. (\ref{phys_field}),
(\ref{true_auxiliary})) and Lorentz-covariant $(p+1)$-forms
\be
\ba{c}
\gl^{s-1}\hat{r}_{(p+1)}\rvac=\bar{v}_1\cdots\bar{v}_{s-1}\hat{{\mathsf R}}_{(p+1)}\rvac\,,
\\
\\
\gl^{s-2}\hat{\cR}_{(p+1)}\rvac=(1-v^1\bar{v}_1)\bar{v}_2\cdots
\bar{v}_{s-1}\hat{{\mathsf R}}_{(p+1)}\rvac\;
\ea
\ee
are the Lorentz components of the curvature $R_{(p+1)}$ associated with the
physical and relevant auxiliary fields, respectively.
Note that the
factor of $\gl^{2s-1}$ in the variation (\ref{fieldEq}) is cancelled by those of
$\hat{r}_{(p+1)}\wedge\breve{\go}_{(p)}$ and
$\hat{\cR}_{(p+1)}\wedge \breve{e}_{(p)}$,
making the flat limit of the variation well defined.

The variation over the relevant auxiliary field $\go_{(p)}$ gives rise to the equation  of motion,
that can be written in the form
\be
\label{equation_w}
r^{a_1[\tilde{h}_1-1],\ldots,a_{s-1}[\tilde{h}_{s-1}-1];\,m[p+1]}=C^{a_1[\tilde{h}_1-1],
\ldots,a_{s-1}[\tilde{h}_{s-1}-1], m[p+1]}\,,
\ee
where the tensor $C^{a_1[\tilde{h}_1-1],\ldots,a_{s-1}[\tilde{h}_{s-1}-1], m[p+1]}(x)$
is either zero if $\tilde{h}_{s-1}=p+1$ or equals to the primary Weyl tensor if
$\tilde{h}_{s-1}> p+1$ \cite{ASV1} of Young symmetry type
\be\label{YC}
Y_{o(d-1,1)}(\underbrace{s,\ldots, s}_{p+1}, s_{p+2},\ldots,s_{\nu})\,.
\ee

The equation (\ref{equation_w}) expresses the relevant auxiliary field
$\go_{(p)}$ in terms of first derivative of the physical field $e_{(p)}$  up to pure gauge
degrees of freedom.
Using the extended Fock notations,  equation (\ref{equation_w}) can be written as
\be
\label{equation_w_fock}
\hat{r}_{(p+1)}\rvac=\hat{C}\rvac\,,
\ee
where the operator $\hat{C}$ is related to the tensor $C$ by (\ref{operF}).
It is convenient to use the $1.5$-formalism with the auxiliary field expressed implicitly
by virtue of its equation of motion (\ref{equation_w}) through the physical field.

To check the invariance under the flat gauge symmetries one needs the explicit form of the
physical
field transformations induced by the corresponding transformations (\ref{fl_gaug})
of the metric-like field
$\Phi$.
In this paper we  assume that the flat gauge parameters
$S_I\,,I=1\,, ...\, , N-1$
are traceless. Note that the gauge parameter $S_N$ associated with the vertical
block of the minimal height $p$ corresponds to the
physical $\ads$ gauge parameter $\varepsilon_{(p-1)}$ (\ref{gaugelaw}).
It follows that flat gauge transformations (\ref{fl_gaug}) of the metric-like field
$\Phi$ result from
the following transformations of the physical field $e_{(p)}$
\begin{align}
\label{di_e}
\gd_{I}\breve{e}_{(p)}\rvac &=\cP_{e} \Big(D^{\nu_I}\breve{{\mathsf S}}_{I}\Big)\rvac\,,
\end{align}
where the operator
$\cP_{e}=\cP_{e}(\psi,\bpsi)$ imposes the necessary trace and Young
symmetry conditions,  projecting to the traceless Young tableau
associated with tangent indices of the field $e_{(p)}$.
The explicit form of $\cP_{e}$ is
complicated but, fortunately, it is not  needed for our analysis.

Substituting (\ref{di_e}) into (\ref{flat_variation}) and neglecting terms
with the variation of the relevant auxiliary field by using the 1.5-order
formalism, one obtains for the gauge variation with the parameters $S_I$
\be
\label{variation_exi}
\gd_{I}\cS_2^{\rm flat}= \int_{\cM^d}\Big(\ga^1_I\Delta^1_{I}+\ga^2_{I}\Delta^2_{I}+
\ga^3_{I}\Delta^3_{I}\Big)\,,
\ee
where $\ga^{1,\,2,\,3}_{I}$ are some coefficients determined by the form
of the projector $\cP_e$ and  function $\cE$, and
\begin{align}
\label{deltas}
\Delta^1_{I}&=\frac{1}{p}\lvac(\Eop)^d\chi \cF_{I}(\bu)
\Big(\bar{D}_1\bar{D}_{\nu_I}\bu_\bullet\Big)\hat{\omega}_{(p)}
\breve{{\mathsf S}}_{i}\rvac\,,\nn\\ \nn\\
\Delta^2_{I}&=\lvac(\Eop)^d\chi \cF_{I}(\bu) \Big(\bar{D}^\bullet\bar{D}_1\bs_{\bullet\nu_I}\Big)
\hat{\omega}_{(p)}\breve{{\mathsf S}}_{i}\rvac\,,\\ \nn\\
\Delta^3_{I}&=\lvac(\Eop)^d\chi \cF_{I}(\bu) \Big(\bar{D}_{\nu_I}\bar D^\bullet\bs_{\bullet 1}\Big)
\hat{\omega}_{(p)} \breve{{\mathsf S}}_{i}\rvac\,\nn
\end{align}
with
\be
\cF_{I}(\bu)=\Big(\prod_{k=1,\,k\neq \nu_I}^{s-1}(\bu_k)^{\tilde{h}_k-1}\Big)
(\bu_{\nu_I})^{\tilde{h}_{\mu(i)}-2}(\bu_\bullet)^{p-1}\,.
\ee
The operators $\Delta_{I}^{1,\,2,\,3}$ represent various types of contractions between the relevant auxiliary field
$\omega_{(p)}$,  traceless flat gauge parameter $S_I$, and two flat space-time derivatives $\d$.
The important fact is that there are only three independent
contractions of the type  $\d\,\d\, \omega_{(p)}\,S_I$.

Indeed relevant auxiliary
field $\go_{(p)}$ has one additional index in the first column and one another index in the $\nu_I$-th
column compared with those of the gauge parameter $S_I$. The number of world indices of $\go_{(p)}$
equals the number of indices in $s$-th column of $S_I$, $I\neq N$. Due to
the Young symmetry and tracelessness conditions of $\go_{(p)}$ and $S_I$
all possible contractions of the type $\d\d\go_{(p)}S_I$ reduce to
\be
\d^{f_1}\d^{f_2}\go_{(p)\;g_1}{}^{a_1\ldots,\ldots,}
{}_{g_2}{}^{a_{\nu_I}\ldots,\ldots;f_3m\ldots}S_{I\;a_1\ldots,\ldots,
a_{\nu_I}\ldots,\ldots,g_3m\ldots}\,,
\ee
where $\cdots$ denotes column-to-column contractions and the indices $f$ and $g$ should be contracted
in all possible ways. Thus, there are three types
of contractions that are represented in Fock space notations by $\Gd_I^{1,\,2,\,3}$.
Using the ambiguity in adding total derivatives without loss of generality we can assume
 that flat derivatives act on $\go_{(p)}$.

\subsection{Proof of invariance}
\label{sec:sketch}

The straightforward  check of the invariance of the flat action,
\textit{i.e.} that $\gd_{I}\cS_2^{\rm flat}=0$, is  complicated
requiring explicit expressions for the relevant auxiliary field
$\omega_{(p)}=\omega_{(p)}(\d e_{(p)})$ and the coefficients $\ga^{1,\,2,\,3}_{I}$.
Fortunately, there is a simpler proof using some relations among
 the operators $\Delta_{I}^{1,\,2,\,3}$ and the coefficients $\ga^{1,\,2,\,3}_{I}$ which
result from Bianchi identities (\ref{bianchi}) and
 the manifest HS gauge symmetries (\ref{gauge}).

Consider the flat limit of the Bianchi identity for the
physical curvature $r_{(p+1)}$
\be
\label{bianchi2}
\extdiff r_{(p+1)} +\sigma^{(1)}_{-}\cR_{(p+1)}+... =0\;
\ee
with
\be
\label{lll}
r_{(p+1)}=\extdiff e_{(p)}+\sigma_-^{(1)}\omega_{(p)}+\cdots\;,
\qquad
\cR_{(p+1)}=\extdiff \go_{(p)}+\cdots\;.
\ee
Here $\sigma_-^{(1)}$ is the operator
that decreases a number of Lorentz indices
of the first vertical block (\ref{sigma-}) by one. Dots denote the
contributions of the  irrelevant auxiliary fields and extra fields
that can be discarded in the variation of the action by virtue of the
decoupling conditions.
In the extended Fock space notations the equations (\ref{bianchi2}) and
(\ref{lll}) read as
\be
\label{bianchi3}
D^\bullet \hat{r}_{(p+1)}\rvac+
\Big(\cP_{e}l^\bullet{}_1\hat{\cR}_{(p+1)}+\cdots\Big)\rvac=0\;,
\ee
\be\label{curv100}
\hat{r}_{(p+1)}\rvac=\Big(D^\bullet \hat{e}_{(p)}+
\cP_{e}l^\bullet{}_1\hat{\go}_{(p)}+\cdots\Big)\rvac\;,
\qquad
\hat{\cR_{(p)}}\rvac=\Big(D^\bullet \hat{\go}_{(p)}+\cdots\Big)\rvac\;.
\ee

Taking into account the equation of motion (\ref{equation_w_fock})
which is a constraint on the relevant auxiliary field,
{\it i.e.} applying the 1.5-order formalism,
 we obtain for (\ref{bianchi3})
\be\label{bi}
D^\bullet \hat{C}\rvac+\cP_{e}l^\bullet{}_1\hat{\cR}_{(p+1)}\rvac=0\,.
\ee
The application of the operator $l^i{}_\bullet l^j{}_\bullet$, $i,j=1,\ldots,s-1$ to
the left-hand-side of (\ref{bi}) annihilates the term with $\hat{C}$.

Indeed, at least one  of the operators $l$ necessarily acts on $\hat{C}$
in
$
l^i{}_\bullet l^j{}_\bullet D^\bullet \hat{C}\rvac
$
 antisymmetrizing one index of the $s$-th column with all indices
of the $i$-th or $j$-th column that is zero by the
 Young symmetry properties of $C$ (\ref{YC}).
It is important  that the
trace properties of the primary Weyl tensor are irrelevant in this analysis.

Taking into account (\ref{curv100}) one obtains the
relation
\be\label{toj_}
l^i{}_\bullet l^j{}_\bullet\cP_{e}l^\bullet{}_1D^\bullet\hat{\go}_{(p)}\rvac=0\,.
\ee
The derivative $D^\bullet$ and the projector $\cP_{e}$
commute to each other since they
are built of oscillators of different types, namely $\kappa^\bullet$ and $\psi, \bpsi$,
respectively.
Multiplying (\ref{toj_}) by the operator $\breve{{\mathsf S}}_I$
associated with the gauge parameter $S_I$, which
is built of the  oscillators that (anti)commute to
those in (\ref{toj_}),  one finally obtains that
\be\label{toj}
l^i{}_\bullet l^j{}_\bullet D^\bullet\cP_{e}l^\bullet{}_1\hat{\go}_{(p)}\breve{{\mathsf S}}_I
\rvac=0\,.
\ee

The identity (\ref{toj_}) expresses specificities of the form of the expression
of the relevant auxiliary field in terms  of the physical field
$\go_{(p)}(e_{(p)})$. {}From (\ref{toj}) we now derive
some useful relations on the operators $\Gd_I^{1,\,2,\,3}$\,.

Let us first consider the case where
either $I\neq 1$  or $I=1$, $m_1>1$ ({\it i.e.} the
first vertical block consists of more than one column). Acting on (\ref{toj})
with $i=j=1$ by the operators
\be\label{op_1}
\cF(\bar u)\bar D^\bullet \bs_{1\bullet}\bs_{1\nu_I}
\ee
and
\be\label{op_2}
\cF(\bar u)\bu_\bullet\bar D_1 \bs_{1\nu_I}\,,
\ee
one finds that
\be
\label{lin1}
\Big(\pi_{I}^{11}\Delta^1_{I}+\pi_{I}^{12}\Delta^2_{I}+\pi_{I}^{13}\Delta^3_{I}\Big)=0\,,
\ee
\be
\label{lin2}
\Big(\pi_{I}^{21}\Delta^1_{I}+\pi_{I}^{22}\Delta^2_{I}+\pi_{I}^{23}\Delta^3_{I}\Big)=0\,
\ee
with some sets of coefficients ${\bf \pi}_I^1=(\pi_{I}^{11}, \pi_{I}^{12}, \pi_{I}^{13})$ and
${\bf \pi}_I^2=(\pi_{I}^{21}, \pi_{I}^{22}, \pi_{I}^{23})$.

Let us show that vectors $\pi_I^1$ and $\pi_I^2$ are linearly independent, namely, that
$\pi_I^{11}=0$ while $\pi_I^{21}\neq 0$.
Indeed, the expression resulting from the combination  of  (\ref{op_1}) and (\ref{toj})
has the form
\be
\label{toj1}
\lvac (\Eop)^d\chi\frac{\cF(\bu)}{\bu_1}\bar D^\bullet\bs^1{}_\bullet\Big(
\bs_{1\bullet}\bar D_{\nu_I}+\bar D_1\bs_{\nu_I\bullet}
\Big)\cP_{e}l^\bullet{}_1\hat{\go}_{(p)}\breve{{\mathsf S}}_I\rvac=0\,,
\ee
One observes that it contains the derivative $\bar D^\bullet$ while
the operator $\Delta_{I}^1$
does not. Recall that contrary to the operator $\Delta_{I}^1$ the operators $\Delta_{I}^2$ and $\Delta_{I}^3$ contain
the derivative $\bar D^\bullet$ (\ref{deltas}).
Therefore, the operator $\Delta_{I}^1$ cannot contribute to (\ref{toj1}),
{\it i.e.} $\pi_{I}^{11}=0$.
One can show that (\ref{lin2}) $\pi_I^{21}\neq 0$
and the coefficients $\pi_I^{12}$ and $\pi_I^{13}$
are not both  zero.

As a result, it follows that operators $\Gd_I^{1,\,2,\,3}$ are
proportional to each other
\be
\Gd_I^1\propto\Gd_I^2\propto\Gd_I^3\,
\ee
so that the variation (\ref{variation_exi}) has the form
\be\label{v}
\gd_{I}\cS_2^{\rm flat}=\gga \int_{\cM^d}\Delta^1_{I}
\ee
with some coefficient $\gga$.

Now let us take into account that the
variation (\ref{v}) is invariant under  the gauge transformation
\be
\label{qwerty}
\delta\hat{\go}_{(p)}\rvac=D^\bullet \hat{\xi}_{(p-1)}\rvac
\ee
with the $(p-1)$-form gauge parameter $\xi_{(p-1)}$
that corresponds to the relevant auxiliary field.
Substituting (\ref{qwerty}) into (\ref{v}) one finds that
the coefficient $\gga$ in (\ref{v}) is
zero thus completing the proof of the fact
that the flat limit of the constructed action is invariant under the
additional flat gauge symmetries in the case $I\neq 1$ or $I=1$, $m_1\neq 1$.

Let us note that from the invariance of
the Bianchi identity (\ref{bianchi})  under the
gauge transformation (\ref{qwerty}) it follows that
\be\label{qsc}
\Gd_I^1=\Gd_I^2=\Gd_I^3\,.
\ee
Actually, taking into account obvious (anti)commutation relations of the operators
$\bs\,,l\,,\bar D$ and $D$ one finds that an
expression of the form
\be
\gb_I^1\Gd_I^1+\gb_I^2\Gd_I^2+\gb_I^3\Gd_I^3
\ee
is invariant under (\ref{qwerty}) provided that
\be\label{qaz}
\gb_I^1+\gb_I^2+\gb_I^3=0\,.
\ee
Along with
(\ref{lin1}), (\ref{lin2}) this implies  (\ref{qsc}).

The case of
$
I=1\,,\qquad m_1=1
$
is special  since the operators (\ref{op_1}) and
(\ref{op_2}) vanish in this case. From the definition of $\Gd_I^{1,\,2,\,3}$
one finds that in this case
\be
\Gd_1^1=0\,,\qquad \Gd_1^2=-\Gd_1^3\,.
\ee
Acting on (\ref{toj}) with $i=j=2$ by the operator
\be
\cF(\bu)\bar D^\bullet \bs_{1\bullet}l^1{}_2\bs_{12}
\ee
one finds that
\be
\frac{\cF(\bu)}{\bu_1}\bar D^\bullet \bar D_1\bs^1{}_\bullet \bs_{1\bullet}
\cP_{e_{(p)}}l^\bullet{}_1\hat{\go}_{(p)}\breve{S}_{(1)}\rvac=0\,,
\ee
from where it follows that
\be
\Gd_1^1=\Gd_1^2=\Gd_1^3=0\,.
\ee
This completes the proof of the invariance of the flat HS action under
enhanced flat gauge symmetries.

\section{Conclusions and outlook}

The  description of massless fields
in $(A)dS$ spacetime of any dimension in terms of gauge connections referred to as
frame-like approach is extended to generic bosonic massless fields.
Let us summarize some of the features of the frame-like formulation of
the bosonic massless HS field dynamics considered in this paper.

\begin{itemize}

\item A given HS massless field of any symmetry type propagating on the $\ads$ background of any dimension $d$
is described as $p$-form gauge field
$$
\Omega_{(p)}^I(x)
$$
that takes values in an appropriate finite-dimensional $o(d-1,2)$-module $I$ .
The module $I$ is some traceless $o(d-1,2)$ tensor representation described by
a Young tableau of the form uniquely defined by a spin of a HS field.

\item With the $p$-form gauge field $\Omega_{(p)}^I(x)$ one associates
gauge-invariant field strength
(curvature) in a standard  fashion as $R_{(p+1)}^I(x)= D_0 \Omega_{(p)}^I(x)$, where
$D_0$ is the of $o(d-1,2)$ covariant derivative that describes  $\ads$ background
via the flatness condition $D_0^2=0$.
The HS gauge symmetries are defined as
$\delta\Omega_{(p)}^I(x) = D_0\xi_{(p-1)}^I(x)$.

\item Being decomposed into $o(d-1,1) \subset o(d-1,2)$ components of the representation $I$,
the $p$-form gauge field reduces to the set of various
symmetry type  Lorentz-covariant fields that have
different dynamical roles.
In particular, the Lorentz-covariant field with a minimal number of
Lorentz tangent indices is the physical field generalizing the frame field in
gravity.

\item The manifestly gauge invariant free HS action functional is constructed in the
form of specific bilinear combinations of the curvatures
$$
{\cal S}_2 = \int R_{(p+1)}R_{(p+1)}\,.
$$
The coefficients in the action are fixed by
the decoupling conditions guaranteeing that the action
is free of higher derivatives and describes the correct number of
degrees of freedom associated with the physical HS field.

\item The flat limit of the $\ads$ theory exhibits the required flat
gauge symmetry enhancement thus providing
the consistency of the HS field dynamics both
in $(A)dS$ and in Minkowski space.

\item The reformulation of the action in terms of an auxiliary fermionic Fock
space  allows us to reduce the problem of reconstruction of a
 free field action to the analysis of an appropriate differential complex,
with the derivation $\cQ$ associated with the variation of the action.

\end{itemize}

Among possible directions for the future research let us mention the
following:

\begin{itemize}

\item the frame-like Lagrangian formulation for fermionic HS  massless fields;

\item the frame-like Lagrangian formulation for the partially massless fields
   of general symmetry type along the lines of \cite{SV};

\item the frame-like Lagrangian formulation for $(A)dS_{4k+1}$ (anti)selfdual
   bosonic and fermionic HS  massless  and partially massless fields;

\item the unfolded formulation of HS (partially) massless fields of mixed-symmetry type
      by working out a structure of the infinite-dimensional modules associated with
       the generalized Weyl tensors.
\end{itemize}

\vspace{5mm}

Accomplishment of this programme  will provide the full
identification of connections, that take values in different finite-dimensional
representations of the $(A)dS$ algebra $(o(d-1,2))o(d,1)$, with the different types
of relativistic fields. Because the frame-like geometric approach makes
 global and local HS symmetries manifest, it plays a fundamental role for
understanding a structure of consistent global  HS symmetries and, at the later stage,
of nonlinear HS gauge theories. Finally, a very interesting direction
that may be important for understanding fundamental symmetries of
string theory would be to extend the proposed formulation to massive HS
fields.


\section*{Appendix}
\addcontentsline{toc}{section}{Appendix}
\label{sec:conc}
\setcounter{equation}{0}\renewcommand{\theequation}{A.\arabic{equation}}

Consider  an expression of the form
$$
\label{M}
\lvac(\wedge\Eop)^{d-m-n}\chi {\mathsf M} \wedge\hat{{\mathsf A}}_{(m)}\wedge \breve{{\mathsf B}}_{(n)}\rvac
$$
defined with respect to any function ${\mathsf M}={\mathsf M}(\bu, \bta,\bv)$ and any
$\hat{{\mathsf A}}_{(m)}, \breve{{\mathsf B}}_{(n)}$
(for definiteness, let these forms have tangent indices described by Young tableau (\ref{pic:ads_YT})).
Then, up to weakly zero terms, the following identity holds
\be
\label{A1}
\ba{c}
\dps  \bs_i{}^j\;{\mathsf M}(\bu, \bta,\bv) \sim \delta_{ij}\,\bu_i\;{\mathsf M}(\bu, \bta,\bv)\,
\\
\\
\dps
- \theta(j-i-1) (\bu_i\frac{\d}{\d \bu_i})^{-1}(\bta^j\frac{\d}{\d \bta^i} +\bv^j\frac{\d}{\d \bv^i})\bu_i\;{\mathsf M}(\bu, \bta,\bv)
\\
\\
\dps
- \theta(i-j-1)(\bu_j\frac{\d}{\d \bu_j})^{-1}(\bta_i\frac{\d}{\d \bta_j} +\bv_i\frac{\d}{\d \bv_j})\bu_j\;{\mathsf M}(\bu, \bta,\bv)\;,

\ea
\ee
where
$$
\theta(n) = \left\{ \ba{l}
1\,,\;\;{n\geq 0} \,, \\ 0\,,\;\;{n<0}\,.
\ea \right.
$$

Let the function ${\mathsf M}$ have the special form
$$
{\mathsf M}(\bs, \bta,\bv)= {\mathsf M}(\bu, n)\; \bta_1\, \bta_k \,\frac{\d}{\d \bv_k}(\bv_2 ... \bv_{s-1}\,\bv^1 ... \bv^{s-1})\;,
$$
where ${\mathsf M}(\bu,n)$ is arbitrary. Then, up to weakly zero terms,
the following identities hold within the interval $\mu_I<k<\mu_{I+1}$ (i.e., for all $k$ enumerating
columns of the $I$-th rectangular block  with exception of the first column):

\be
\label{A2}
 {\mathsf M}(\bu, n)\; \bta_1\, \bta_k \,\frac{\d}{\d \bv_k}(\bv_2 ... \bv_{s-1}\,\bv^1 ... \bv^{s-1})
\sim \Big(\frac{1}{\bar{N}_{I}+1}{\mathsf M}(\bu, n)\Big)\; \bta_1\, \bta_{\mu_I} \,\frac{\d}{\d \bv_{\mu_I}}(\bv_2 ... \bv_{s-1}\,\bv^1 ... \bv^{s-1})\;,
\ee

\be
\label{A3}
 {\mathsf M}(\bu, n)\; \bta^1\, \bta_k \,\frac{\d}{\d \bv_k}(\bv_1 ... \bv_{s-1}\,\bv^2 ... \bv^{s-1})
\sim
\Big(\frac{1}{\bar{N}_{I}+1}{\mathsf M}(\bu, n)\Big)\; \bta^1\, \bta_{\mu_I} \,\frac{\d}{\d \bv_{\mu_I}}(\bv_1 ... \bv_{s-1}\,\bv^2 ... \bv^{s-1})\;,
\ee

\be
\label{A4}
 {\mathsf M}(\bu, n)\; \bta_1\, \bta^k \,\frac{\d}{\d \bv^k}(\bv_2 ... \bv_{s-1}\,\bv^1 ... \bv^{s-1})
\sim
\Big( \frac{1}{\bar{N}_{I}+1}{\mathsf M}(\bu, n)\Big)\; \bta^1\, \bta^{\mu_I} \,\frac{\d}{\d \bv^{\mu_I}}(\bv_2 ... \bv_{s-1}\,\bv^1 ... \bv^{s-1})\;.
\ee


\begin{center}
{\large\textbf{Erratum:\; Frame-like formulation for free mixed-symmetry
bosonic massless higher-spin fields in $AdS_d$}}

\end{center}
\begin{center}
hep-th/0601225
\end{center}
\begin{center}
K.B. Alkalaev, O.V. Shaynkman and M.A. Vasiliev
\end{center}


In this paper and in hep-th/0501108 the Lagrangian formulation of mixed-symmetry
field dynamics in $AdS_d$ space-time was proposed.  Our
construction contained  the conjecture expressed by the formula (\ref{equation_w})
of Section 8 that torsion-like components
of the linearized curvature are zero. However, recently, it was shown that for fields of general
type this conjecture is not true \cite{Skvortsov:2009nv}.
Hence, we cannot claim that the action proposed in this paper works
properly for general massless mixed symmetry fields.
At this stage we can only claim that it does work for the particular
class of mixed symmetry fields described by rectangular Young diagrams of
arbitrary length $s$ and height $h_1\leq [(d-1)/2]$ and
fields with spins described by Young diagrams composed of two horizontal
rectangular blocks, the upper block is of arbitrary length $s$ and height
$h_1$, the second block is a column of height $h_2$ provided $h_1+h_2 \leq [(d-1)/2]$.
In the first case  equation (8.16) is true because torsion-like components
of the linearized curvature are absent. In the second case these components are non-zero but
nonetheless
they are consistently eliminated
by virtue of Bianchi identities. The analysis of the general case is more
complicated and requires further investigation.

Let us stress that despite the aforementioned problems,
the set of gauge fields, their gauge
symmetries and linearized curvatures suggested  in our paper provide correct
setting for the analysis of on-shell higher spin dynamics, as was checked in particular in
\cite{Boulanger:2008up,Skvortsov:2009zu}.

\vspace{3mm}

We thank N. Boulanger and E. Skvortsov for useful  communications
and discussions.

\providecommand{\href}[2]{#2}\begingroup\raggedright
\addtolength{\baselineskip}{-3pt} \addtolength{\parskip}{-1pt}


\begin{thebibliography}{99}
\addcontentsline{toc}{section}{References}

\parindent=0pt
\parskip=0pt


\bibitem{ASV1} K.B. Alkalaev, O.V. Shaynkman and  M.A. Vasiliev,
Nucl.\ Phys.\ B {\bf 692} (2004) 363,
\href{http://arXiv.org/abs/hep-th/0311164}{{\tt hep-th/0311164}}.

\bibitem{ASV2}  K.B.Alkalaev, O.V. Shaynkman and M.A. Vasiliev, JHEP 0508:069 (2005),
\href{http://arXiv.org/abs/hep-th/0501108}{{\tt hep-th/0501108}}.

\bibitem{fronsdal_flat}
C.~Fronsdal, Phys.\ Rev.\ D {\bf 18} (1978) 3624.
\bibitem{fronsdal_ads}
C.~Fronsdal, Phys.\ Rev.\ D {\bf 20} (1979) 848.


\bibitem{WF} B. de Wit and D.Z. Freedman,  Phys. Rev.  {\bf D21}
(1980) 358.

\bibitem{Curtright:1979uz}
T.~Curtright,
Phys.\ Lett.\ B {\bf 85} (1979) 219.

\bibitem{vas_yadfiz} M.A. Vasiliev, Yad.\ Fiz.\  {\bf 32} (1980) 855.

\bibitem{ardes}C.~Aragone and S.~Deser,
Nucl.\ Phys.\ B {\bf 170} (1980) 329.

\bibitem{V1} M.A. Vasiliev, Fortsch.\ Phys.\ {\bf 35} (1987) 741.

\bibitem{LV}
V.E.~Lopatin and M.A.~Vasiliev, Mod.\ Phys.\ Lett.\ A {\bf 3}
(1988) 257.

\bibitem{vf} M.A. Vasiliev, Nucl.~Phys.  B \textbf{301} (1988) 26.

\bibitem{Siegel:1986zi}
W.~Siegel and B.~Zwiebach, Nucl.\ Phys.\ B {\bf 282} (1987) 125.

\bibitem{Pashnev}
A.~Pashnev and M.~M.~Tsulaia, Mod.\ Phys.\ Lett.\ A {\bf 12}
(1997) 861, \href{http://arXiv.org/abs/hep-th/9703010}{{\tt
hep-th/9703010}}; Mod.\ Phys.\ Lett.\ A {\bf 13} (1998) 1853,
\href{http://arXiv.org/abs/hep-th/9803207}{{\tt hep-th/9803207}}.


\bibitem{francia}
D. Francia and  A. Sagnotti, Phys. Lett. B \textbf{543} (2002)
303, \href{http://arXiv.org/abs/hep-th/0207002}{{\tt
hep-th/0207002}}; Class. Quant. Grav. {\bf 20} (2003) S473, {\tt hep-th/0212185};
  Phys.\ Lett.\ B {\bf 624} (2005) 93, \href{http://arXiv.org/abs/hep-th/0507144}{{\tt
hep-th/0507144}}; ``Higher-Spin Geometry and String Theory'', {\tt hep-th/0601199}.


\bibitem{SS1}
E.~Sezgin and P.~Sundell, JHEP {\bf 0109}, 036 (2001), \href{http://arXiv.org/abs/hep-th/0109067}{{\tt hep-th/0105001}}.

\bibitem{alk1} K.B. Alkalaev, Phys.\ Lett.\ B {\bf 519} (2001) 121,
\href{http://arXiv.org/abs/hep-th/0107040}{{\tt hep-th/0107040}}.

\bibitem{Buchbinder:2001bs}
I.L.~Buchbinder, A.~Pashnev and M.~Tsulaia,
Phys.\ Lett.\ B {\bf 523} (2001) 338,
\href{http://arXiv.org/abs/hep-th/0109067}{{\tt hep-th/0109067}}.

\bibitem{Buchbinder:2004gp}
I.~L.~Buchbinder, V.~A.~Krykhtin and A.~Pashnev, Nucl. Phys. B \textbf{711} (2005) 367,
\href{http://arXiv.org/abs/hep-th/0410215}{{\tt hep-th/0410215}}.




\bibitem{AKO} C.~S.~Aulakh, I.~G.~Koh and S.~Ouvry,
Phys. Lett. B \textbf{173} (1986) 284.


\bibitem{LabastidaMorris} J.M.F. Labastida and T.R. Morris,
Phys. Lett. B {\bf 180} (1986) 101.

\bibitem{Labastida2} J.M.F. Labastida,
Phys. Rev. Lett. \textbf{58} (1987) 531; Nucl. Phys. B \textbf{322} (1989) 185.


\bibitem{Pashnev2}
C. Burdik, A. Pashnev and M. Tsulaia,
Mod.\ Phys.\ Lett.\ A {\bf 16} (2001) 731,
\href{http://arXiv.org/abs/hep-th/0101201}{{\tt hep-th/0101201}};
Nucl.\ Phys.\ Proc.\ Suppl.\  {\bf 102} (2001) 285,
\href{http://arXiv.org/abs/hep-th/0103143}{{\tt hep-th/0103143}}.

\bibitem{Evans}
 N.~T.~Evans,
J.\ Math.\ Phys.\ {\bf 8},  170 (1967).

\bibitem{Metsaev}
R.R. Metsaev, Phys.\ Lett. B \textbf{354} (1995) 78;
Phys.\ Lett.\ B {\bf 419} (1998) 49,
\href{http://arXiv.org/abs/hep-th/9802097}{{\tt hep-th/9802097}};
Talk given at International Seminar on Supersymmetries and Quantum
Symmetries (Dedicated to the Memory of Victor I. Ogievetsky),
Dubna, Russia, 22-26 Jul 1997,
\href{http://arXiv.org/abs/hep-th/9810231}{{\tt hep-th/9810231}}.

\bibitem{BMV} L.~Brink, R.R.~Metsaev and M.A.~Vasiliev,
Nucl.\ Phys.\ B {\bf 586} (2000) 183,
\href{http://arXiv.org/abs/hep-th/0005136}{{\tt hep-th/0005136}}.


\bibitem{DW}
  S.~Deser and A.~Waldron,
  Phys.\ Rev.\ Lett.\  {\bf 87}, 031601 (2001)
  [arXiv:hep-th/0102166];
   Nucl.\ Phys.\ B {\bf 607} (2001) 577
   [arXiv:hep-th/0103198];
  Phys.\ Lett.\ B {\bf 508}, 347 (2001)
  [arXiv:hep-th/0103255];
  Phys.\ Lett.\ B {\bf 513}, 137 (2001)
  [arXiv:hep-th/0105181];
   Nucl.\ Phys.\ B {\bf 662} (2003) 379
   [arXiv:hep-th/0301068];
   ``Conformal invariance of partially massless higher spins'',
   arXiv:hep-th/0408155.

\bibitem{Zin}
   Y.~M.~Zinoviev,
   ``On massive high spin particles in (A)dS'',
   arXiv:hep-th/0108192.


\bibitem{SV} E.D. Skvortsov and M.A. Vasiliev, hep-th/0601094.

\bibitem{bekaert} X. Bekaert and N. Boulanger, Comm. Math. Phys. \textbf{245} (2004) 27,
\href{http://arXiv.org/abs/hep-th/0208058}{{\tt hep-th/0208058}}.

\bibitem{bekaert2}
X.~Bekaert and N.~Boulanger, Phys.\ Lett.\ B {\bf 561} (2003) 183,
\href{http://arXiv.org/abs/hep-th/0301243}{{\tt hep-th/0301243}}.

\bibitem{MedHull} P.~de Medeiros and  C.~Hull, JHEP {\bf 0305} (2003) 019,
\href{http://arXiv.org/abs/hep-th/0303036}{{\tt
hep-th/0303036}}.


\bibitem{SS2}
E. Sezgin and P. Sundell, JHEP {\bf 0109}, 025 (2001),
\href{http://arXiv.org/abs/hep-th/0203115}{{\tt hep-th/0107186}}.

\bibitem{siegel}
T.~Biswas and W.~Siegel, JHEP {\bf 0207} (2002) 005,
\href{http://arXiv.org/abs/hep-th/0203115}{{\tt hep-th/0203115}}.

\bibitem{Zinoviev}
Y.M. Zinoviev, ``On massive mixed symmetry tensor fields in
Minkowski space and (A)dS",
\href{http://arXiv.org/abs/hep-th/0211233}{{\tt hep-th/0211233}};
``First Order Formalism for Mixed Symmetry Tensor Fields",
\href{http://arXiv.org/abs/hep-th/0304067}{{\tt hep-th/0304067}}.


\bibitem{Metsaev_d5}
R.R.~Metsaev, Phys.\ Lett.\ B {\bf 531} (2002) 152,
\href{http://arXiv.org/abs/hep-th/0201226}{{\tt hep-th/0201226}}.

\bibitem{A2} K.B. Alkalaev, Theor.\ Math.\ Phys.\  {\bf 140} (2004) 1253
[Teor.\ Mat.\ Fiz.\  {\bf 140} (2004) 424],
\href{http://arXiv.org/abs/hep-th/0311212}{{\tt hep-th/0311212}}.

\bibitem{deMedeiros:2003px}
P.~de Medeiros,
Class.\ Quant.\ Grav.\  {\bf 21} (2004) 2571
\href{http://arXiv.org/abs/hep-th/0304067}{{\tt hep-th/0311254}}.

\bibitem{Alkalaev:2005kt}
  K.~B.~Alkalaev, ``Mixed-symmetry massless gauge fields in AdS(5)'',
  arXiv:hep-th/0501105.


\bibitem{MM}
S.W.~MacDowell and F.~Mansouri, Phys.\ Rev.\ Lett.\  {\bf 38}
(1977) 739 [Erratum-ibid.\  {\bf 38} (1977) 1376]; F. Mansouri,
Phys. Rev. D \textbf{16} (1977) 2456.

\bibitem{fradvas} E.S. Fradkin and M.A. Vasiliev,
Phys. Lett. B {\bf 189} (1987) 89; Nucl. Phys. B {\bf 291} (1987)
141.

\bibitem{nonlineq} M.~A.~Vasiliev, Phys.\ Lett.\ B {\bf 243} (1990) 378; Phys.\ Lett.\ B {\bf 285} (1992) 225;
Phys.\ Lett.\ B {\bf 567} (2003) 139 [arXiv:hep-th/0304049].

\bibitem{V_obz2} M.A. Vasiliev,
``Higher Spin Gauge Theories: Star-Product and AdS Space";
Contributed article to Golfand's Memorial Volume``Many faces of
the superworld", ed. by M.Shifman, World Scientific Publishing Co
Pte Ltd, Singapore, 2000,
\href{http://arXiv.org/abs/hep-th/9910096}{{\tt hep-th/9910096}};
Int.\ J.\ Mod.\ Phys.\ D {\bf 5} (1996) 763,
\href{http://arXiv.org/abs/hep-th/9611024}{{\tt hep-th/9611024}}.

\bibitem{V_obz3} X. Bekaert, S. Cnockaert, C. Iazeolla and M. A. Vasiliev,
 \href{http://arXiv.org/abs/hep-th/0503128}{{\tt hep-th/0503128}}.

\bibitem{Flato:1986uh}
  M.~Flato and C.~Fronsdal,
  Commun.\ Math.\ Phys.\  {\bf 108} (1987) 469.


\bibitem{SW} K. Stelle and P. West, Phys.\ Rev.\ D {\bf 21}
(1980) 1466.

\bibitem{compensator} C.R. Preitschopf and M.A. Vasiliev, The Superalgebraic Approach to
Supergravity, in Proceedings of 31st International Ahrenshoop
Symposium On The Theory Of Elementary Particles, Berlin, Wiley-VCH,
1998, 496p, \href{http://arXiv.org/abs/hep-th/9805127}{{\tt
hep-th/9805127}}.


\bibitem{VD5}
M.~A.~Vasiliev, Nucl.\ Phys.\ B {\bf 616} (2001) 106,
\href{http://arXiv.org/abs/hep-th/0109067}{{\tt hep-th/0106200}}.

\bibitem{BarutRonchka} A.O. Barut and R. Raczka,
Theory of Group Representations and Applications
(PWN-Polish Scientific Publishers, Warszawa 1977)

\bibitem{wigner} E.~P.~Wigner,
Annals Math.\  {\bf 40} (1939) 149 [Nucl.\ Phys.\ Proc.\ Suppl.\  {\bf 6} (1989) 9].

\bibitem{dirac}
  P.~A.~M.~Dirac,
  Can.\ J.\ Math.\  {\bf 2} (1950) 129.

\bibitem{GT}
  D.~M.~Gitman and I.~V.~Tyutin,
  ``Quantization Of Fields With Constraints'', Berlin, Germany: Springer (1990) 291 p. (Springer series in nuclear and particle physics).



\end{thebibliography}

\begin{thebibliography}{99}





\bibitem{Skvortsov:2009nv}
  E.~D.~Skvortsov,
  JHEP {\bf 1001} (2010) 106
  [arXiv:0910.3334 [hep-th]].



\bibitem{Boulanger:2008up}
N.~Boulanger, C.~Iazeolla and P.~Sundell,
JHEP {\bf 0907} (2009) 013, [arXiv:0812.3615 [hep-th]];
JHEP {\bf 0907} (2009) 014, [arXiv:0812.4438 [hep-th]].




\bibitem{Skvortsov:2009zu}
E.~D. Skvortsov, {\em J. Phys.} {\bf A42} (2009) 385401,
  [arXiv:0904.2919 [hep-th]].



\end{thebibliography}
\end{document}